\documentclass[useAMS,usenatbib]{mn2e}

\usepackage{lscape}
\usepackage{longtable}
\usepackage{multicol}
\usepackage{graphicx}
\usepackage{url}
\pdfoutput=1
\title{The Violent Youth of Bright and Massive Cluster Galaxies and their Maturation over 7 Billion Years}
\author[B. Ascaso et al.]{B. Ascaso$^{1}$\thanks{E-mail:
ascaso@iaa.es},  B.C. Lemaux$^{2}$, L.M. Lubin$^{3}$, R.R. Gal$^{4}$, D.D. Kocevski$^{5}$,  \newauthor N. Rumbaugh$^{3}$, \& G. Squires$^{6}$\\
$^{1}$Instituto de Astrof\'isica de Andaluc\'ia, Glorieta de la Astronom\'ia s/n, 18008, Granada, Spain\\
$^{2}$Aix Marseille Universit$\acute{e}$, CNRS, LAM (Laboratoire d'Astrophysique de Marseille) UMR 7326, 13388, Marseille, France \\
$^{3}$Department of Physics, University of California, Davis, One Shields Ave., Davis, CA 95616, USA \\
$^{4}$ University of Hawai'i, Institute for Astronomy, 2680 Woodlawn Drive, Honolulu, HI 96822, USA \\ 
$^{5}$ University of Kentucky, Department of Physics \& Astronomy, 505 Rose Street, Lexington, KY 40506, USA \\
$^{6}$ Spitzer Science Center, California Institute of Technology, M/S 220-6, 1200 E. California Blvd., Pasadena, CA 91125, USA}
\begin{document}

\date{Accepted . Received }


\maketitle

\label{firstpage}

\begin{abstract}


In this study we investigate the formation and evolution mechanisms of the brightest cluster galaxies (BCGs) over cosmic time. At high redshift ($z\sim0.9$), we selected BCGs and most massive cluster galaxies (MMCGs) from the Cl1604 supercluster and compared them to low-redshift  ($z\sim0.1$) counterparts drawn from the MCXC meta-catalog, supplemented by SDSS imaging and spectroscopy. We observed striking differences in the morphological, color, spectral, and stellar mass properties of the BCGs/MMCGs in the two samples. High-redshift BCGs/MMCGs were, in many cases, star-forming, late-type galaxies, with blue broadband colors, properties largely absent amongst the low-redshift BCGs/MMCGs. The stellar mass of BCGs was found to increase by an average factor of $2.51\pm0.71$ from $z\sim0.9$ to $z\sim0.1$. Through this and other comparisons we conclude that a combination of major merging (mainly wet or mixed) and \emph{in situ} star formation are the main mechanisms which build stellar mass in BCGs/MMCGs. The stellar mass growth of the BCGs/MMCGs also appears to grow in lockstep with both the stellar baryonic and total mass of the cluster. Additionally, BCGs/MMCGs were found to grow in size, on average, a factor of $\sim3$, while their average S\'ersic index increased by $\sim$0.45 from $z\sim0.9$ to $z\sim0.1$, also supporting a scenario involving major merging, though some adiabatic expansion is required. These observational results are compared to both models and simulations to further explore the implications on processes which shape and evolve BCGs/MMCGs over the past $\sim$7 Gyr.

\end{abstract}


\section{Introduction}

Brightest Cluster Galaxies (BCGs) are the largest ($r_{e}>10$ kpc) and most massive  ($\mathcal{M}_s \ge 10^{12}\mathcal{M}_{\odot}$) galaxies in the 
universe. A number of early studies at lower redshift ($z\sim0.1$, e.g. \citealt{vonderlinden07,liu09}) have reported a largely homogeneous set of properties for these galaxies. Such galaxies are typically located at the center of the cluster potential well, exhibit early type morphologies, sometimes with a large halo, and contain old stellar populations. 

Early formation and evolution theories of these galaxies, such as galactic cannibalism (e.g. \citealt{ostriker75}), cooling flows (e.g. \citealt{fabian94}), tidal stripping  (e.g. \citealt{gallagher72}), or rapid merging of galaxies during cluster collapse (e.g. \citealt{merritt85}) were proposed. At present, the hierarchical structure formation scenario is widely accepted.  \cite{delucia07} used the Millenium Run simulation \citep{springel05}  to show that BCGs could be formed simply through a combination of major and minor merging. This simulation reproduced the color evolution consistent with a passively evolving stellar population formed at high redshift ($z_f \sim$2-5). However, new works have reported cases where BCGs possess bluer stellar populations than the typical red galaxies in the cluster and, in addition, are sometimes observed with high star formation rates  \citep{wen11,liu12,postman12}. While efforts have been made to improve simulations to account for these observations (e.g., \citealt{tonini12}), the  main mechanisms needed to produce these properties are not fully understood. 
 
Complementarily, a vast number of results have been published in the last decade about the size increase in massive ellipticals (e.g., \citealt{daddi05,trujillo06,buitrago08,vikram09,vandokkum10,ryan12,postman12,huertas-company13}), in which a size increase of a factor of $\ga2$ is observed from redshifts $\ga1$ to the present day. Other works have measured the size evolution only in BCGs, and found larger rates of increase compared to z$\sim 0$ BCGs (e.g., $\sim2$ from $z\sim0.5$, \citealt{nelson02}; $\sim1.7$ from $z\sim0.25$, \citealt{bernardi09}; $\sim2$ from $z\sim0.6$, \citealt{ascaso11}). Different mechanisms have been suggested from numerical simulations \citep{conroy07,ruszkowski09,hopkins10,dubois13,shankar13} and from observations to explain these size increases, such as major/minor mergers \citep{bernardi07,vonderlinden07,liu09,edwards12,liu13,burke13,lidman13} or adiabatic expansion 
\citep{fan08,collins09,stott11,ascaso11}. In a recent study of 160 BCGs spanning a large range in redshift ($0.03<z<1.63$), \citet{lidman12} found that BCGs increase in stellar mass considerably from $z\sim1$ to the present day, appealing to major dry merging as the primary transformative mechanism. As such, the picture is far from complete, and the conditions that arise in cluster environments which allow such mechanisms to be efficient, as well as the epoch in which these mechanisms are most instrumental in transforming BCGs/MMCGs in clusters, is still far from understood. 

In this paper we compare the BCGs/MMCGs in the constituent clusters and groups of the Cl1604 supercluster \citep{oke98,lubin00,gal04,gal08} at $z\sim0.9$ to a matched sample of BCGs/MMCGs drawn from local clusters with similar spectroscopic coverage and photometric data. We observe striking differences between the high- and low-redshift samples with respect to the BCG/MMCG morphology, color, spectral properties, and luminosity gaps. Using the 531 spectroscopically confirmed members of the Cl1604 supercluster we investigate the amount of stellar mass surrounding the Cl1604 BCGs/MMCGs and contrast this to that of the low-redshift clusters. These comparisons, in conjunction with comparisons of our observational results to semi-analytic models and numerical simulations, allow us to isolate the mechanisms responsible for the growth of stellar mass and size of BCGs/MMCGs from $z\sim0.9$ to the present day. 

The structure of the paper is as follows. In Section 2 we describe the observational datasets used in this paper along with the method used to derive secondary parameters from the photometry. Section 3 is devoted to the analysis of the BCG/MMCG properties, such as the cluster stellar mass distribution, the color-magnitude diagram of the parent clusters/groups, and the color, size, spectral, and morphological evolution of the BCGs/MMCGs in our samples. In Section 4, we compare and contrast our results against the backdrop of simulations and discuss the implications of our results. Section 5 contains the final conclusions of the paper. Throughout this paper we adopt $H_0$=70 km s$^{-1}$ Mpc$^{-1}$, $\Omega_M$ =0.27, $\Omega_{\Lambda}$=0.73. All magnitudes are given in the AB system \citep{oke83,fukugita96} and all equivalent width measurements are presented in the rest-frame.

\section{Data}

In this work, we consider the BCGs and MMCGs of the $z\sim0.9$ Cl1604 supercluster, which contains clusters and groups that span a wide range in halo mass, and a comparison sample of comparable clusters at low-redshift ($z\sim0.1$) observed as part of the Sloan Digital Sky Survey (SDSS). In this section we describe the characteristics of both samples as well as the observational datasets available for each. 

\subsection{The Cl1604 supercluster at $z\sim0.9$}
\label{cl1604}

The Cl1604 supercluster, located at $z\sim0.9$, is one of the most well-characterized superclusters in  the high-redshift universe. As part of the Observations of Redshift Evolution in Large Scale Environment survey (ORELSE; \citealt{lubin09}), the galaxy populations of both the constituent clusters and groups of the Cl1604 supercluster as well as that of the intermediate density environments connecting the various structures have been studied in detail \citep{gal08,lemaux09,kocevski09a,kocevski09b,lemaux10,kocevski11a,kocevski11b,rumbaugh12,lemaux12}. The galaxies that comprise the Cl1604 complex reside in a large range of environments that span from massive, virialized clusters to small groups and chains of starbursting and active galaxies located in environments with densities comparable to that of typical field galaxies at $z\sim1$. Similarly, the clusters and groups housed in the supercluster have markedly different properties, ranging in velocity dispersion from $300-800$ km $\rm{s}^{-1}$ and exhibiting a large variety of evolutionary states (see \citealt{lemaux12}). As such, the Cl1604 supercluster is an ideal structure to investigate the transformation of massive cluster and group galaxies at high redshift. 

The wealth of observations available for the Cl1604 supercluster have been described in depth elsewhere (\citealt{gal08,kocevski09a,kocevski11a,lemaux12}). As such, we only briefly summarize here those observations which are utilized in this study. Ground-based imaging in six bands ($Vr^{\prime} i^{\prime} z^{\prime} JK$) was obtained on the entire Cl1604 field with a variety of different telescopes (Subaru, Palomar Hale 5-m, \& UKIRT). Accompanying deep \emph{Spitzer} InfraRed Array Camera (IRAC; \citealt{fazio04}) imaging was obtained in four bands (3.6/4.5/5.8/8.0$\mu$m). This 10-band imaging was primarily used in this study to calculate stellar masses for the Cl1604 member\footnote{When referring to specific clusters or groups, a ``member galaxy" throughout the paper is defined as a galaxy at a projected distance $R<2R_{\rm{vir}}$ from the center of
a given cluster/group and $\delta v<3\sigma_{v}$, where $\sigma_{v}$ is the line-of-sight velocity dispersion measured for that cluster/group. This is different than a Cl1604 member galaxy, referred to here, which is defined simply as a galaxy within the redshift range $0.84<z<0.96$.} galaxies, as described in detail in \citet{lemaux12}. We adopt the r.m.s. difference between masses derived from our SED-fitting process and those derived using our $K$-band imaging (0.23 dex) as an estimate of the uncertainty in our stellar masses. While this underestimates systematic uncertainties resulting from our choice of stellar templates, $\mathcal{M}/L$ ratios, and initial mass function (IMF), these uncertainties are of little interest for this study as the mass comparisons made for Cl1604 galaxies are relative and the conclusions drawn from these comparisons are unaffected by the mass ``zero point''. This will also be true when we compare the Cl1604 galaxies to those selected by SDSS at low-redshift, as the mass fitting process for the SDSS galaxies is similar 
to the one chosen for the Cl1604 galaxies. 

The primary datasets of interest for this paper are the Hubble Space Telescope (\emph{HST}) Advanced Camera for Surveys (ACS; \citealt{ford98}) imaging and the comprehensive Keck I/II spectroscopic campaign undertaken in the Cl1604 field. The \emph{HST} data consist of a 17-pointing mosaic in the $F606W$ and $F814W$ bands, covering a large fraction of the constituent members and reaching 5$\sigma$ point source completeness limits of 27.2 \& 26.8 mags, respectively, in the shallowest regions. Through the combination of the Low-Resolution Imaging Spectrometer (LRIS; \citealt{oke95}) and the DEep Imaging Multi-Object Spectrograph (DEIMOS; \citealt{Faber03}) on the Keck I/II telescopes, nearly 2500 spectra have been obtained in the field, resulting in the spectroscopic confirmation of 531 member galaxies (for details see \citealt{lemaux12}). In the vicinity of the Cl1604 clusters and groups ($R<2R_{\rm{vir}}$) the spectroscopic data are roughly complete to a limit of $F814W\sim23.5$ for red-sequence galaxies, corresponding to a rest-frame limit of $M_{g\prime} =-20.35$\footnote{This limit is calculated by $k$-correcting the observed-frame $F814W$ band at $z=0.9$ to the rest-frame $g^{\prime}$ band using a 5 Gyr old elliptical template from \citet{maraston05}, broadly appropriate for our red-sequence galaxies. No evolutionary $k$-correction was applied.}. 

For each cluster or group the BCG\footnote{Here and throughout the paper the terms ``BCG'' and ``MMCG'' will be used for groups as well as clusters in order to bring the acronym down to a reasonable size.} was selected as the member galaxy with the brightest $F814W$ magnitude within $r_{\rm{proj}}< 1 h_{70}^{-1}$ Mpc of the luminosity-weighted cluster/group center (see \S\ref{clustermass} for a definition of this center). In principle, selecting a BCG in the $K$ band rather than the $F814W$ band is preferable, as the latter can be sensitive to star-formation processes at these redshifts, which may lead to biases in the BCG selection. However, because of the drastic difference in precision between our \emph{HST} ACS imaging and that of our UKIRT $K$-band imaging and in order to be consistent with the method used to select BCGs at low-redshift, we chose to exclusively rely on $F814W$ magnitudes to select the BCG. In practice, for six of the eight groups and clusters in Cl1604, the BCG selected in the $F814W$ band is either identical or consistent within the photometric errors to the BCG selected in the $K$ band. In addition, we trivialize issues related to the selection of BCGs in different bands by including MMCGs, whose stellar masses are estimated through the SED-fitting process or $K_s$-band imaging as explained below, in both the low-redshift and high-redshift samples. We therefore ignore any such biases for the remainder of the paper. 

\begin{figure*}
\includegraphics[clip,angle=0,width=1.6\columnwidth]{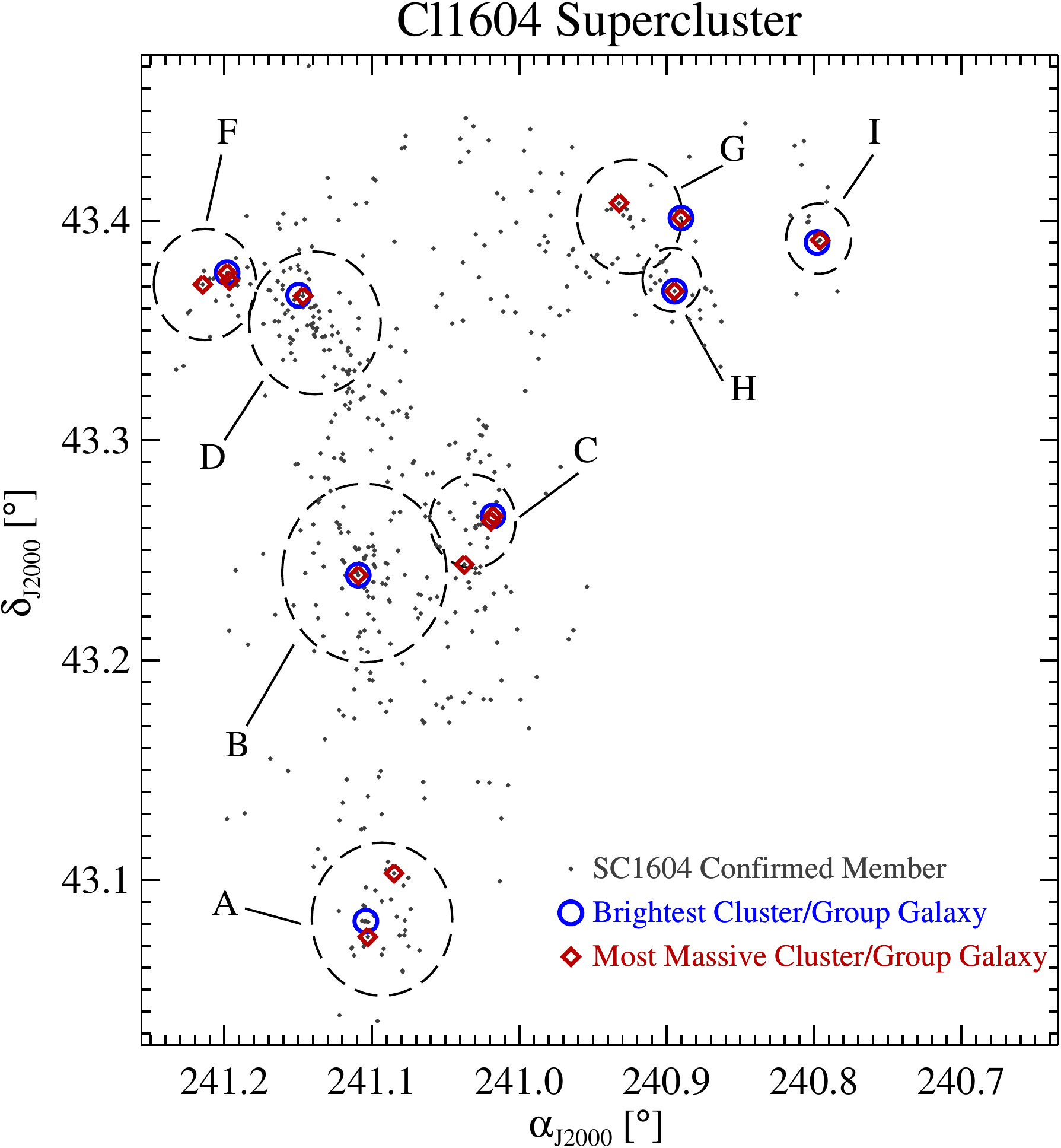}
\caption{Sky plot of the Cl1604 supercluster. The locations of each of the constituent clusters and groups of Cl1604 are denoted by circles centered and the luminosity-weighted mean location of member galaxies. The radius of each circle corresponds to the virial radius of that  cluster/group. While a better comparison to the X-Ray selected MCXC/SDSS sample would be to use the X-Ray centers of each structure, only two are available for the high-redshift sample (those of clusters A \& B) and these two are not appreciably different from the luminosity-weighted centers (see \citealt{rumbaugh13}). Small points indicate all spectroscopically confirmed member galaxies of the Cl1604 supercluster. The BCG of each cluster/group is shown by a circumscribed blue circle, while MMCG candidates are shown by circumscribed red diamonds. Multiple MMCG candidates exist for some clusters/groups (see text). A majority of the Cl1604 BCGs/MMCGs appear to be significantly offset from the luminosity-weighted center of their parent structure}
\label{fig:sc1604radec}
\end{figure*}

The MMCGs for the Cl1604 clusters and groups were selected in a nearly identical manner as the BCGs, with the obvious exception that the stellar mass derived through SED-fitting process or through the $K$-band imaging was used in place of the $F814W$ magnitude (for more details on how stellar masses for Cl1604 member galaxies were determined see \citealt{lemaux12}). Because the stellar masses derived for the Cl1604 member galaxies are considerably less precise than the $F814W$ magnitudes used to select BCGs, there existed a few cases where several galaxies in the cluster/group had stellar masses consistent within the errors of the measured MMCG. To mitigate this ambiguity, in all cases galaxies with stellar masses within $1\sigma$ of that of the galaxy with the highest measured mass were also selected and our analysis was repeated for each potential MMCG (setting a maximum of three 
MMCG ``candidates" for each cluster/group). In Figure \ref{fig:sc1604radec} we show the full extent of the Cl1604 supercluster complex along with the spatial location of all BCGs and MMCG candidates selected in this section.

While the galaxies of the Cl1604 structure constitute the whole of our high-redshift comparison sample in this study, it is essential to emphasize here that the supercluster is comprised of hundreds of member galaxies that are situated in a large variety of environments. The constituent clusters and groups are spread in such a way throughout the supercluster that they are not currently (at $z\sim0.9$) experiencing major interactions between each other (see, e.g., Figures 11 \& 12 in \citealt{gal08}). As such, for the study presented here, the eight clusters and groups of the Cl1604 supercluster can essentially be considered as isolated and independent structures at $z\sim0.9$. A complication seemingly arises when comparing these clusters and groups to lower redshift  isolated clusters. By virtue of the LSS in which they reside, both the global and local conditions of clusters  and groups could vary appreciably over the course of their lifetime relative to those of isolated clusters at $z\sim0$. However, given the uncertain dynamics and evolution of the supercluster and the large line-of-sight and projected distances between the Cl1604 clusters and groups, distances which are likely large enough to prevent major interactions by $z\sim0$, we ignore any complication of the LSS to the evolution of the galaxy populations of the Cl1604 clusters and groups and instead treat them in this study as isolated structures. 

\subsection{MCXC clusters at $z\sim0.1$}
\label{mcxc}

Many subtleties exist when attempting to compare BCGs/MMCGs at high-redshift to those at low-redshift (see, e.g., the discussion in \citealt{lidman12}). Astrophysical biases, such as those that arise due to the relationship between BCG properties and the mass of the host cluster (e.g., \citealt{lidman12,lin13,burke13}), must be carefully accounted for in order to make valid comparisons. Additionally, biases can be induced by comparing inhomogenous datasets or by making inhomogenous selections. Incomplete spectroscopy, morphological comparisons using bands that probe significantly different (rest-frame) wavelength regimes, and differences in cluster selection methods all have non-trivial effects on the comparisons made in this study. As such, we require a low-redshift sample that both astrophysically and observationally mimics the constituent clusters and groups of the Cl1604 supercluster. 

To this end we drew a sample of low-redshift clusters from the Meta-Catalog of the compiled properties of X-ray detected galaxy Clusters (MCXC; \citealt{piffaretti11}). The clusters contained within this catalog fulfilled several criteria important to this study. First, the catalog draws on all public data available at the time of publishing, which results in a large number of galaxy clusters. This is important because, while we will make an attempt later in this paper to evolve the Cl1604 clusters/groups to the present day (see \S\ref{clustermass}), there is much uncertainty involved in this process. Having a larger sample at low-redshift allows us to explore a larger range of physical properties and masses of the ``descendant'' clusters with which to compare the Cl1604 clusters and groups. Another virtue of this catalog is its overlap with the SDSS. This is a tremendous virtue, as it allows access to an enormous dataset of well-calibrated and internally consistent multiwavelength imaging and spectroscopy. Such an overlap allows us to select galaxy clusters with both $g^{\prime}$-band imaging (the $g^{\prime}$ band having roughly equivalent rest-frame coverage at $z\sim0.1$ to the $F814W$ band at $z\sim0.9$) and magnitude-limited spectroscopy of potential cluster members. The depth of the SDSS imaging slightly exceeds the depth of the ACS imaging in the Cl1604 cluster (in units of $L_{\odot}/pc^2$), important so that we are able to observe low (inherent) surface brightness features in galaxies at low-$z$ and high-$z$ (see \S\ref{morphevo} for a detailed discussion on this issue). 

A final virtue of the MCXC catalog is the well-defined and homogeneously measured X-Ray luminosities of all MCXC clusters. Cross-correlating these X-Ray detections with overdensities of galaxies in the SDSS ensures a high level of purity in our sample, i.e., that each X-Ray detection is truly associated with a galaxy cluster. For the comparisons made in this paper, we do not require an extremely large sample of low-redshift galaxy clusters, but rather a sample which spans completely the properties of the potential descendants of the Cl1604 clusters and groups. As will be shown later, the clusters selected from the MCXC catalog fulfill these requirements, and thus we value purity over completeness. Additionally, because each cluster in the MCXC catalog has, by definition, a hot intracluster medium (ICM), the level of virialization in each MCXC can be, in concert with SDSS spectroscopy, directly probed. In Appendix A we discuss the importance of this particular aspect of the MCXC catalog and the consequences for our results.

The process of selecting a sample of galaxy clusters from the MCXC catalog was as follows. Galaxy clusters which appeared in the MCXC catalog were cross-referenced with the SDSS DR8 public database\footnote{http://www.sdss3.org/dr8/data\_access.php} to look for overlap. For those clusters which fell within the SDSS footprint we required that any potential comparison cluster have $i)$ imaging in the $g^{\prime}$ band to sufficient depth to make valid morphological comparisons ($\sim 23.5$ mag/arcsec$^2$), $ii)$ sufficient photometry to estimate stellar masses (i.e., measured magnitudes in all SDSS bands), $iii)$ spectroscopy of at least 80\% of potential member galaxies brighter than $M_{g\prime}=-20.35$\footnote{The MCXC galaxies are $k$-corrected to $z=0$ in a similar fashion to the Cl1604 galaxies in the previous section.} that lie within a projected distance of $1 h_{70}^{-1}$ Mpc from the cluster center, and $iv)$ spectroscopy of the 10 brightest objects in the $g^{\prime}$ band within a projected distance of $1 h_{70}^{-1}$ Mpc from the cluster center as defined by the MCXC X-Ray centroid. Imposing these criteria resulted in a sample of 100 low-redshift clusters out of the 1743 clusters contained within the MCXC catalog. A color-magnitude diagram (CMD) of the members of each of the 100 clusters was inspected visually. Those clusters that did not have well-defined red sequence such that the presence of a galaxy cluster could not be confirmed (only two clusters), had more than one red sequence (i.e., projected with another cluster), clusters with questionable velocity dispersions (see \S\ref{clustermass}), or clusters for which the brightest cluster galaxy was in some way ambiguous were rejected. These cuts resulted in a final sample of 81 clusters with a median redshift of $\langle z\rangle=0.08$ and a redshift range of $z=0.02-0.21$. 

\begin{figure}
\includegraphics[clip,angle=0,width=1.0\hsize]{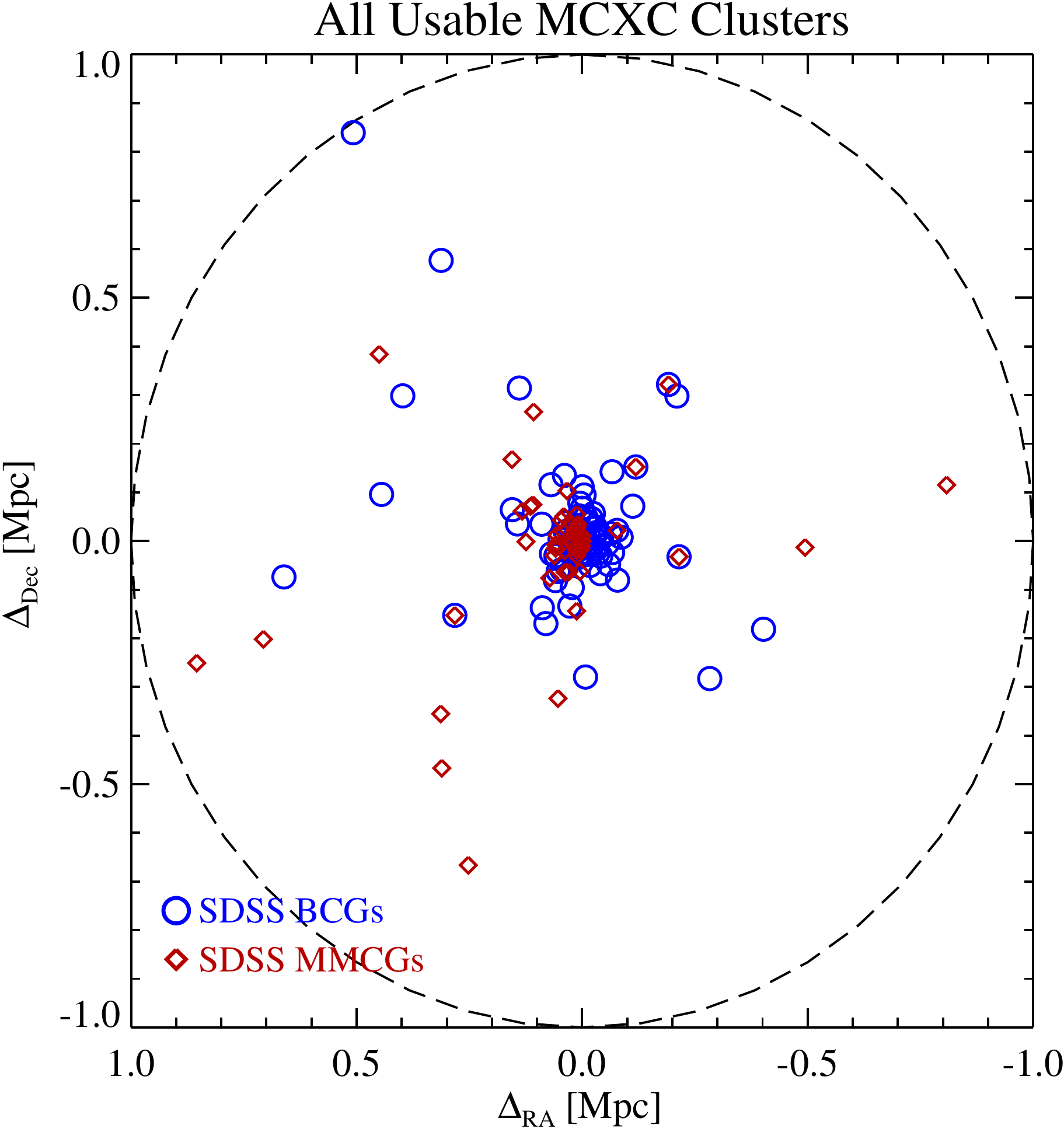}
\caption{Sky distribution of the 81 BCGs and 60 MMCGs of the final MCXC/SDSS sample, the latter only selecting cases where the MMCG was unambiguous. All galaxies are shown on a single plot, with the center of the plot corresponding to the X-Ray center of each of the MCXC clusters. The radius of the circle overplotted is 1 $h_{70}^{-1}$ Mpc. By construction this is the largest possible distance a BCG/MMCG can be from the X-Ray center of its parent cluster. In contrast to the BCGs/MMCGs of the Cl1604 supercluster, the BCGs and MMCGs of the low-redshift MCXC/SDSS sample appear largely at the center of the cluster potential well, though there are a few obvious exceptions.} 
\label{fig:SDSSradec}
\end{figure}

The BCG for each cluster was selected in a manner identical to that of the Cl1604 BCGs, with the selection being performed in the $g^{\prime}$ band. ``Total" stellar masses for all SDSS galaxies were taken from SDSS DR7\footnote{http://www.mpa-garching.mpg.de/SDSS/DR7/Data/stellarmass.html}, determined in a method similar to that of \citet{salim07}. As was the case for the Cl1604 members, the fitting used for SDSS DR7 employed stellar population synthesis models from \citet{bc03} employing a \citet{chabrier03} IMF.  The MMCG for each cluster was selected as the most massive member galaxy within $1 h_{70}^{-1}$ Mpc of the X-Ray cluster center. However, it was not possible to definitively determine a MMCG in all of the MCXC clusters. Of the 81 clusters in our final sample, 21 contained galaxies whose $i^{\prime}$ magnitudes rivaled that of the BCG that went untargeted in SDSS, but which had redshifts consistent with their parent cluster from various literature. In such cases, the SDSS routine was not used to fit a stellar mass, and while the BCG could still be definitively determined, we could generally not determine a MMCG. Exceptions were made for those few clusters where the $i^{\prime}$ magnitude of the BCG exceeded the brightness of any galaxy untargeted by SDSS by greater than 1.5 magnitudes. For the remaining 60 clusters where a MMCG could be definitively determined, only 13 had a BCG that differed from the MMCG. In the remaining 47 clusters the $g^{\prime}$-band selection was effective at isolating the MMCG. In Figure \ref{fig:SDSSradec} we show the spatial positions of the BCGs and MMCGs selected for all MCXC clusters.

One final discussion is necessary on the nature of the structures studied in the MCXC and Cl1604 samples. As we will show in the next section, the structures that have been selected span a wide range in line-of-sight velocity dispersions and virial masses for both samples. This is potentially problematic, as it is well known that different processes have different effects in clusters which are in different dynamical states (e.g., \citealt{moran07}) and, furthermore, that galaxies embedded in group systems experience certain processes more frequently than do their cluster counterparts (e.g., \citealt{zabludoff98}). During the presentation of our analysis, where relevant, we will match the MCXC and Cl1604 host structures in virial mass, such that the BCGs/MMCGs of the two samples can be compared, on average, fairly. However, the question still remains as to what term to use to refer to the MCXC structures. In general, the dividing line between clusters and groups is fairly arbitrary. In principle, the delineation point is set primarily using the criterion that the latter have galaxy velocity dispersions that are roughly comparable to the velocity dispersion of stars in an individual galaxy. This is done so that galaxy-galaxy merging is, by definition, efficient in ``group environments" and inefficient in ``cluster environments". In practice, groups are defined using a wide variety of galaxy velocity dispersions, dispersions which in many cases exceed the stellar dispersions of even the most massive galaxies. For the Cl1604 supercluster, we adopt a commonly used threshold of 600 km s$^{-1}$ \citep{mulchaey00,osmond04}, which allows us to define the constituent structures of the supercluster in a consistent manner with previous works (e.g., \citealt{gal08,lemaux12}). Such a choice is, however, for this study purely a pedagogical one and does not have an effect on any of our results. Though we will show in the next section that some of the MCXC structures have line-of-sight velocity dispersions below this dividing line, we will continue to refer to the MCXC structures simply as ``clusters" throughout the paper for the sake of brevity, noting again that this choice has no effect on our results. 

\section{Sample properties}

\subsection{Cluster mass distribution}
\label{clustermass}

As noted in the previous section, galaxy clusters which were selected from the MCXC catalog contained a large number of spectroscopically confirmed members from SDSS. The median number of spectroscopically confirmed members in the MCXC clusters was 36, ranging from 11 in the sparsest cases ($\la10\%$ of the clusters) to greater than 325 members in the richest cases. These clusters were also supplemented with additional redshifts from the literature. With this number of members it is feasible, in principle, to calculate line-of-sight velocity dispersions, $\sigma_v$, for each cluster, which could then be used in turn as a mass proxy. Though the velocity dispersions measured for the most sparsely sampled clusters will be of questionable accuracy (see, e.g., \citealt{girardi93}), especially in those cases where only the brightest galaxies are sampled, the large MCXC/SDSS sample employed here mitigates a poorly sampled velocity dispersion for any given cluster. While cluster masses can also be derived through the X-Ray quantities,  a typical method used to derived total cluster mass relies on the X-Ray temperature ($T_X$) of the ICM, a quantity which is not provided by default in the MCXC catalog. The MCXC catalog does provide a $M_{500}$ value that is based on the X-Ray luminosity for each cluster, but we preferred not to use this value except for internal comparisons between MCXC clusters. Furthermore, more than half of the Cl1604 clusters and  groups do not have an associated X-Ray measurement due to the high-redshift of the structure and the relatively shallow \emph{Chandra} imaging, making a self-consistent comparison in this respect impossible. Cluster (and group) line-of-sight velocity dispersions were measured in a manner nearly identical to that of \citet{rumbaugh13} (see also \citealt{gal04}) and are translated into a virial mass via \citep{carlberg97,biviano06,poggianti09}: 

\begin{equation}
M_{\rm{vir}} = \frac{3\sqrt{3}\sigma_{v}^3}{11.4 G \, H(z)}
\label{eqn:Mvir}
\end{equation}

\noindent where $G$ is Newton's gravitational constant and $H(z)$ is the value of the Hubble parameter at the median redshift of the cluster members. While the velocity dispersion was measured on 89 of the MCXC clusters which passed all of our other criteria, eight were rejected at this point due to extreme non-Gaussian velocity distributions, prohibitively sparse spectroscopic sampling, or convergence issues. Of the remaining 81 MCXC clusters with well-measured velocity dispersions, the average cluster galaxy velocity dispersion was $\langle \sigma_{v} \rangle = 688$ km s$^{-1}$ corresponding to $\langle M_{vir} \rangle = 6.7\times10^{14} \mathcal{M}_{\odot}$. For the bulk of our analysis the galaxies in these clusters will comprise our low redshift comparison sample. However, later in the paper a comparison will be made between the radial distributions of stellar mass surrounding both the high- and low-redshift BCGs/MMCGs (see \S\ref{massrad}). For this sample it is necessary to define a sample of galaxy clusters which have not only a definitively determined MMCG, but for which the stellar mass of all massive nearby companions has been measured. For 21 of the 81 MCXC clusters, the first criterion is not fulfilled (see \S\ref{mcxc}). Of the remaining 60 clusters, we required that all galaxies within a projected radius of $R_{\rm{proj}}<0.5R_{vir}$ which were within 1.5 magnitudes of the $i^{\prime}$ magnitude of the BCG/MMCG have both a measured redshift and a measured stellar mass. It is important to note that all Cl1604 clusters and groups also satisfied these criteria. Imposing these additional criteria on the 81 clusters in our full MCXC/SDSS sample resulted in 53 clusters. These clusters have a median velocity dispersion of $\langle \sigma_{v} \rangle = 625$ km s$^{-1}$, which corresponds to $\langle M_{vir} \rangle = 3.55\times10^{14} \mathcal{M}_{\odot}$. These values are not appreciably different from those values of the full sample of 81 MCXC clusters and, as we will show later in this section, the distributions of $\sigma_{v}$ and $M_{vir}$ between the two samples are not significantly different. 

The line-of-sight velocity dispersions and virial masses of the Cl1604 clusters and groups were calculated in a manner identical to those of the MCXC clusters with one exception. Since six out of the eight Cl1604 clusters/groups do not have an X-Ray centroid, it is necessary to define another center from
which to calculate the velocity dispersion. These were determined in the following manner. The $F814W$ luminosity-weighted center was calculated for each cluster and group from all members within 1 $h_{70}^{-1}$ Mpc of the centers defined from red galaxy density maps generated following the method of \citet{gal08}.  All member galaxies within 1 $h_{70}^{-1}$ Mpc of this  $F814W$ luminosity-weighted center were then used to calculate a new $F814W$ luminosity-weighted center. This process was repeated as many times as necessary to achieve convergence, i.e., where further iterations of this exercise resulted in centers which differed by less than 2$\arcsec$. The line-of-sight velocity dispersion for each cluster and group was then calculated from all member galaxies within 1 $h_{70}^{-1}$ Mpc of the final luminosity-weighted center using identical methodology to that of \citet{rumbaugh13}. In such a way, information from all galaxy populations is utilized rather than just the red sequence galaxies, important since many of the Cl1604 clusters and groups are still in the process of formation. Under the limit of a traditional virialized cluster whose galaxy population is dominated by bright red-sequence galaxies, this exercise reduces to the original center defined by the red galaxy density peaks. Since the line-of-sight velocity dispersions of the MCXC clusters were calculated from populations defined relative to their X-Ray centers, this latter point is important for any relative bias between the two methods, as a large fraction of the MCXC clusters exhibit properties of traditional virialized clusters (see \S\ref{colormag}). Indeed, if we instead choose the MCXC BCG center as a centroid, which is a good approximation of the luminosity-weighted center, the average change in the measured velocity dispersions is $<$10\% relative to those calculated using the X-Ray center. This is also the case for the two Cl1604 clusters (A \& B) where this check can be performed. Such a change has no effect on our results and we thus ignore any possible differential bias induced on our two samples as a result of these choices. 

The line-of-sight velocity dispersion values (and, as a consequence, $M_{vir}$) calculated for the Cl1604 clusters and groups, as well as their central coordinates and redshifts, differ slightly from those given in \citet{lemaux12} due to the improved methodology used in calculating $\sigma_{v}$. 
For the Cl1604 groups and clusters, the average velocity dispersion was $\langle \sigma_{v} \rangle = 431$ km s$^{-1}$ and $\langle \sigma_{v} \rangle = 742$ km s$^{-1}$, respectively, translating to average virial masses of $\langle M_{vir} \rangle = 8.8\times10^{13} \mathcal{M}_{\odot}$ and 
$\langle M_{vir} \rangle = 3.9\times10^{14} \mathcal{M}_{\odot}$, respectively. Table \ref{tab:globalproperties1604} lists the cluster/group name, $\alpha_{\rm{J}2000}$, $\delta_{\rm{J}2000}$, $\langle z \rangle$, number of members within 1 $h^{-1}$ Mpc, velocity dispersion, and virial masses of the Cl1604 clusters and groups.

\begin{table*}
\caption{Properties of the Galaxy Groups and Clusters in the Cl1604 Supercluster} 
\begin{tabular}{cccccccc}
\hline 
 & & & & & & $\sigma_{v}$ & $\mathcal{M}_{vir}$ \\
Name & ID & $\alpha_{\rm{J}2000}$ & $\delta_{\rm{J}2000}$ & $\langle z \rangle$ & N$_{\rm{mem}} ^{1}$ & (km s$^{-1}$)$^{2}$ & ($\times10^{14} h_{70}^{-1} \mathcal{M}_{\odot})^{3}$ \\ [0.5ex]
\hline \hline 
A & Cl1604+4304 & 241.0931 & 43.0821 & 0.898 & 35 & 722$\pm$135 & 3.54$\pm$1.32 \\[4pt]
B & Cl1604+4314 & 241.1080 & 43.2397 & 0.865 & 49 & 818$\pm$74  & 5.26$\pm$0.95 \\[4pt] 
C & Cl1604+4316 & 241.0314 & 43.2679 & 0.934 & 32 & 454$\pm$40 & 0.86$\pm$0.15 \\[4pt]
D & Cl1604+4321 & 241.1409 & 43.3539 & 0.923 & 70 & 688$\pm$88 & 3.03$\pm$0.78 \\[4pt]
F & Cl1605+4322 & 241.2010 & 43.3684 & 0.933 & 20 & 542$\pm$110 & 1.47$\pm$0.60 \\[4pt] 
G & Cl1604+4324 & 240.9274 & 43.4030 & 0.902 & 18 & 539$\pm$124  & 1.47$\pm$0.67 \\[4pt] 
H & Cl1604+4322 & 240.8989 & 43.3670 & 0.853 & 10 & 287$\pm$68 & 0.29$\pm$0.11 \\[4pt] 
I & Cl1603+4323 & 240.7975 & 43.3915 & 0.902 & 7 & 333$\pm$129 & 0.35$\pm$0.27  \\[1pt] 
\hline
\end{tabular}
\begin{flushleft}
1: Within $1\rm{h_{70}}^{-1}$ Mpc \\
2: Calculated from all member galaxies within 1 $h_{70}^{-1}$ Mpc of the luminosity-weighted cluster/group center \\
3: Errors in $M_{\rm{vir}}$ are calculated from errors in $\sigma_v$ 
\end{flushleft}
\label{tab:globalproperties1604}
\end{table*}

\begin{figure}
\centering
\includegraphics[clip,angle=0,width=1.0\hsize]{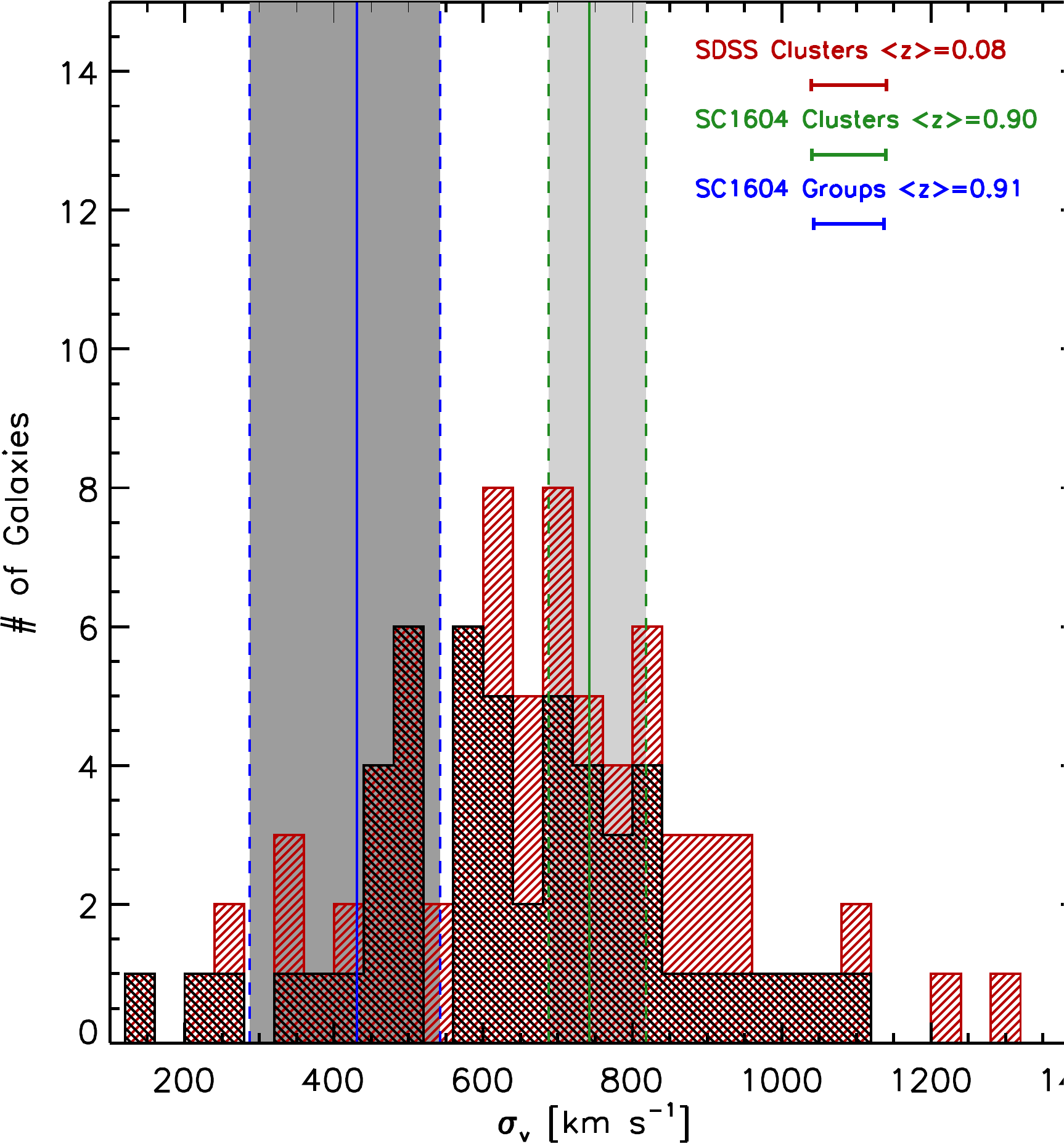}
\caption{Cluster galaxy line-of-sight velocity dispersion distribution of the 81 MCXC clusters with well-measured dispersions (red hashed histogram) 
compared with the sub-sample of 53 clusters used in \S\ref{massrad} (black hashed histogram). Also plotted are those distributions of the Cl1604 clusters (light shaded) and groups (dark shaded). The average redshift and error on the dispersion of the main MCXC/SDSS sample and the Cl1604 samples is shown in the upper right hand corner.}
\label{fig:sigmadist}
\end{figure}

\begin{figure}
\centering
\includegraphics[clip,angle=0,width=1.0\hsize]{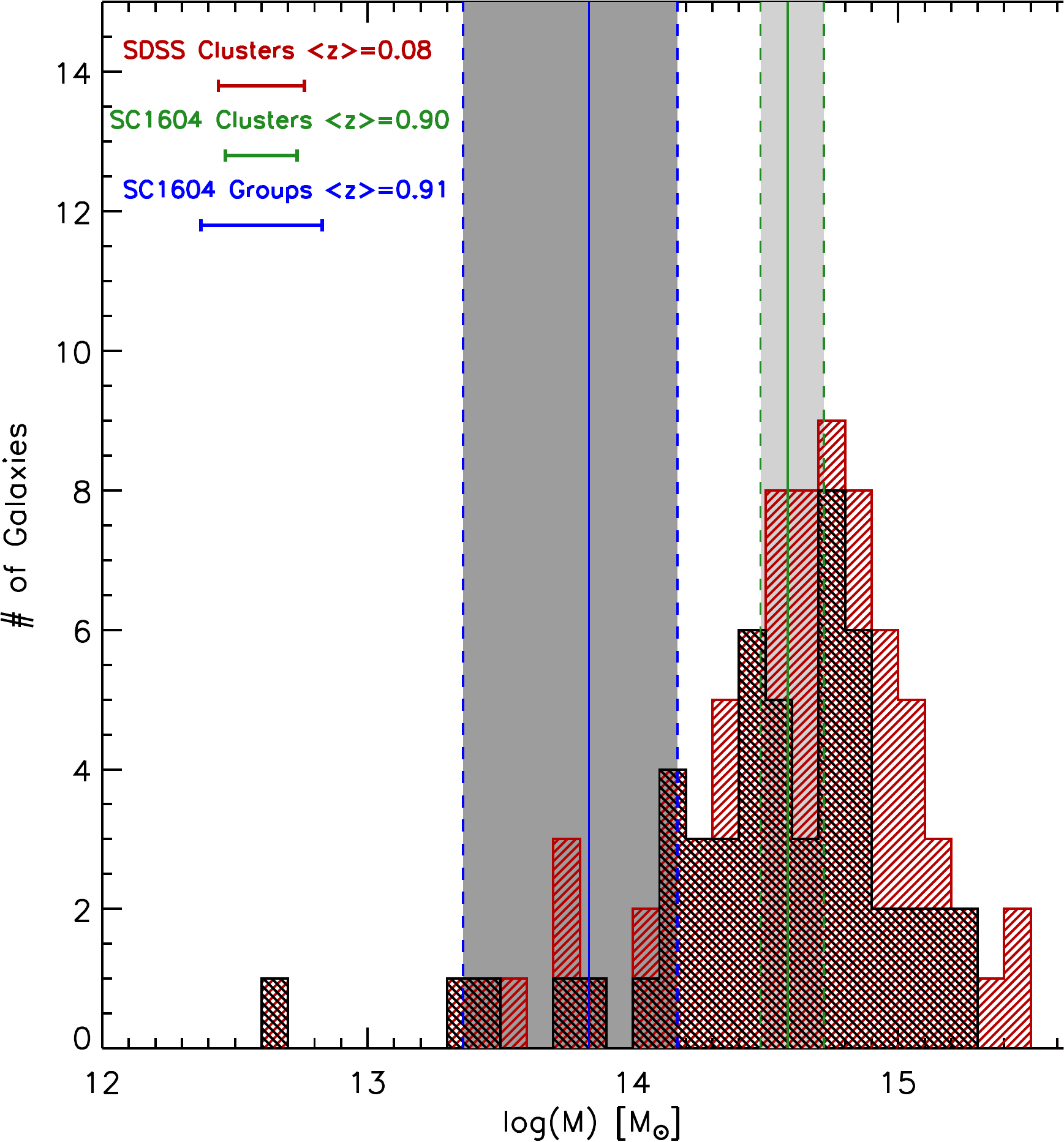} 
\caption{Virial mass distribution of the 81 MCXC clusters with well-measured dispersions and the sub-sample of 53 MCXC clusters used in \S\ref{massrad} compared with those of Cl1604 clusters and groups. The meanings of colors and lines are the same as in Figure \ref{fig:sigmadist}.}
\label{fig:Mvirdist}
\end{figure}

In Figure \ref{fig:sigmadist} we plot the galaxy velocity dispersions for the 81 MCXC clusters with measured dispersions along with the distribution of 53 MCXC clusters from which a valid comparison sample will be drawn for the analysis presented in \S\ref{massrad}. Overplotted is the range of velocity dispersions of the Cl1604 clusters and groups. In Figure \ref{fig:Mvirdist} we plot virial mass distributions, derived from the velocity dispersions, for all the samples shown in the previous figure. In both cases, the distribution of the two MCXC/SDSS samples appear extremely similar by eye, and this similarity is confirmed by a Kolmogorov-Smirnov (KS) test in which the two samples are found to be statistically indistinguishable in both parameters. This is an important point. The two samples are used as bases of comparison for the galaxy populations of the Cl1604 supercluster at different stages in this paper. As will be discussed in detail later, strong correlations have been observed between the total cluster mass and the stellar mass of the BCG (see \S\ref{massrad}). The extreme similarity observed between the two MCXC/SDSS samples ensures that no differential bias in this regard has been introduced by our selection methods. This still allows for absolute bias relative to the Cl1604 clusters and groups sample. However, as we will discuss later, the bulk of our results are unaffected by the specific choice of subsamples used for comparison, meaning that the absolute effect of such a bias is generally negligible for this study.

\begin{figure}
\centering
\includegraphics[clip,angle=0,width=1.0\hsize]{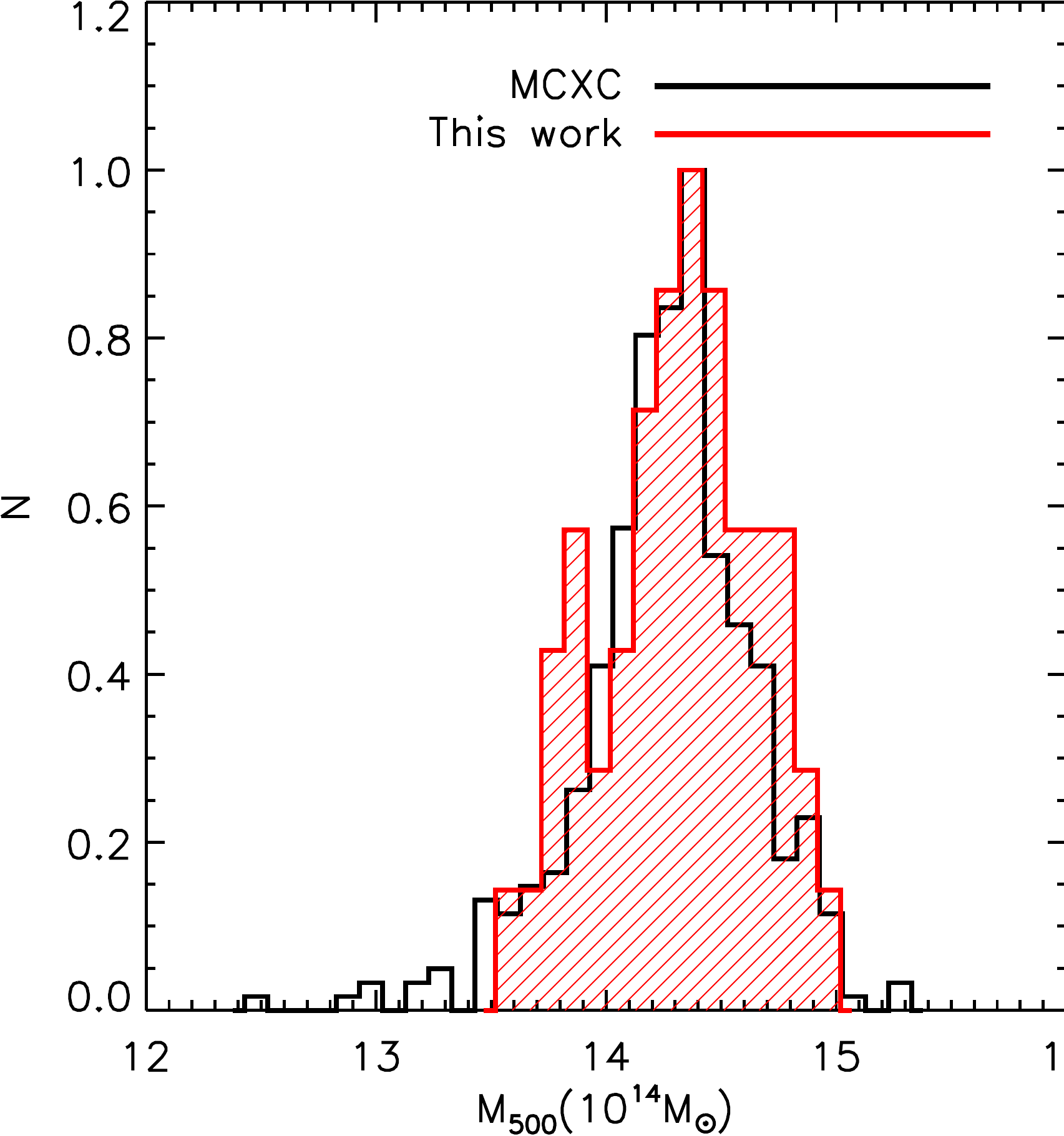} 
\caption{Normalized distribution of $M_{500}$ as given by the MCXC catalog for the full MCXC cluster sample (open black histogram) plotted against the distribution of the 81 MCXC galaxy clusters in our final sample. The two distributions are statistically indistinguishable, a result which remains unchanged if the 53 MCXC clusters selected in \S\ref{clustermass} are instead compared. Neither selection appears to select clusters that are biased (in $M_{500}$) with respect to the full MCXC cluster sample.}
\label{fig:checkmass}
\end{figure}

Noticeable in Figures \ref{fig:sigmadist} and \ref{fig:Mvirdist} is the lack of extremely massive ($\ga$10$^{15} M_{\odot}$) or ($>$1200 $km/s$) MCXC clusters, surprising for a sample of clusters selected at low redshift. In order to investigate if our selection method is biased with respect to the full MCXC catalog we compare in Figure \ref{fig:checkmass} a scaled distribution of $M_{500}$ (taken from the MCXC catalog) of the full MCXC/SDSS sample with the distribution of the 81 MCXC clusters in our full MCXC/SDSS sample. Visually, the two distributions appear to trace each other, a similarity which is again confirmed by performing a Kolmogorov-Smirnov (KS) test. This result is unchanged if we instead use the 53 galaxy clusters with measured velocity dispersion selected above. Additionally, there is no reason to believe that the MCXC catalog is preferentially selecting lower mass clusters at low redshift. Though the selection function of the MCXC clusters is extremely complicated (see \citealt{piffaretti11}), the wide variety of X-Ray samples used to generate the MCXC catalog essentially results in a random sampling of the sky to varying depths in $L_{X}$. Indeed, because galaxy clusters in the catalog must, by definition, be bright in X-Rays, clusters selected in the MCXC catalog are, if anything, biased towards the higher mass end of the cluster mass function at $z\sim0.1$. The fact that so few massive clusters exist in the MCXC catalog is likely due to the small volume probed at low redshift (due to the angular to physical scale conversion and the limited sky coverage of sufficiently deep X-Ray data) and the relative rarity of massive galaxy clusters rather than selection effects.

\subsection{Color-magnitude diagrams}
\label{colormag}

We begin the study of the constituent galaxy populations of the MCXC clusters and the Cl1604 clusters and groups by investigating the broadband colors of the BCGs and MMCGs in these structures and quantify their colors relative to the overall galaxy population. For this section, and indeed for all analysis save that of \S\ref{massrad}, the full 81 clusters selected in the previous section are used as a basis of comparison. While it was briefly discussed  in the previous section that bias can be introduced from cluster samples with improperly matched masses, due to the broad homogeneity of the properties of the MCXC BCGs/MMCGs, the results presented in our analysis are insensitive to the choice of comparison samples. The fractions presented in this section, as well as the results presented in every section except \S\ref{massrad} do not change appreciably ($<1\%$) if the sample of 53 clusters defined in the previous section is used or if a cluster total mass-matched sample is used (as is done in \S\ref{massrad}). 
 
\begin{figure}
\centering
\includegraphics[clip,angle=180,width=1.0\hsize]{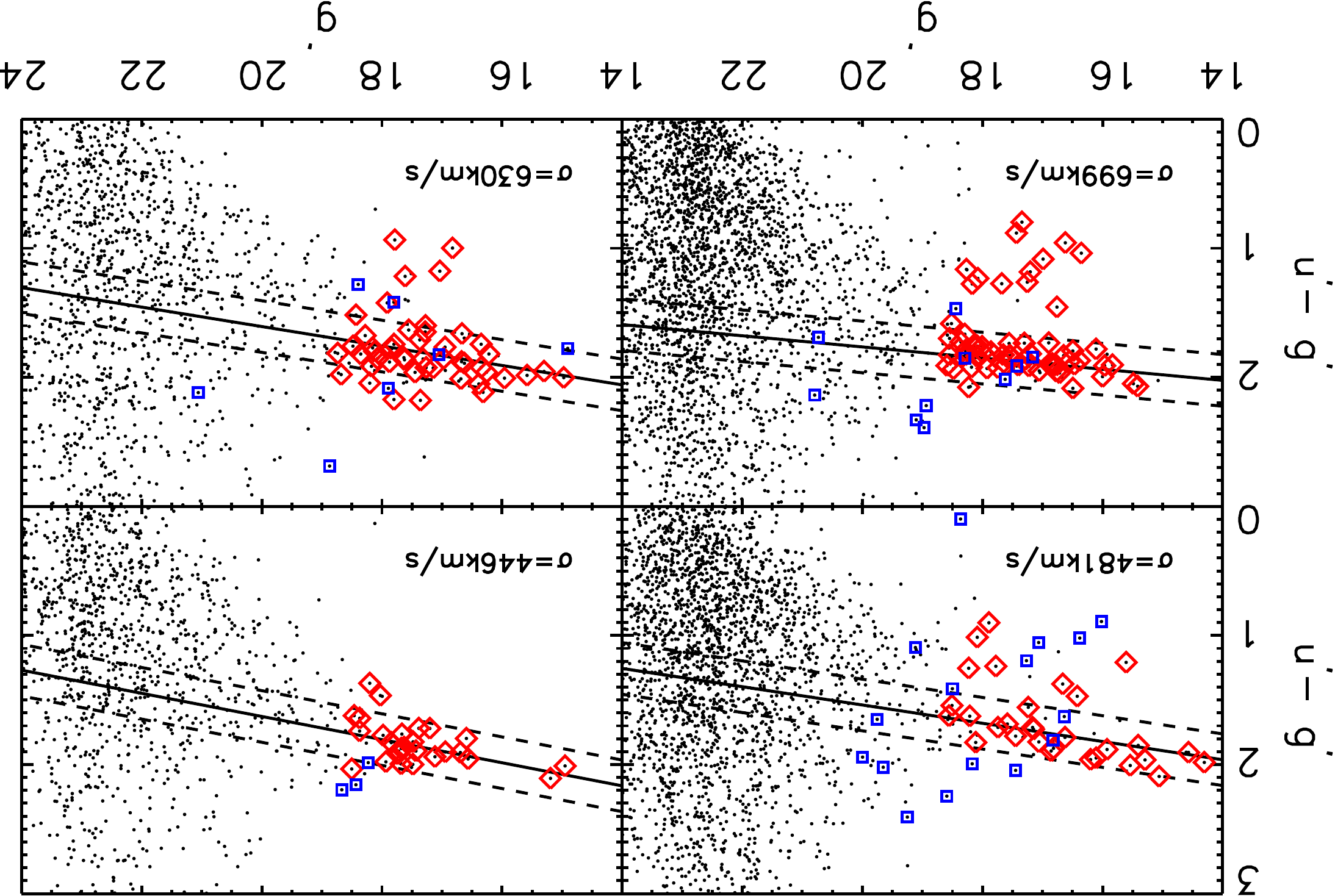}
\includegraphics[clip,angle=0,width=1.0\hsize]{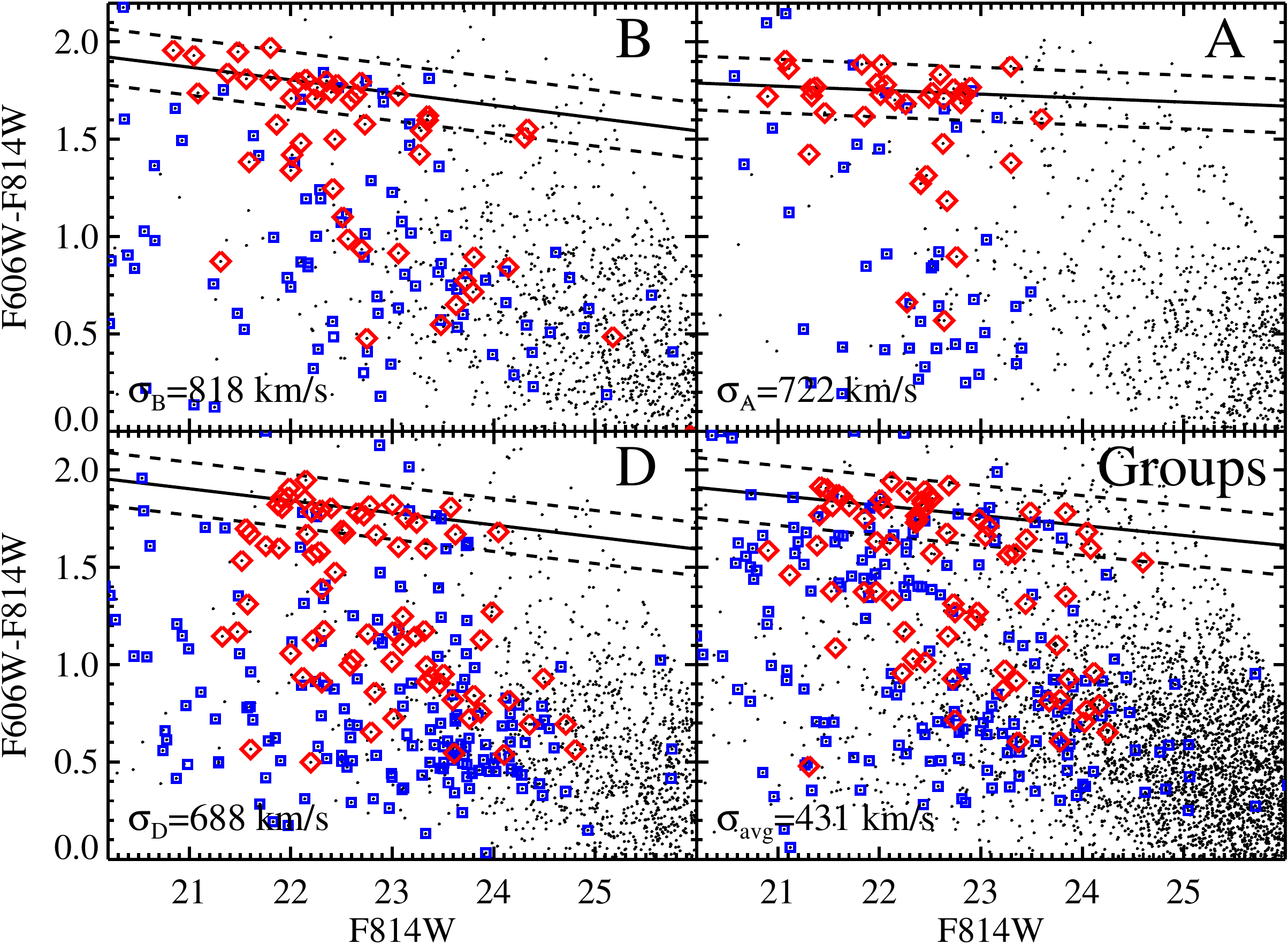}
\caption{Top panel: SDSS $g^{\prime}$ vs. $u^{\prime}-g^{\prime}$ color magnitude diagram (CMD) for four randomly selected low-redshift MCXC clusters used in this study. Only objects within 1 $h_{70}^{-1}$ of the X-Ray center are shown in each CMD. Symbols circumscribed by red diamonds refer to spectroscopically confirmed member galaxies whereas symbols circumscribed by blue squares are galaxies with spectroscopic redshifts inconsistent with the redshift of the cluster. Points that are not circumscribed refer to galaxies that were either untargeted by spectroscopy or galaxies with low-quality redshifts. The two bright red-sequence objects in cluster D that went untargeted are clearly stars from the imaging data. The best-fit red sequence and 3$\sigma$ width is shown by solid black and dotted black lines in each panel. The velocity dispersion of the members of each cluster is shown in the bottom left of each CMD. Bottom panel: ACS $F814W$ vs. $F606W-F814W$ CMD of the galaxies comprising the clusters and groups of the Cl1604 supercluster at $z\sim0.9$. Galaxies of the five groups of the Cl1604 supercluster have been combined into one sample. The meanings of symbols and lines are identical to the top panel. Though the spectral coverage between the two datasets is extremely similar, the color-magnitude properties of the constituent galaxies are decidedly different.} 
\label{fig:CMR}
\end{figure}

In Figure \ref{fig:CMR} we show the SDSS color-magnitude diagrams (CMDs) for the galaxy populations of four randomly selected MCXC clusters and \emph{HST}/ACS CMDs for the galaxies inhabiting the Cl1604 clusters and groups for all photometric objects within 1 $h_{70}^{-1}$ Mpc from the cluster/group centers. Objects circumscribed by red diamonds indicate spectroscopically confirmed member galaxies of a particular cluster or group (see \S\ref{cl1604}), while objects circumscribed by blue squares indicate galaxies at redshifts inconsistent with the cluster/group redshift. Small black points that are not circumscribed indicate objects which either remain untargeted by spectroscopy or which were targeted by spectroscopy, but did not have a sufficiently high-quality redshift measurement. As was done in \citet{lemaux12}, a combined CMD is plotted for the Cl1604 groups. For both samples, red-sequence 
fitting was performed in a manner identical to that of \citet{lemaux10} using only spectroscopically confirmed members. 

The color-magnitude properties of the galaxies in the Cl1604 groups/clusters is discussed extensively in \citet{lemaux12} and, as such, we mention them only briefly here as a contrast to those of the low-redshift clusters. Not unexpectedly, the galaxies which comprise the Cl1604 clusters and groups exhibit a wider variety of  colors than those of the low-redshift clusters, typical of structures in the process of formation (e.g., \citealt{mei09}). To the magnitude limit of both surveys, the average fraction of blue galaxies is immensely higher in the high-redshift clusters and groups than that of the low-redshift clusters. There also exist a large number of galaxies with ``transitional colors" (i.e., just blueward of the red sequence, sometimes referred to as the ``green valley") in the Cl1604 clusters and groups, galaxies which are largely absent in the MCXC clusters. With the exception of the two most massive clusters in Cl1604,  these galaxies also comprise some of the brightest galaxies in the Cl1604 supercluster. Indeed, in cluster D and the Cl1604 groups, the BCG often has colors blueward of the red sequence, a phenomenon rarely observed in the low-redshift clusters: only three out of 81 MCXC clusters have a BCG that is bluer than the red sequence, two of which are not the MMCG in the cluster. Table \ref{tab:fractions} lists the fraction of blue BCGs and MMCGs for both the MCXC/SDSS sample and the Cl1604 galaxies. 
 
Another noticeable feature of the CMDs is the disparity between the luminosity gap of the two samples, i.e., the difference in magnitudes between the BCG and second brightest cluster member. The average luminosity gap between the BCG and the next brightest member galaxy in the eight Cl1604 clusters/groups is $\langle \Delta m_{F814W}\rangle=0.30\pm0.02$. This value is identical to the value determined by \citet{fassbender11} for a large sample of X-Ray selected clusters in the redshift range $0.9<z<1.6$, suggesting the value is typical for BCGs still under formation. In the MCXC clusters, the average value is significantly higher $\langle \Delta m_{g^\prime}\rangle=0.70\pm0.12$, a value consistent with other such measurements at low redshift (e.g., \citealt{smith10}). The lower luminosity gap observed in the high-redshift clusters is again a clear indicator of these BCGs being in the process of evolving via major/minor mergers in the high-redshift clusters. I.e., in cases where no other extremely bright galaxies are accreted from the field from $z\sim0.9$ to $z\sim 0$, an unlikely scenario given the high level of spectroscopic completeness of our sample and the magnitude distribution of galaxies surrounding the Cl1604 clusters/groups (see \citealt{lemaux12}), the luminosity gap is directly related to the dynamical age of the galaxy cluster/group \citep{dariush07,dariush10}. Moreover, poorer structures at high redshift are more likely to have multiple bright galaxies merge to form the final BCG within the cosmic time difference between the Cl1604 and MCXC/SDSS samples, which leads in turn to a larger observed luminosity gap by the present day \citep{milosavljevic06,dariush10}. Thus, the observed disparity in the luminosity gap between the Cl1604 and MCXC BCGs hints at significant merging activity
of bright members of the Cl1604 clusters/groups which has largely subsided by $z\sim0.1$. We will return to this point repeatedly throughout the next several sections.

\subsection{Visual Morphology}
\label{vismorph}

\begin{figure*}
\centering
\includegraphics[clip,angle=0,width=0.3\hsize]{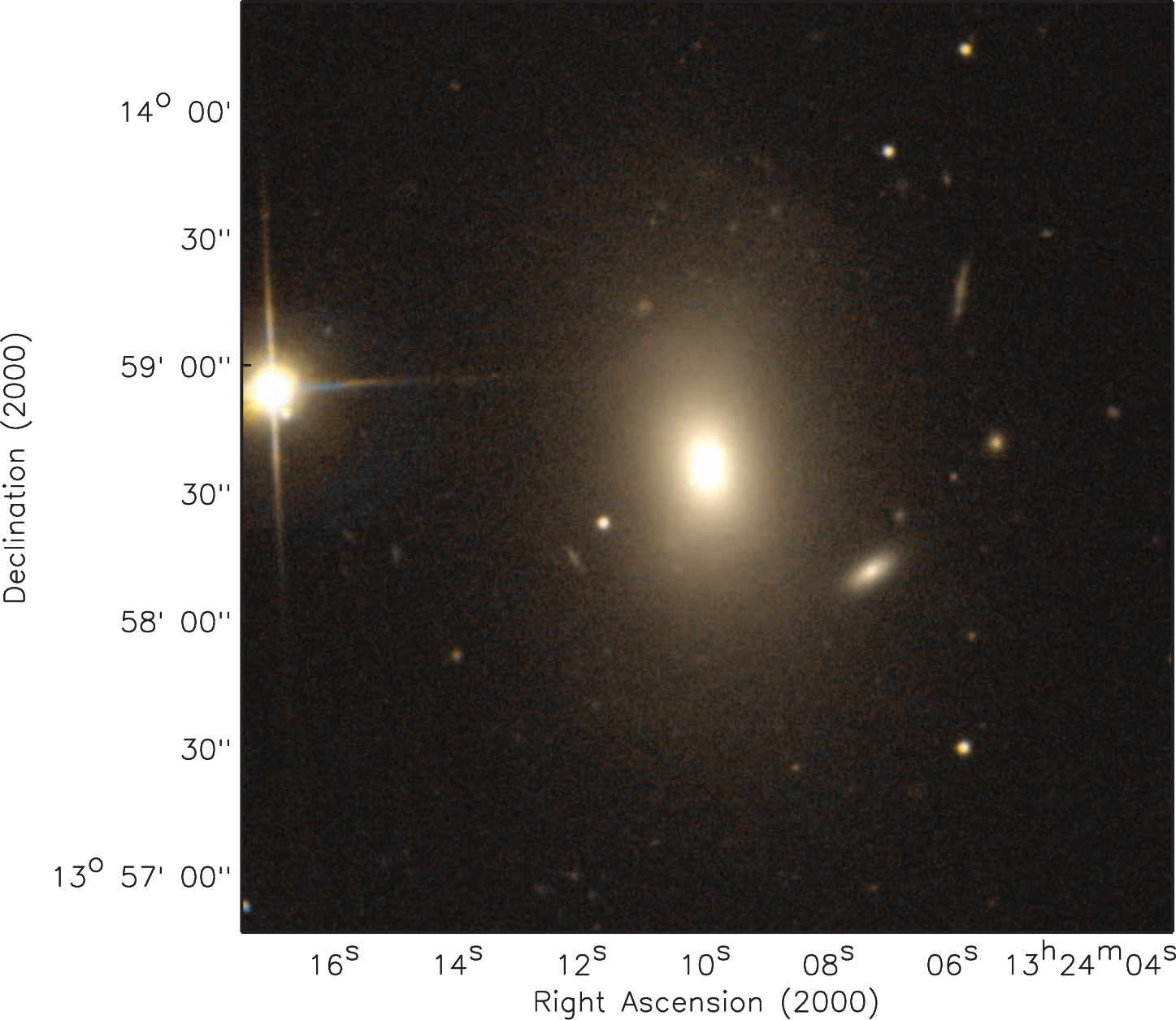} 
\includegraphics[clip,angle=0,width=0.3\hsize]{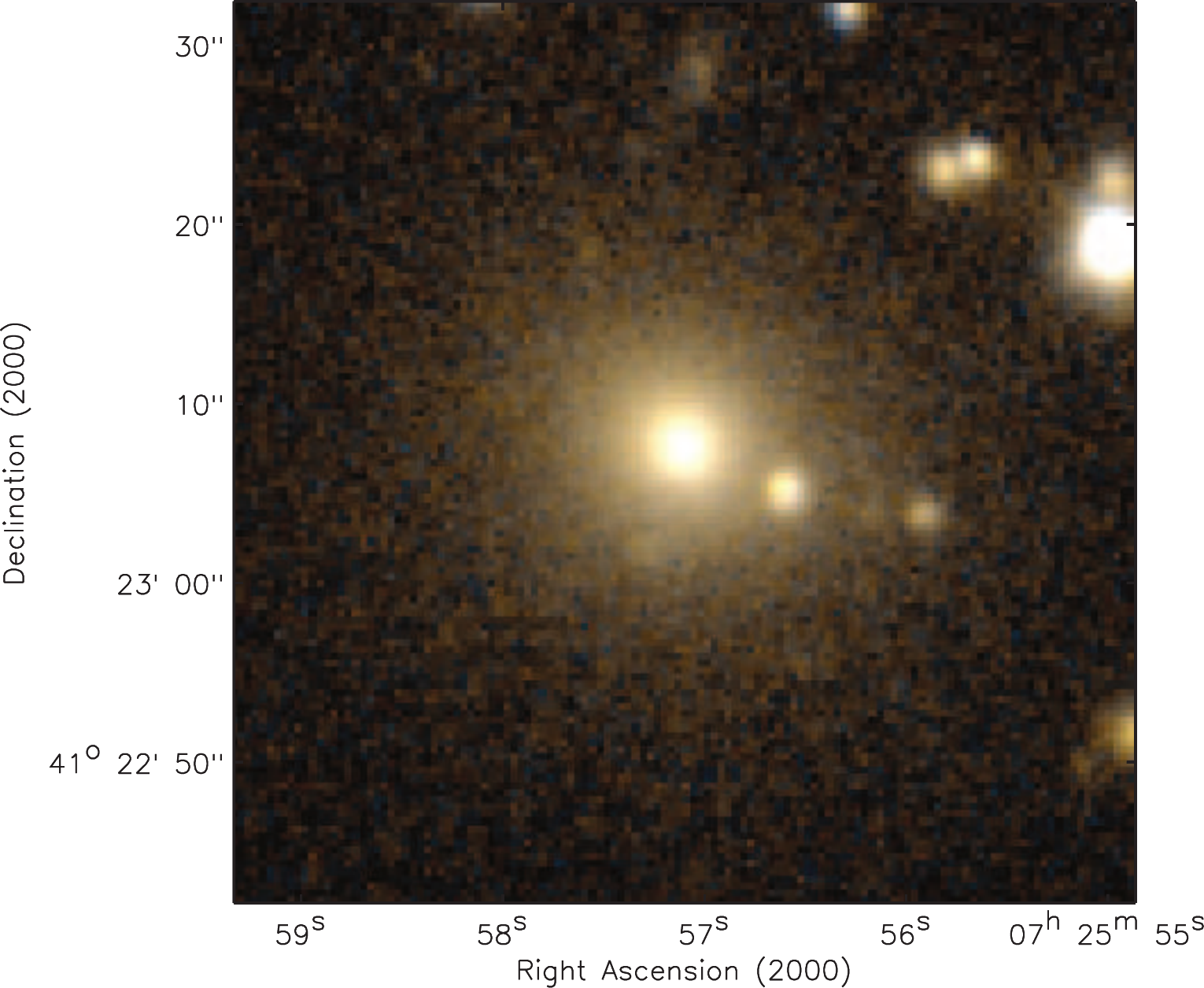} 
\includegraphics[clip,angle=0,width=0.3\hsize]{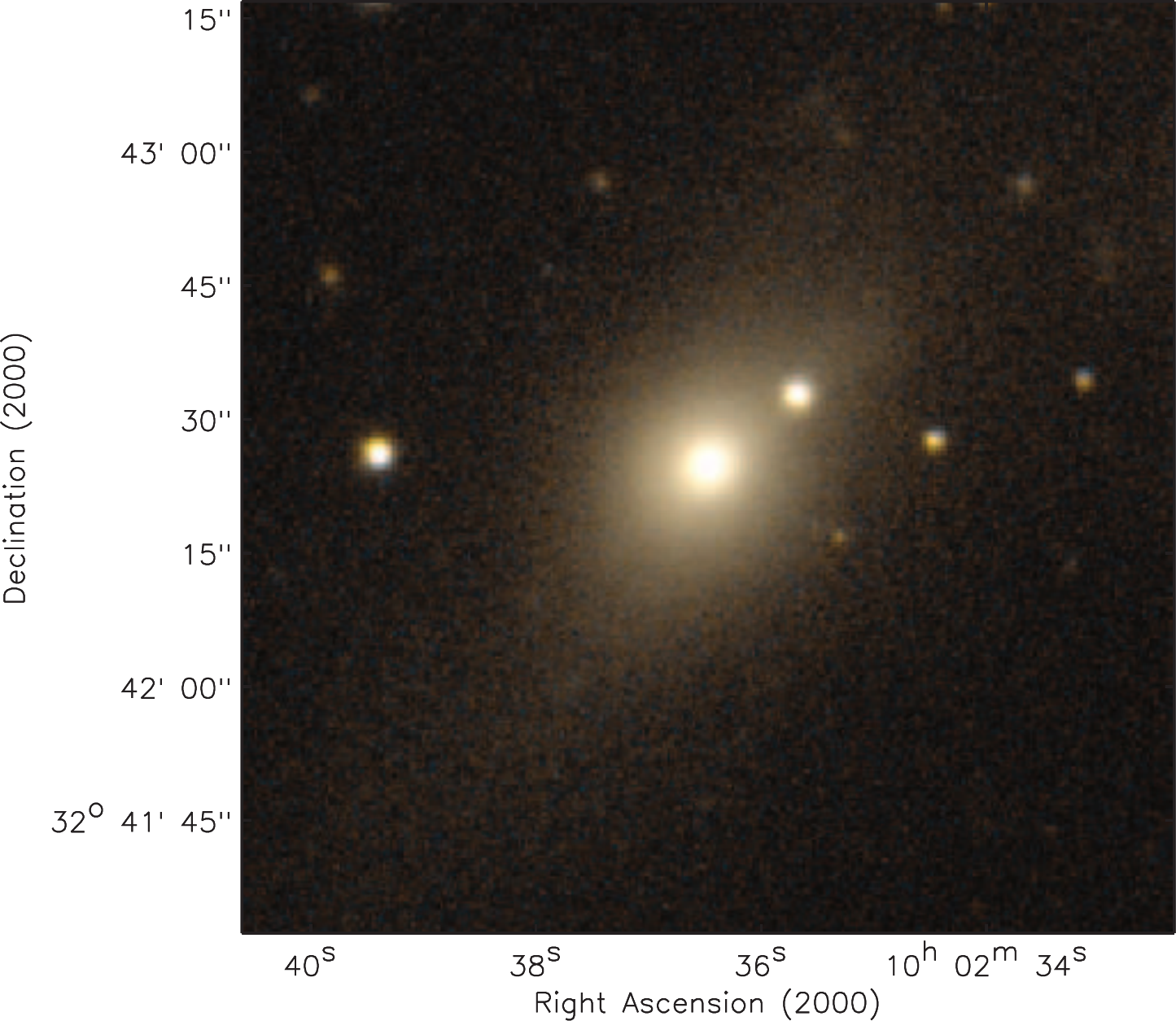} 
\includegraphics[clip,angle=0,width=0.3\hsize]{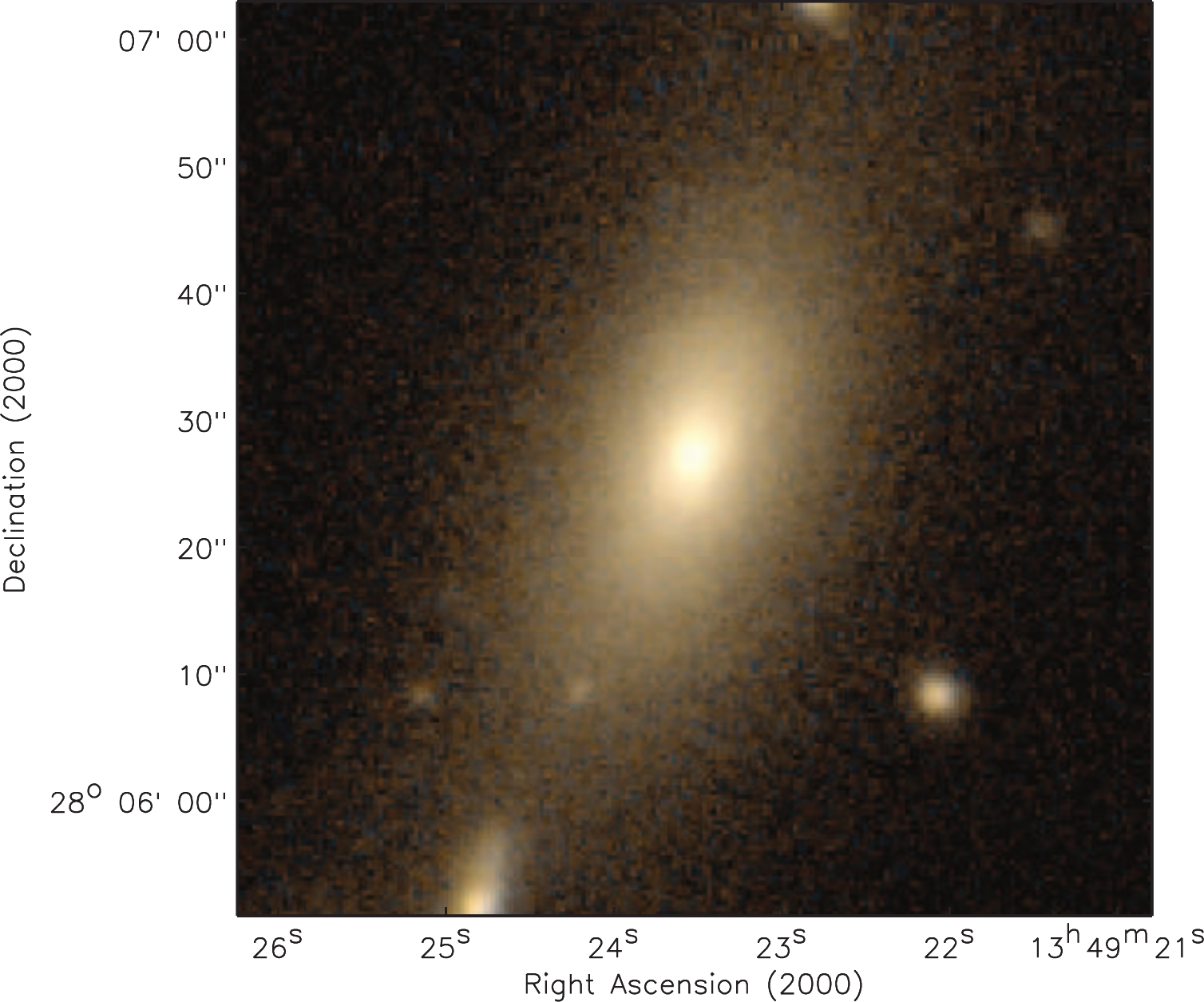} 
\includegraphics[clip,angle=0,width=0.3\hsize]{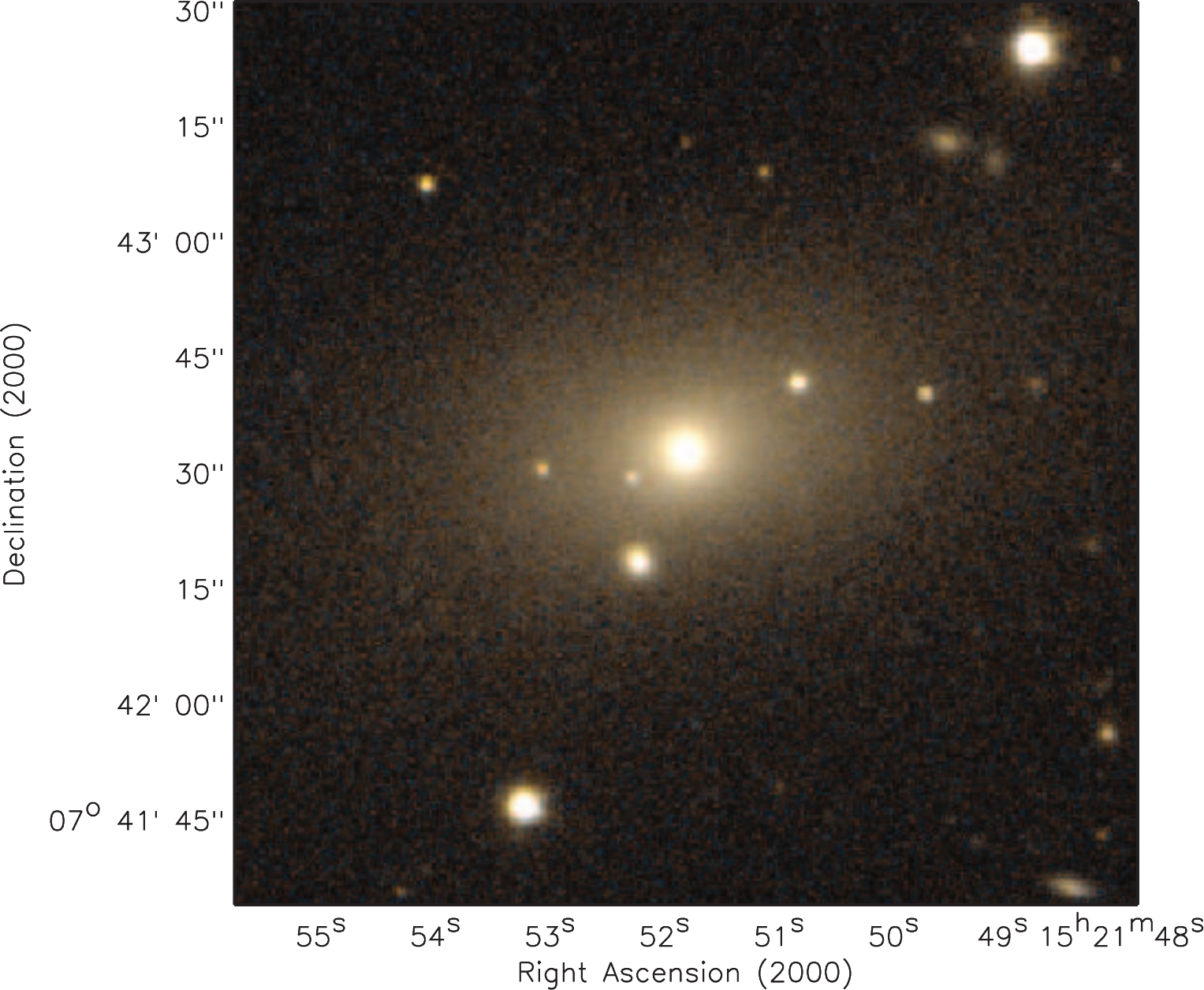} 
\includegraphics[clip,angle=0,width=0.3\hsize]{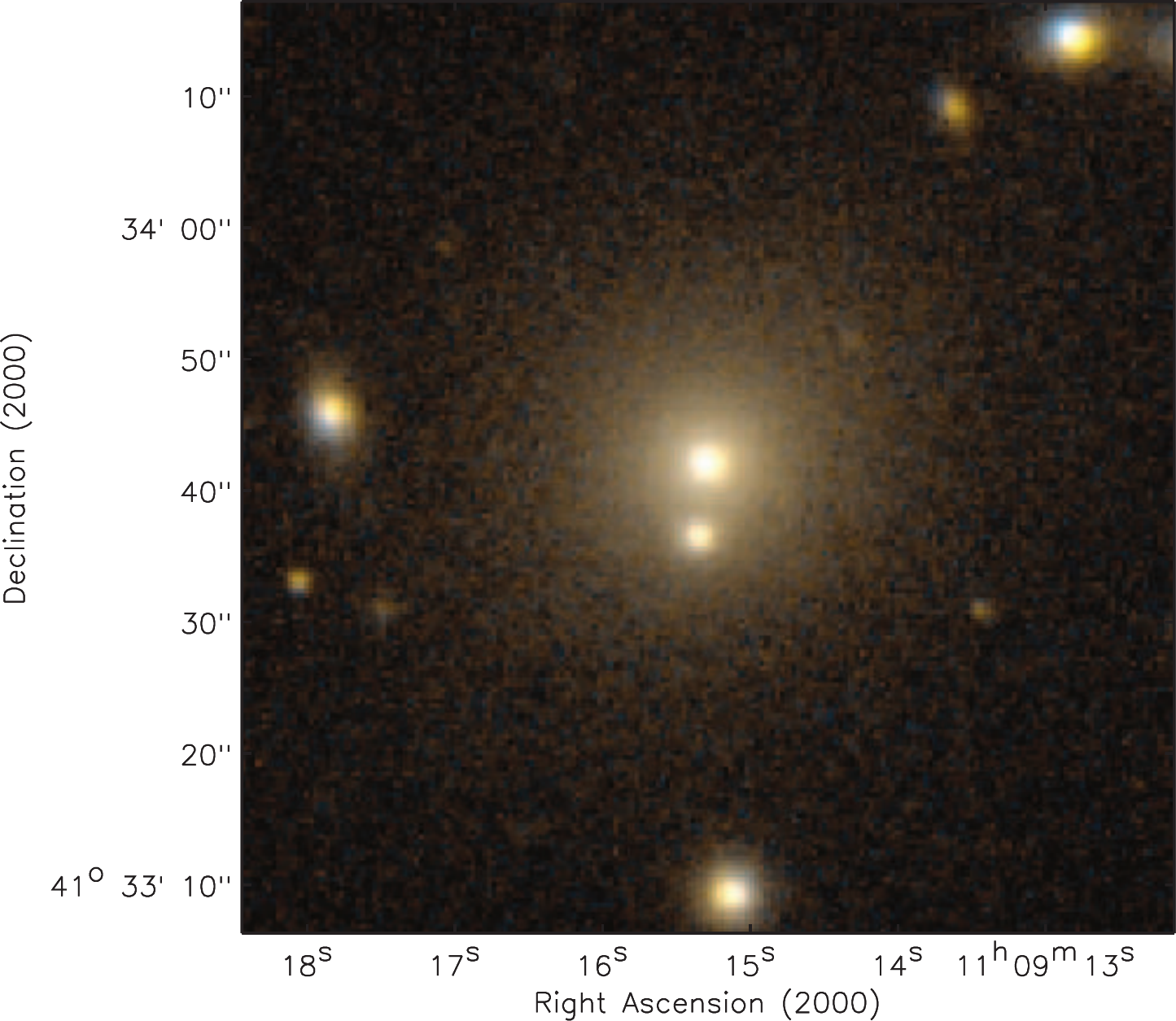} 
\includegraphics[clip,angle=0,width=0.3\hsize]{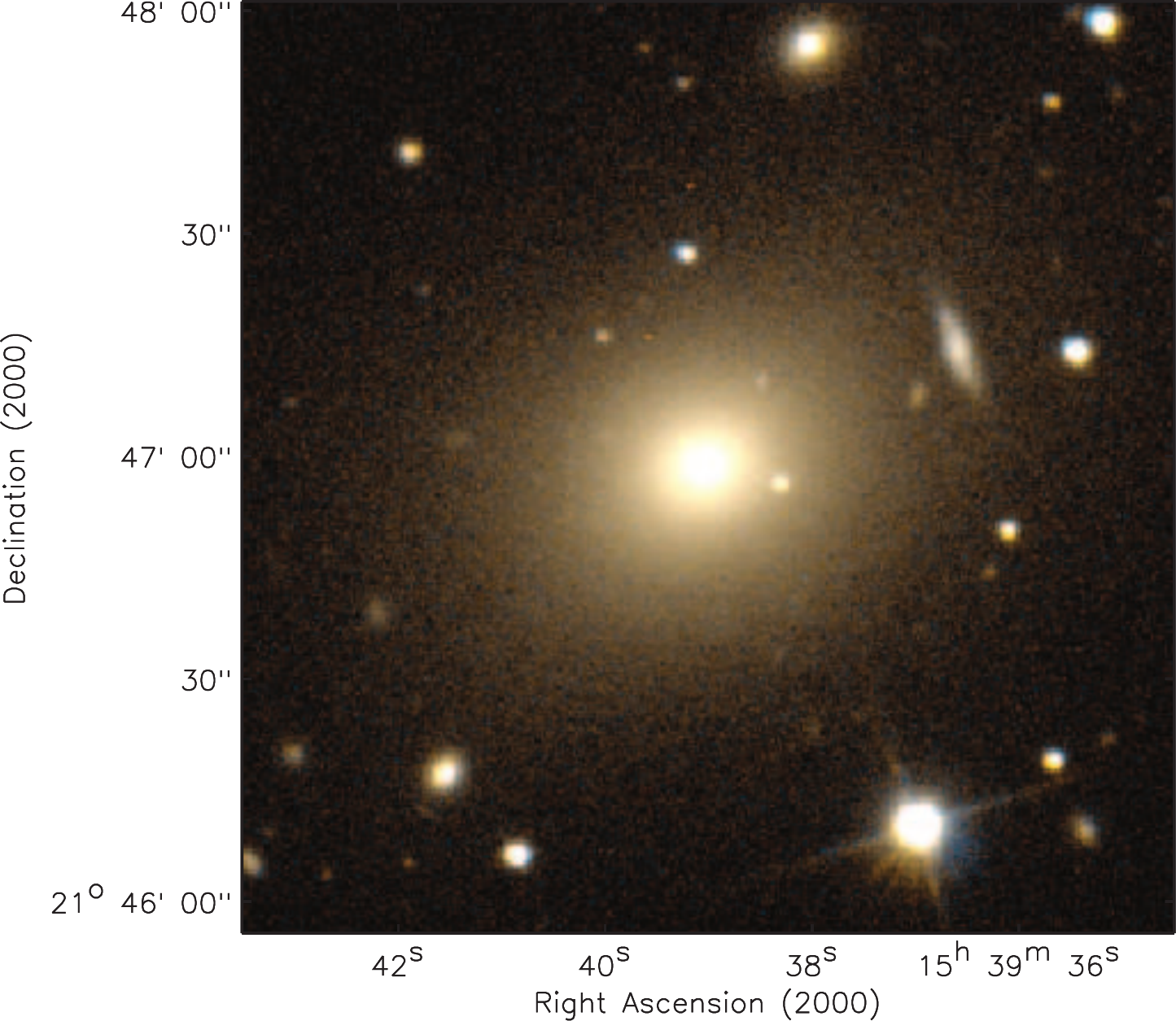} 
\includegraphics[clip,angle=0,width=0.3\hsize]{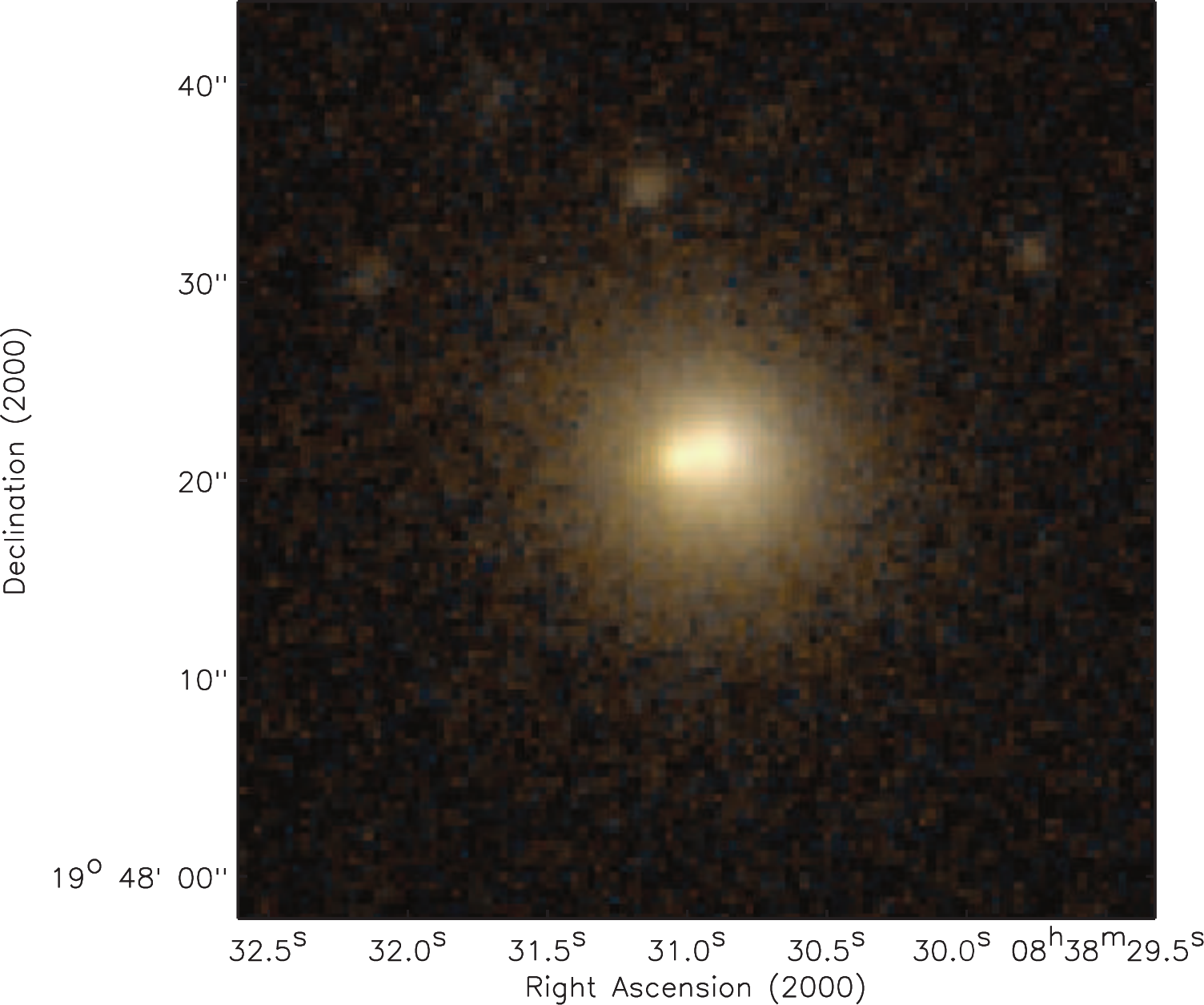} 
\includegraphics[clip,angle=0,width=0.3\hsize]{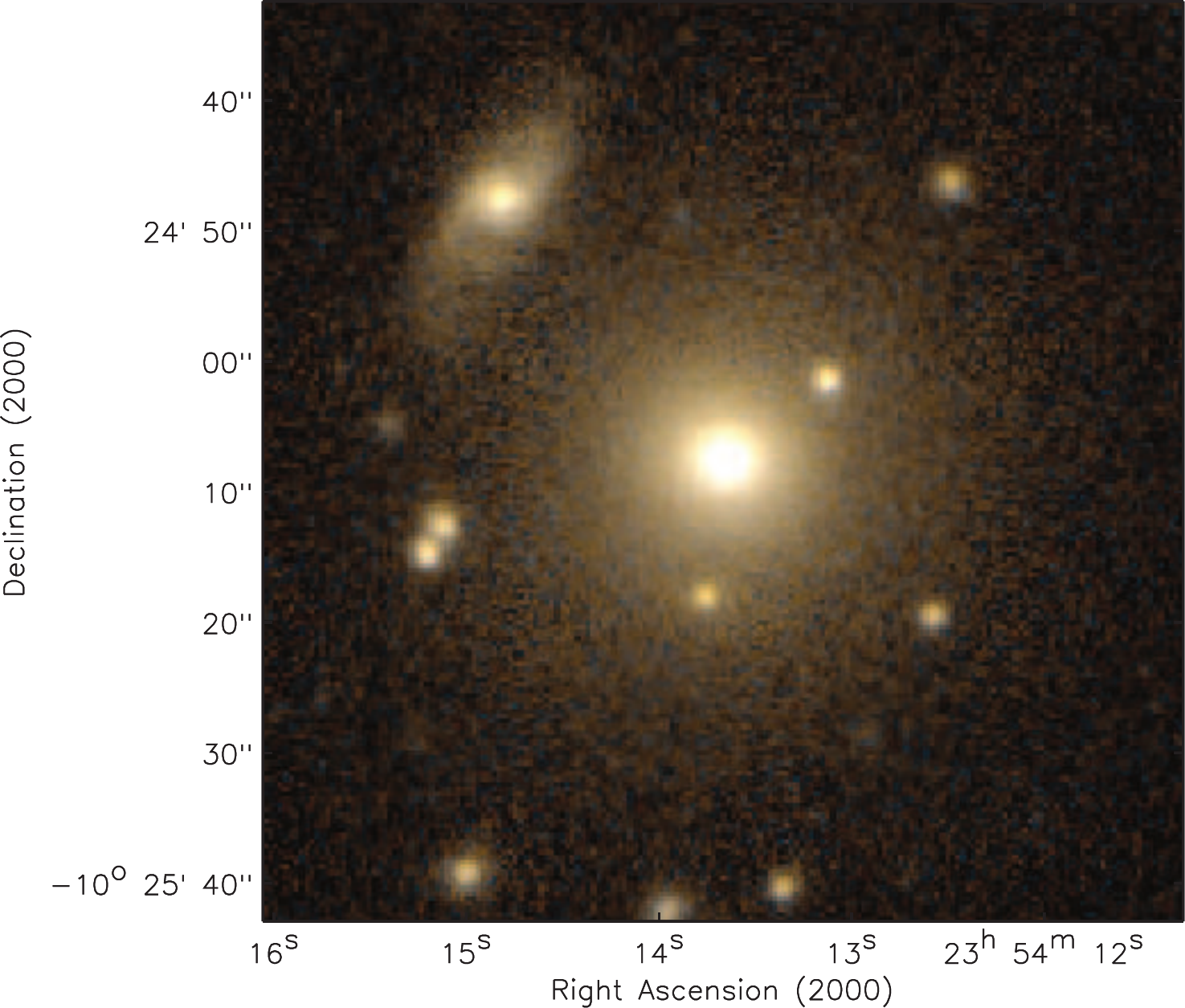} 
\caption{Nine examples of BCGs in the low-redshift MCXC/SDSS sample. Color images were generated from SDSS $u^{\prime}g^{\prime}r^{\prime}$ imaging. The physical size of all postage stamps is 50$kpc$ on a side. The BCG associated with a random MCXC cluster in the lowest (virial) mass bin is shown in the top left postage stamp. The mass of each cluster associated with each BCG increases from left to right followed by top to bottom, with the  BCG associated with a random MCXC cluster in the highest mass bin shown in the bottom right panel (see text). The morphology of the MCXC BCG sample is more homogeneous than the Cl1604 BCGs (see Figure \ref{fig:highzBCG}), as the former are comprised almost entirely of elliptical galaxies which appear largely undisturbed.} 
\label{fig:lowzBCG}
\end{figure*}

\begin{figure*}
\centering
\includegraphics[clip,angle=0,width=0.3\hsize]{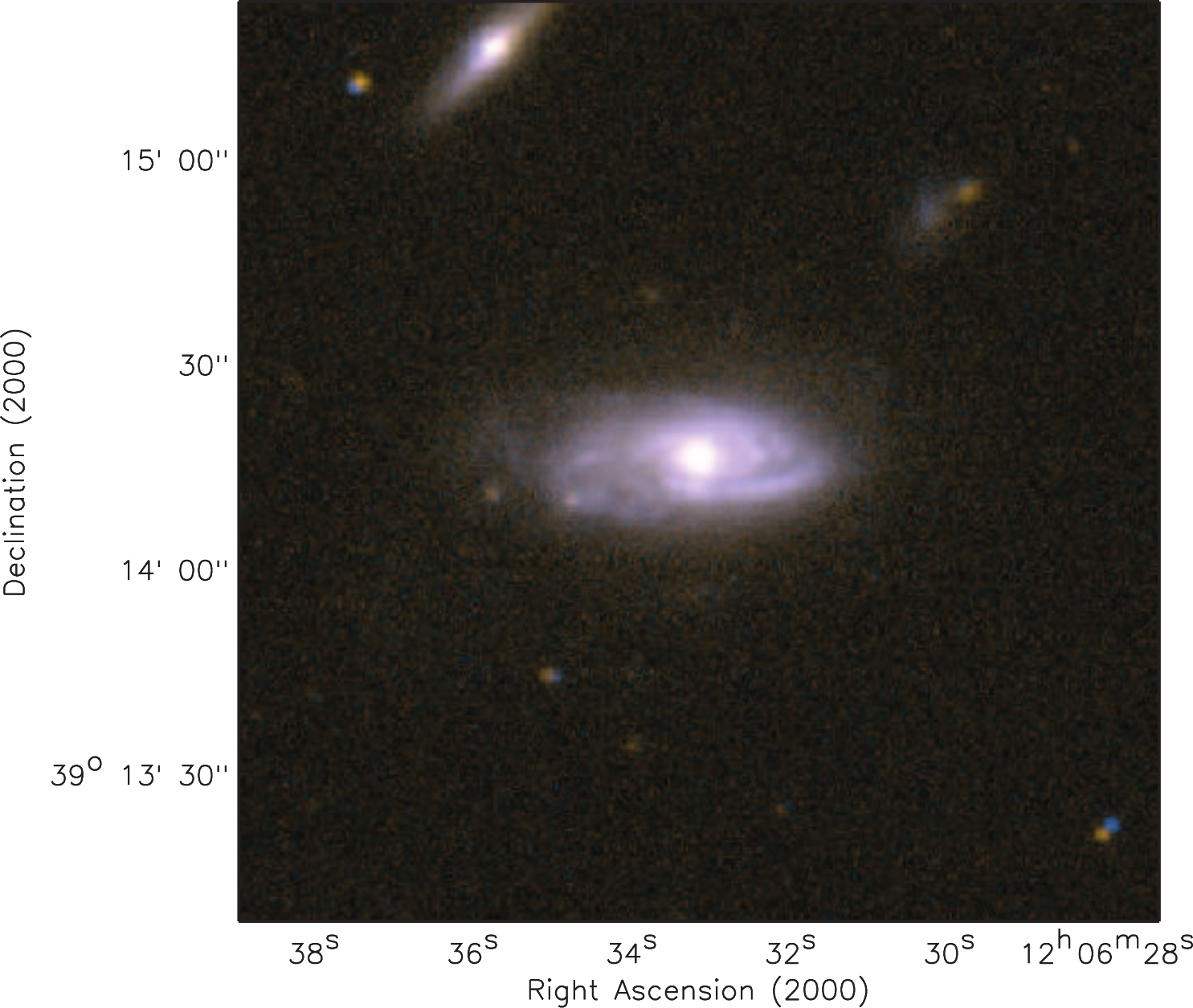} 
\includegraphics[clip,angle=0,width=0.3\hsize]{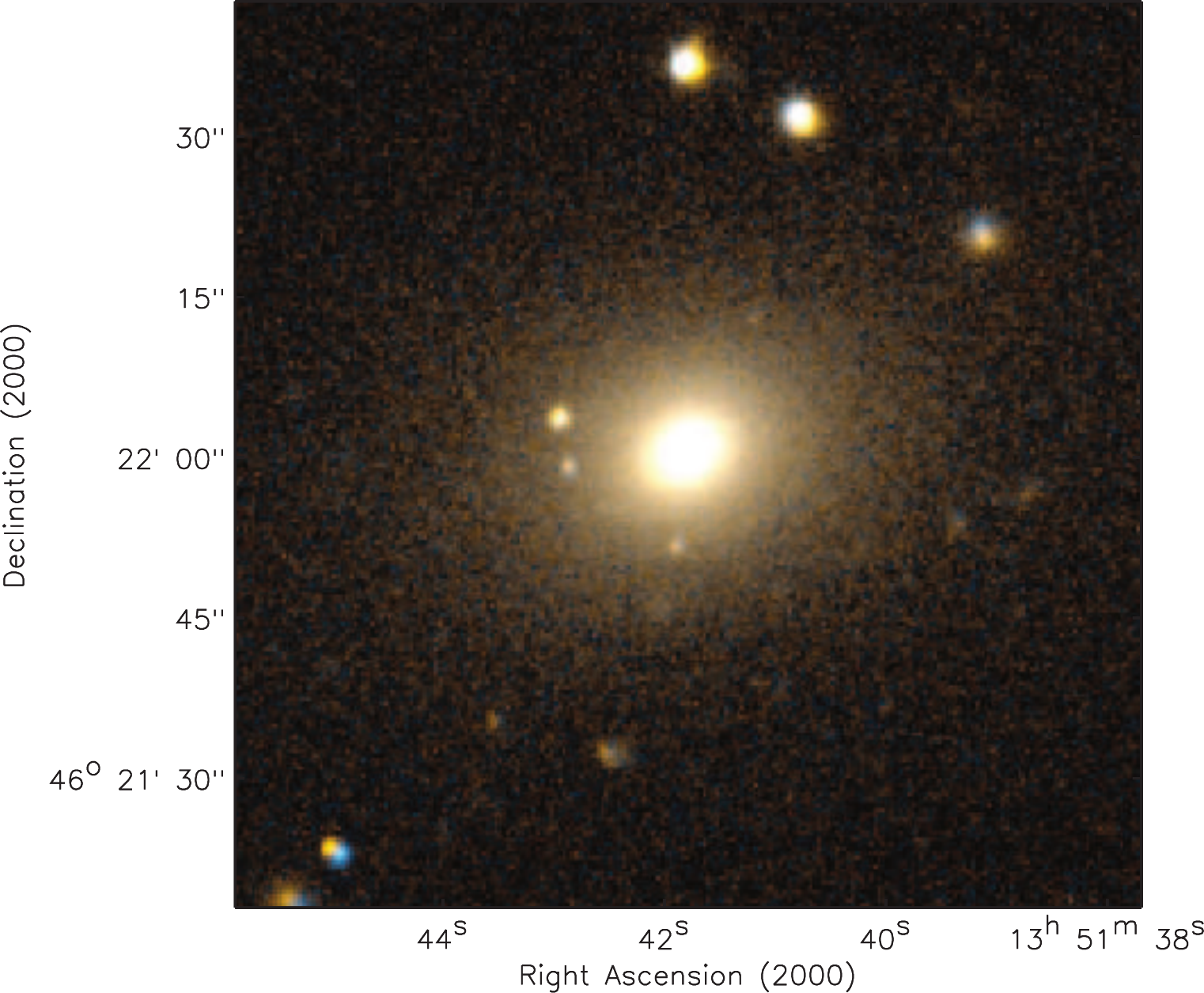} 
\includegraphics[clip,angle=0,width=0.3\hsize]{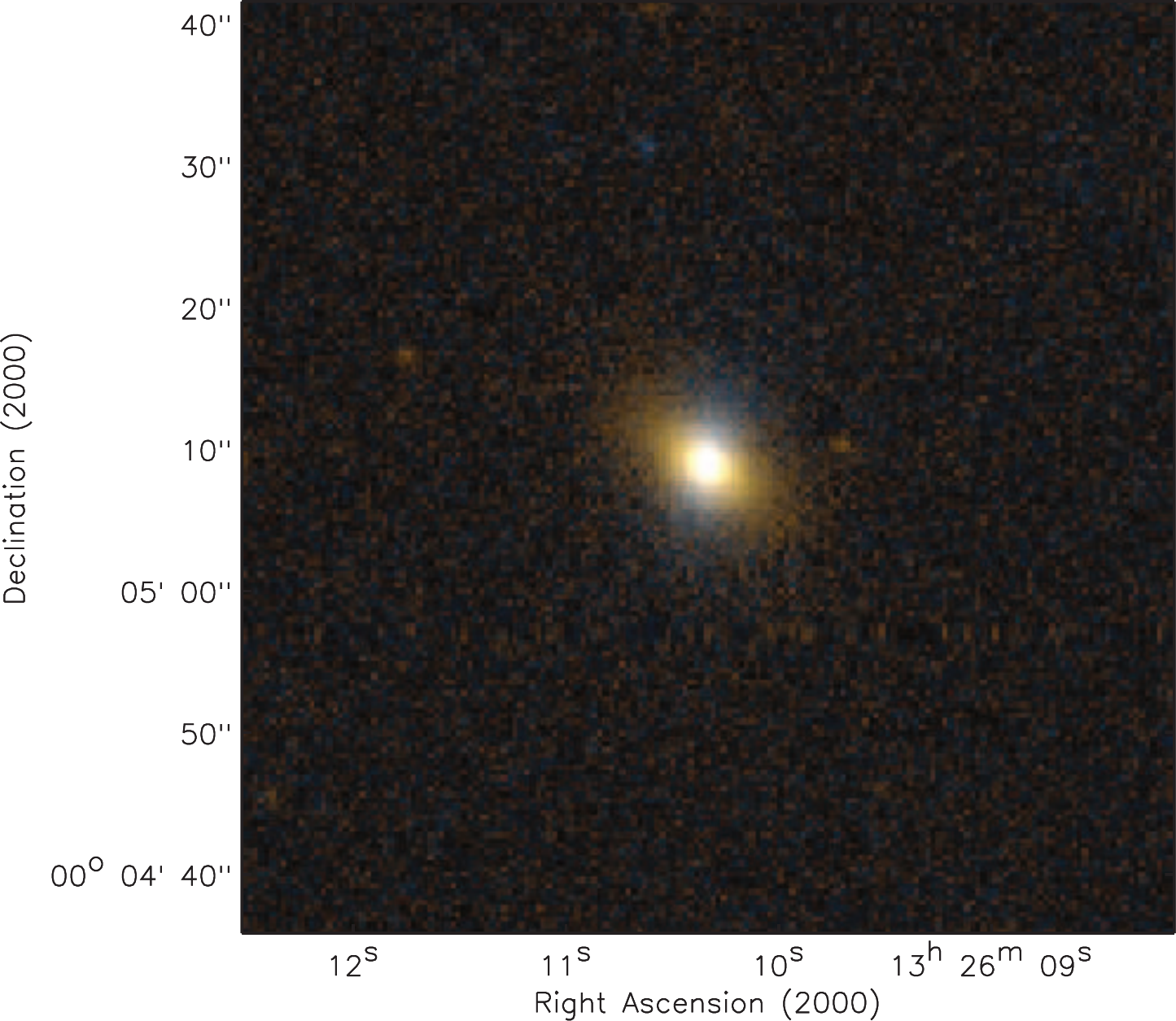} 
\includegraphics[clip,angle=0,width=0.3\hsize]{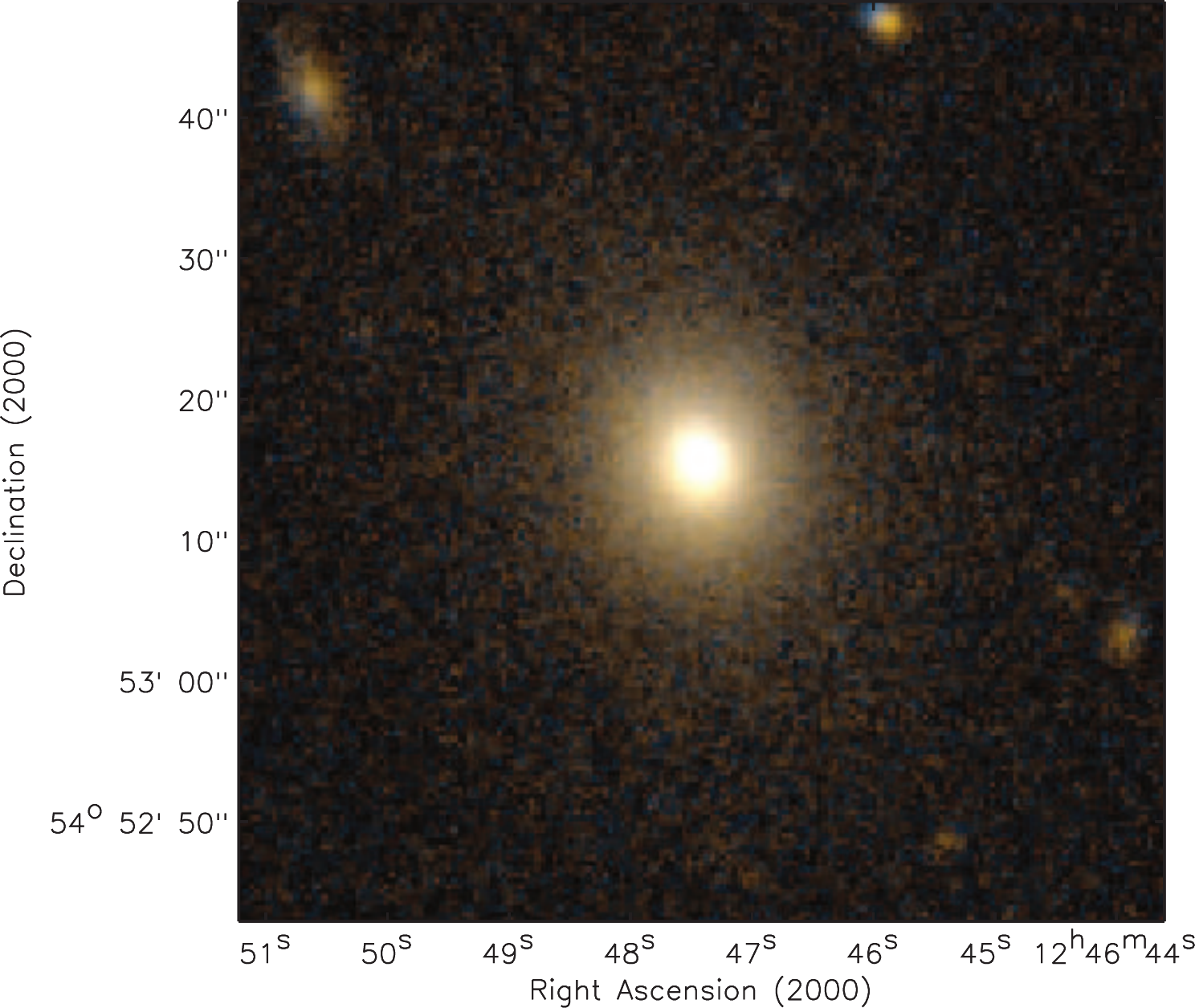} 
\includegraphics[clip,angle=0,width=0.3\hsize]{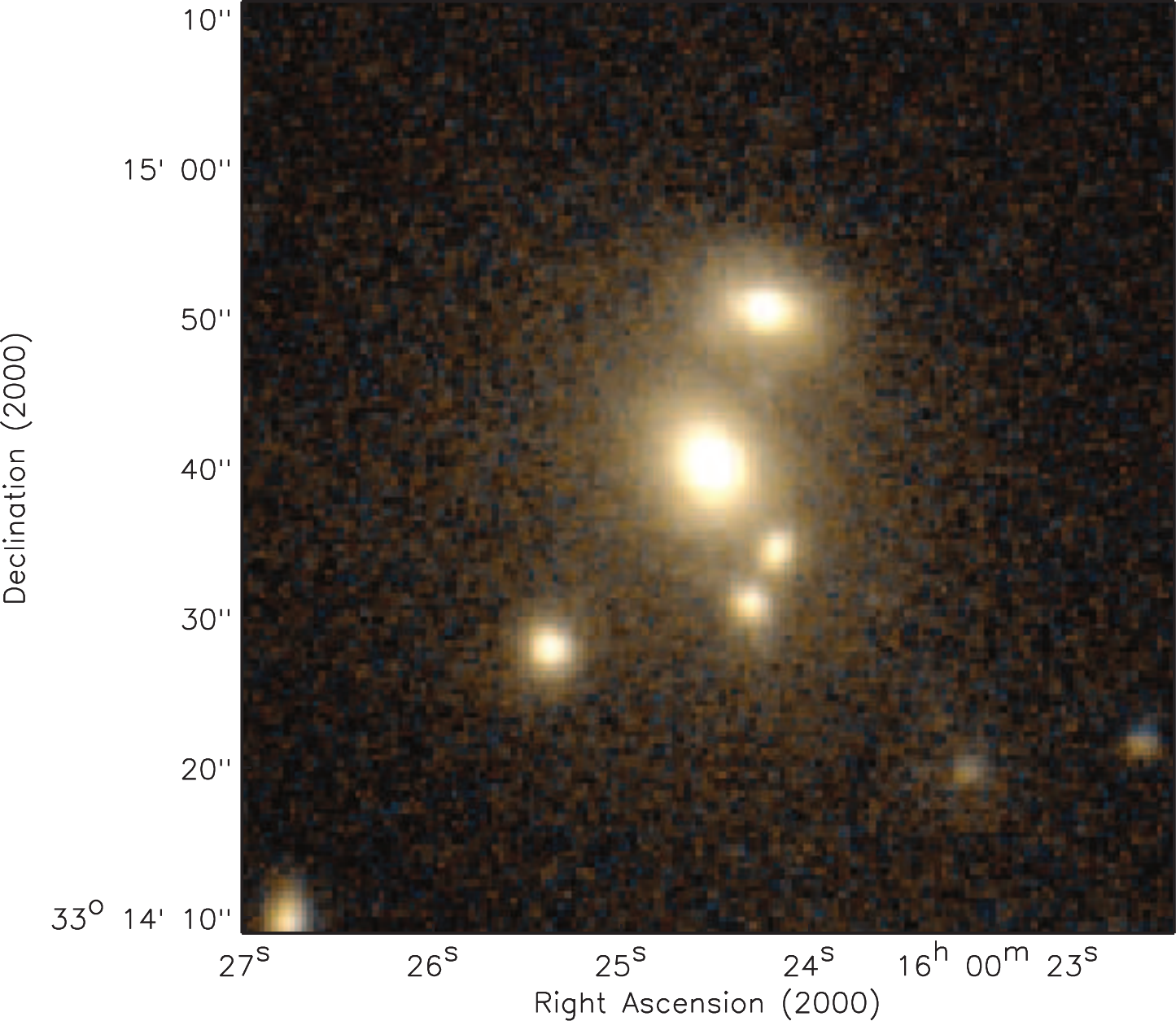} 
\includegraphics[clip,angle=0,width=0.3\hsize]{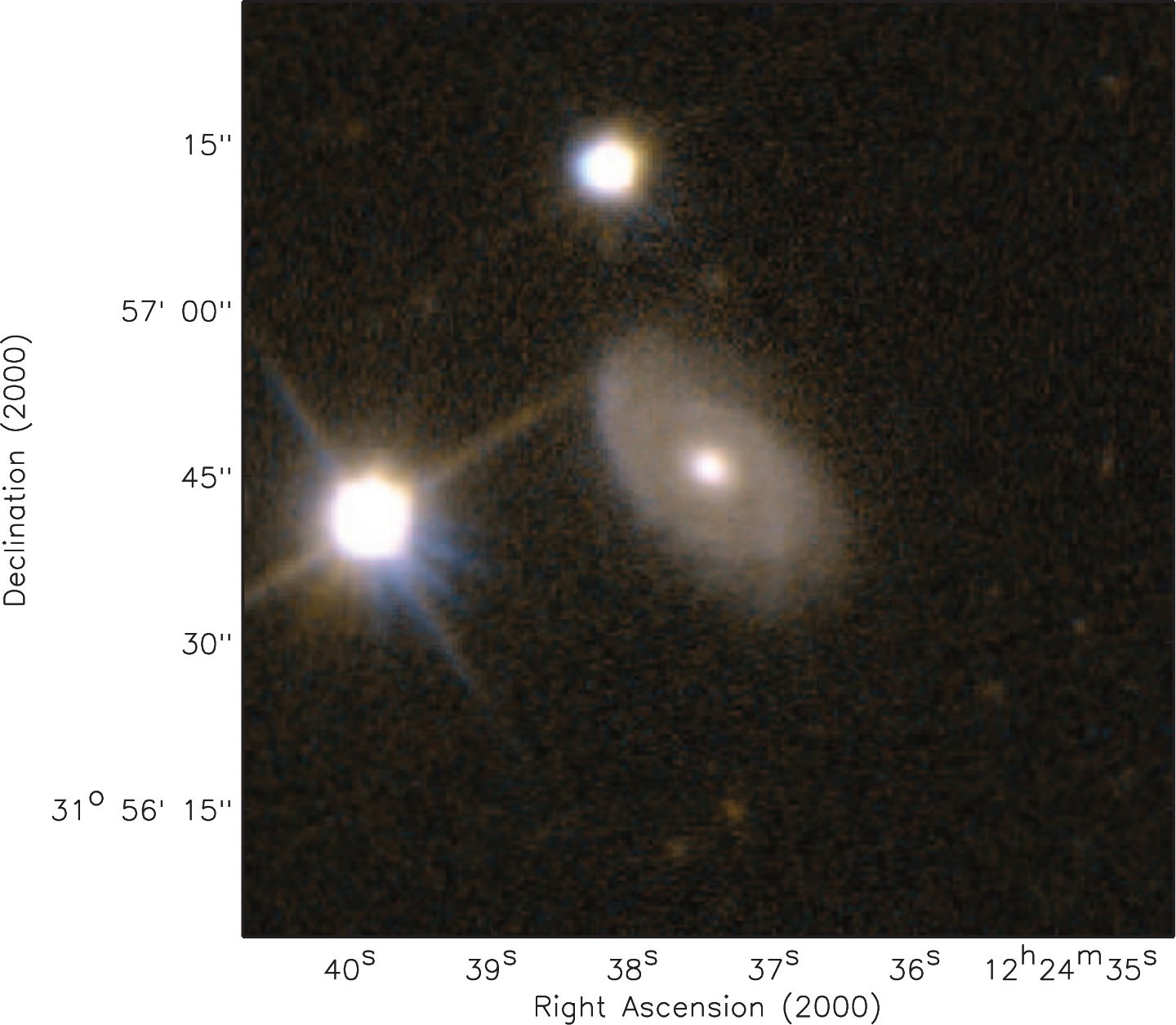} 
\includegraphics[clip,angle=0,width=0.3\hsize]{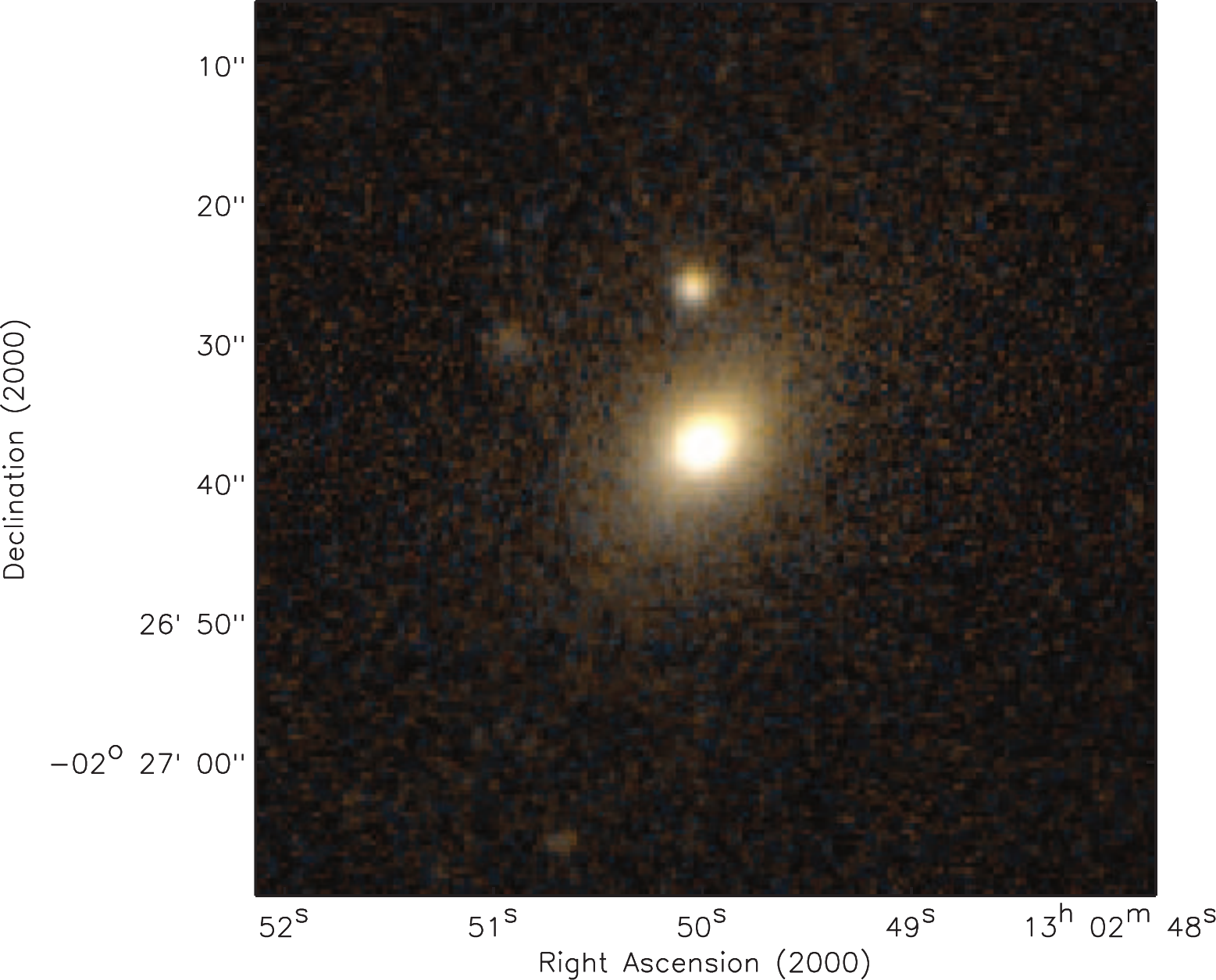} 
\includegraphics[clip,angle=0,width=0.3\hsize]{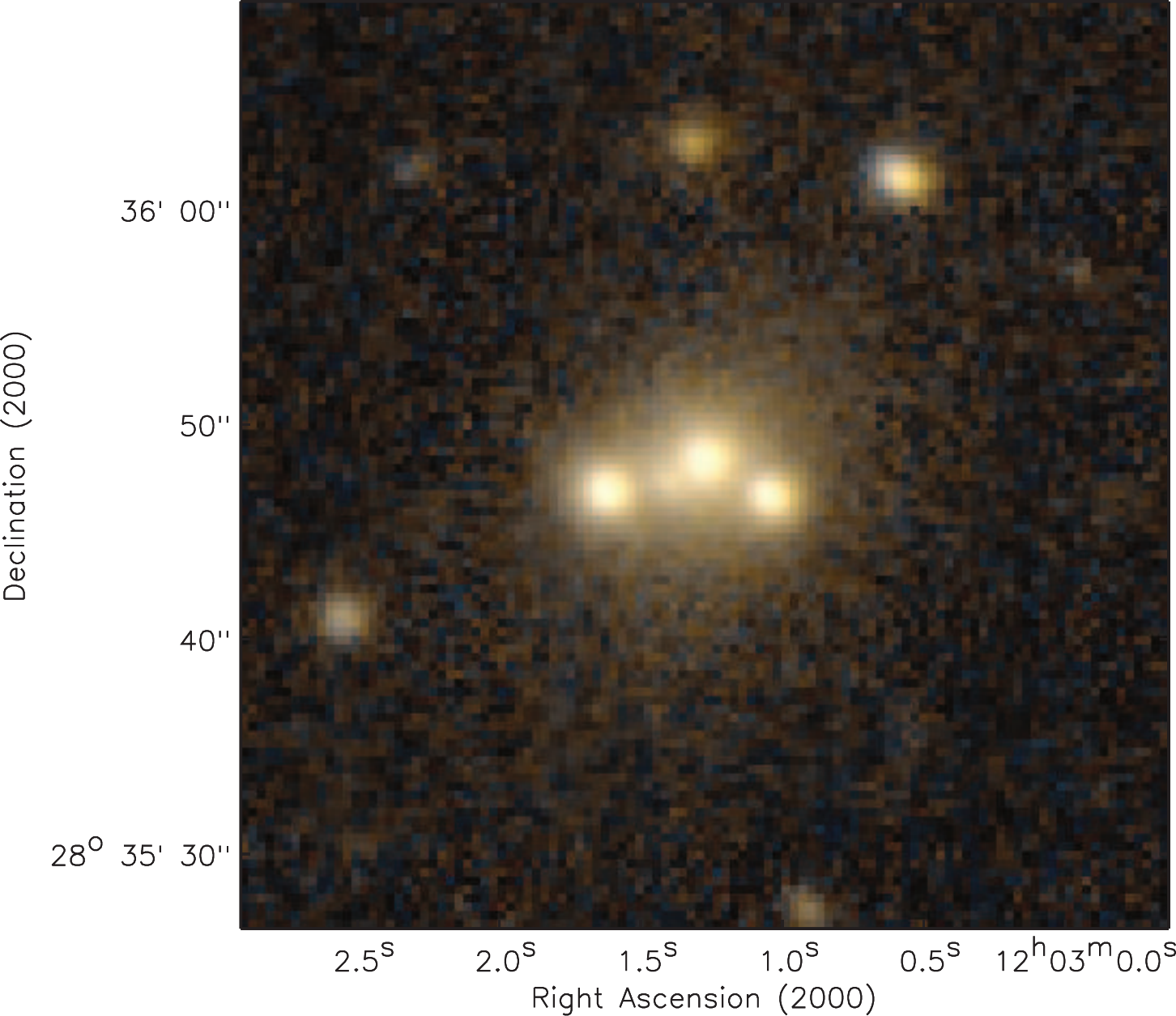} 
\includegraphics[clip,angle=0,width=0.3\hsize]{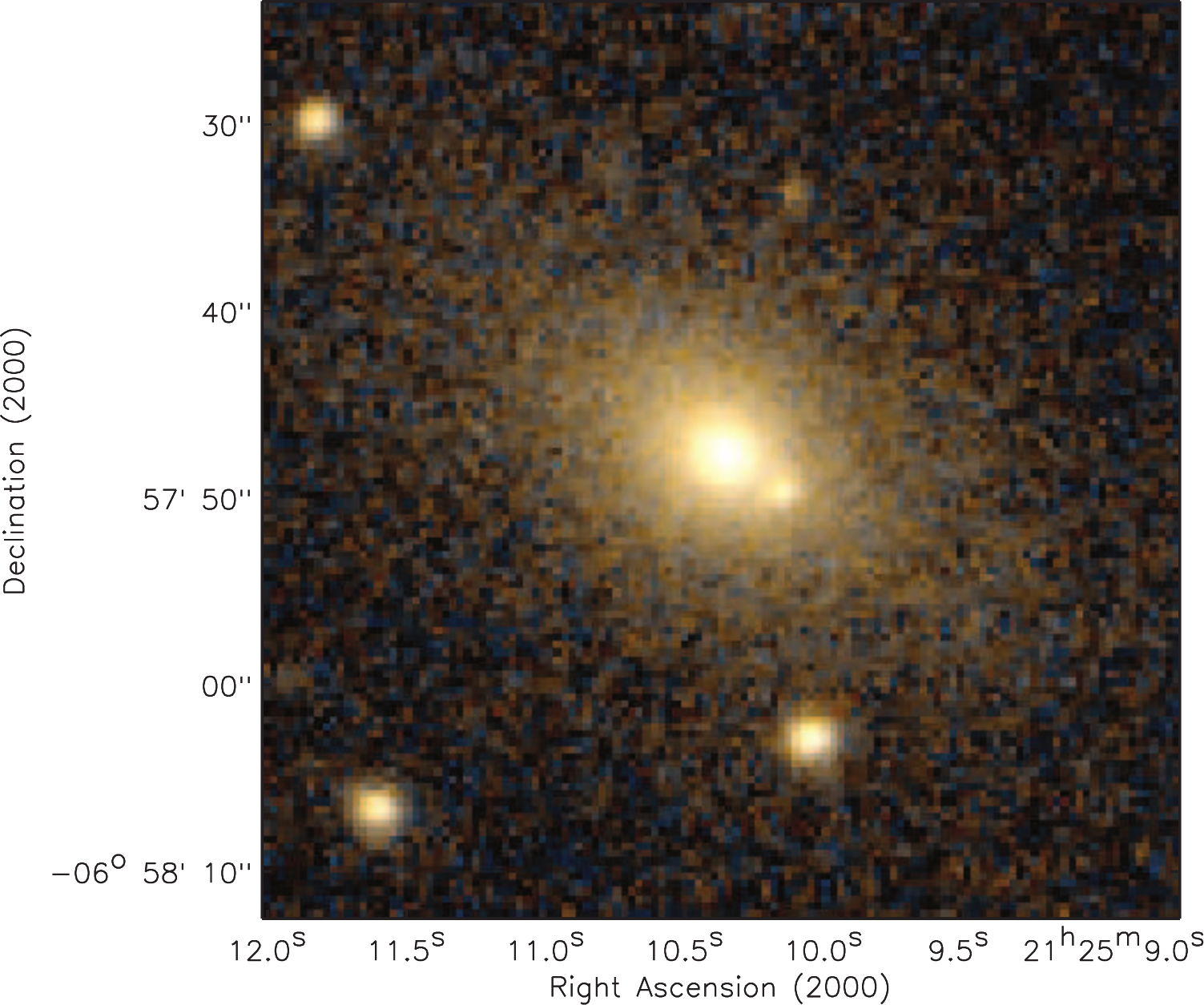} 
\caption{Nine examples of MCXC MMCGs which were not also selected as MCXC BCGs. The size and method used to generate the postage stamps are identical to Figure \ref{fig:lowzBCG}. While the morphology of this sample is less homogeneous than the MCXC BCG sample, only 13 of the 81 MCXC MMCGs were different from the BCG, which means that the overall MMCG sample remains largely homogeneous. As in Figure \ref{fig:lowzBCG}, the MMCG associated with the lowest (virial) mass cluster is shown in the top left postage stamp and the MMCG associated with the highest mass cluster is shown in the bottom right.}
\label{fig:lowzMMCG}
\end{figure*}

\begin{figure*}
\centering
\includegraphics[clip,angle=0,width=0.3\hsize]{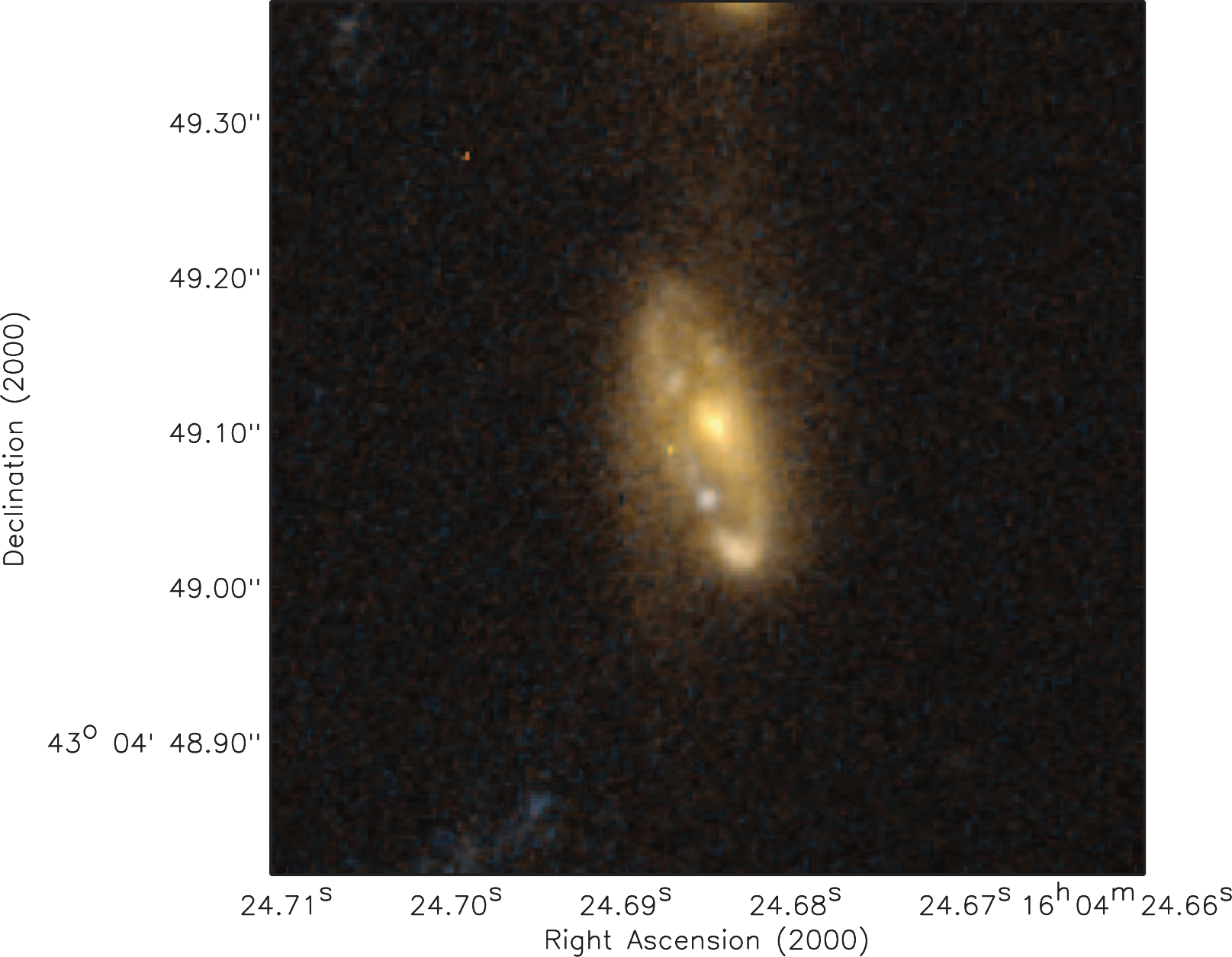}
\includegraphics[clip,angle=0,width=0.3\hsize]{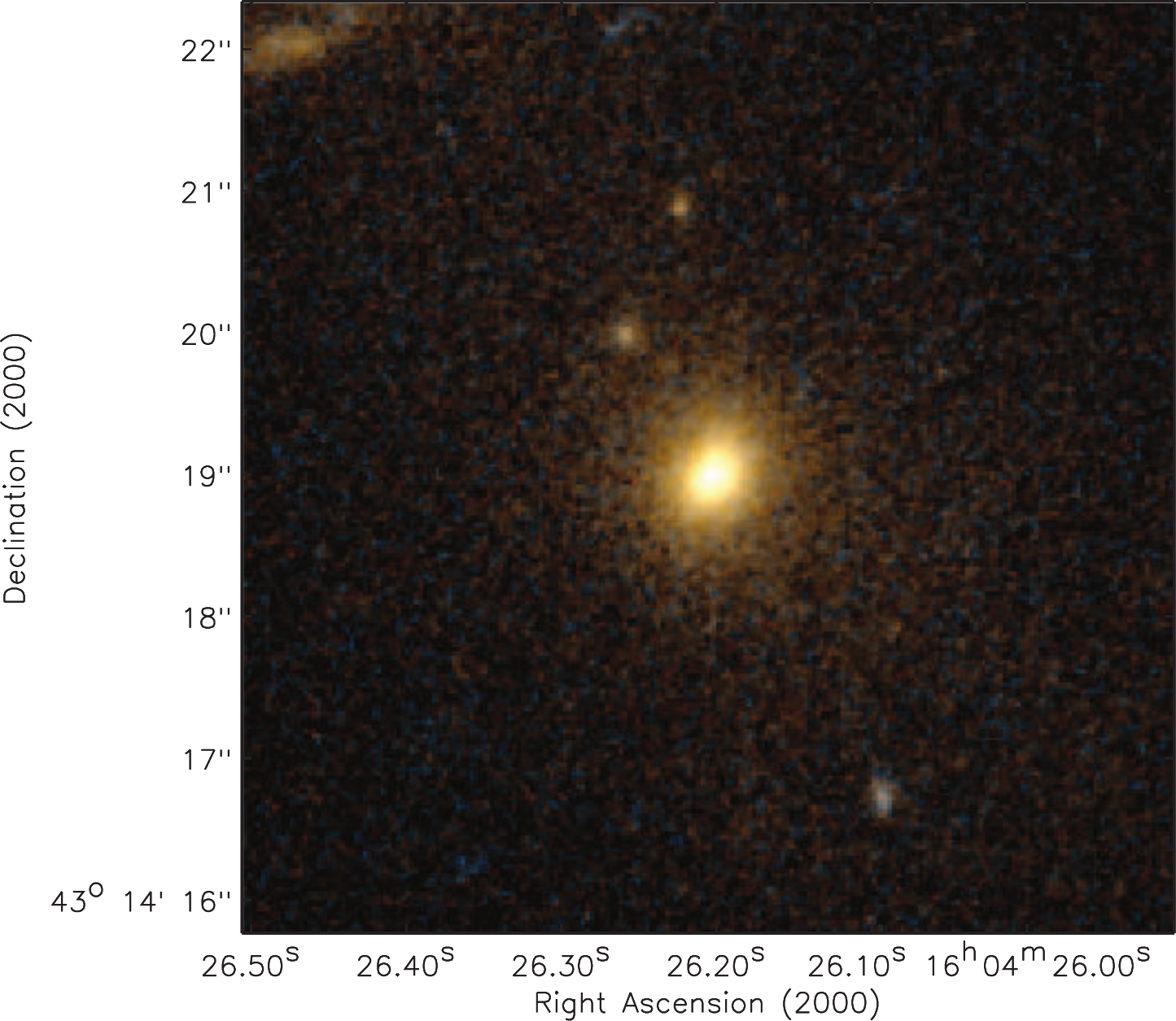}
\includegraphics[clip,angle=0,width=0.3\hsize]{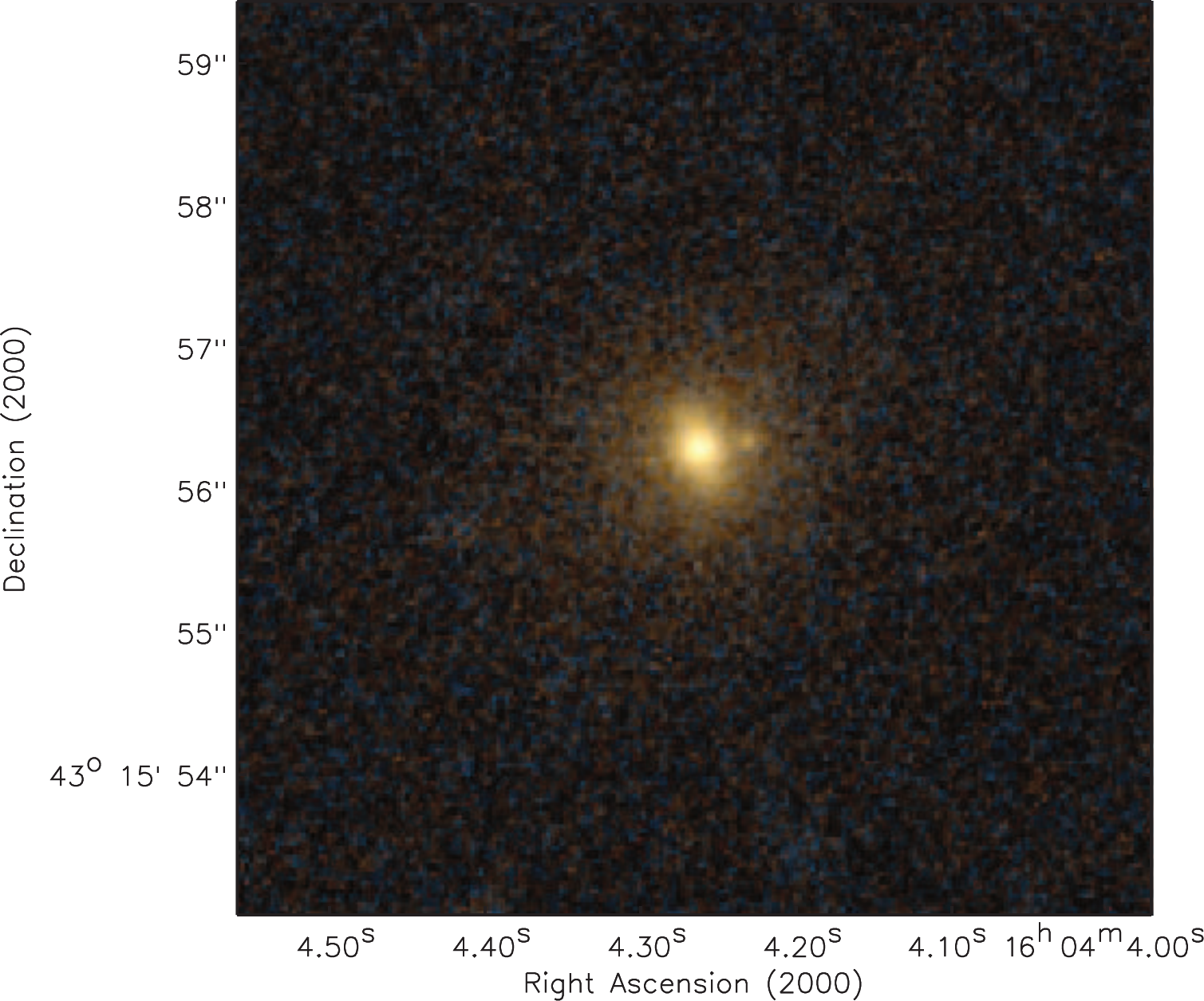} 
\includegraphics[clip,angle=0,width=0.3\hsize]{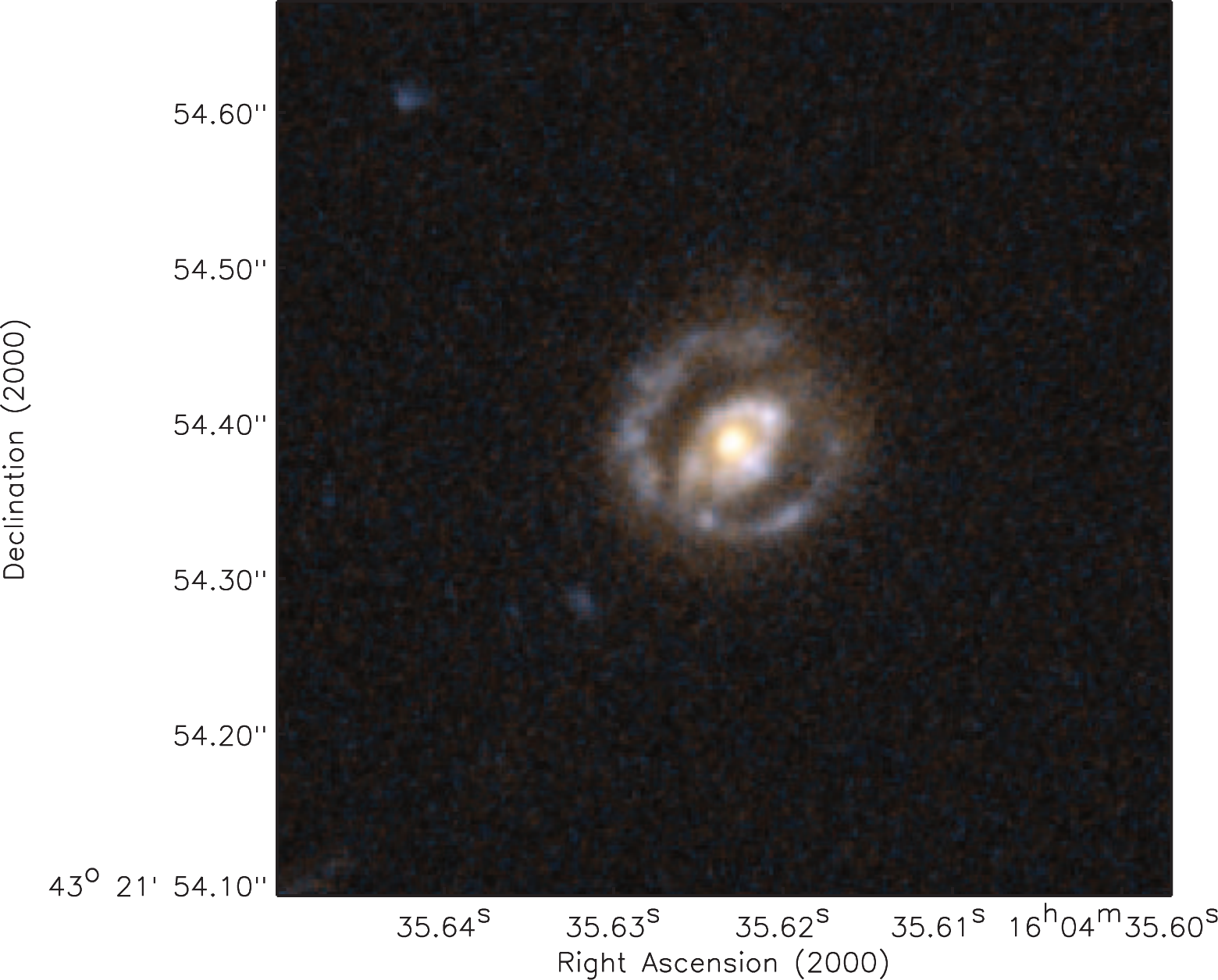} 
\includegraphics[clip,angle=0,width=0.3\hsize]{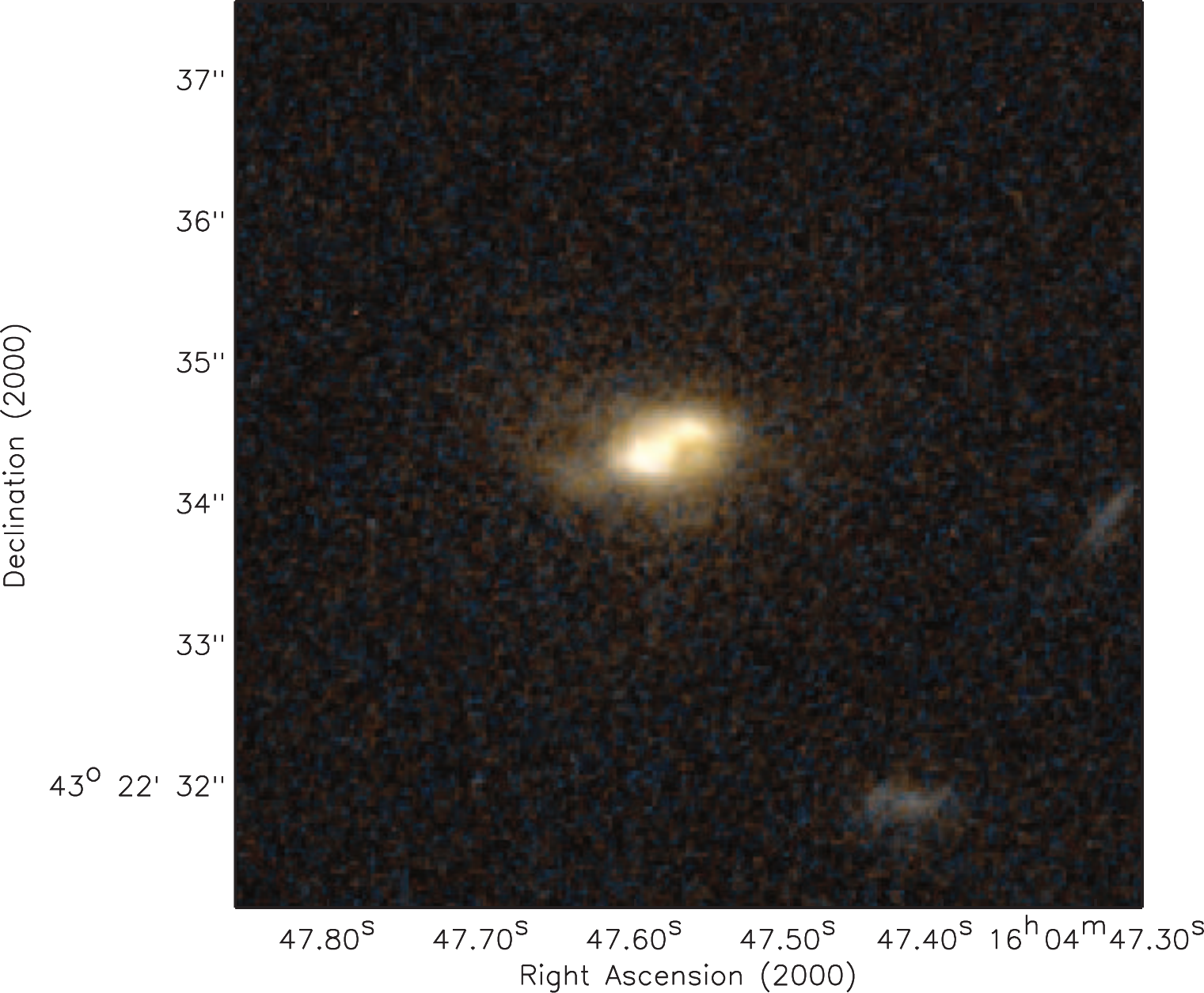} 
\includegraphics[clip,angle=0,width=0.3\hsize]{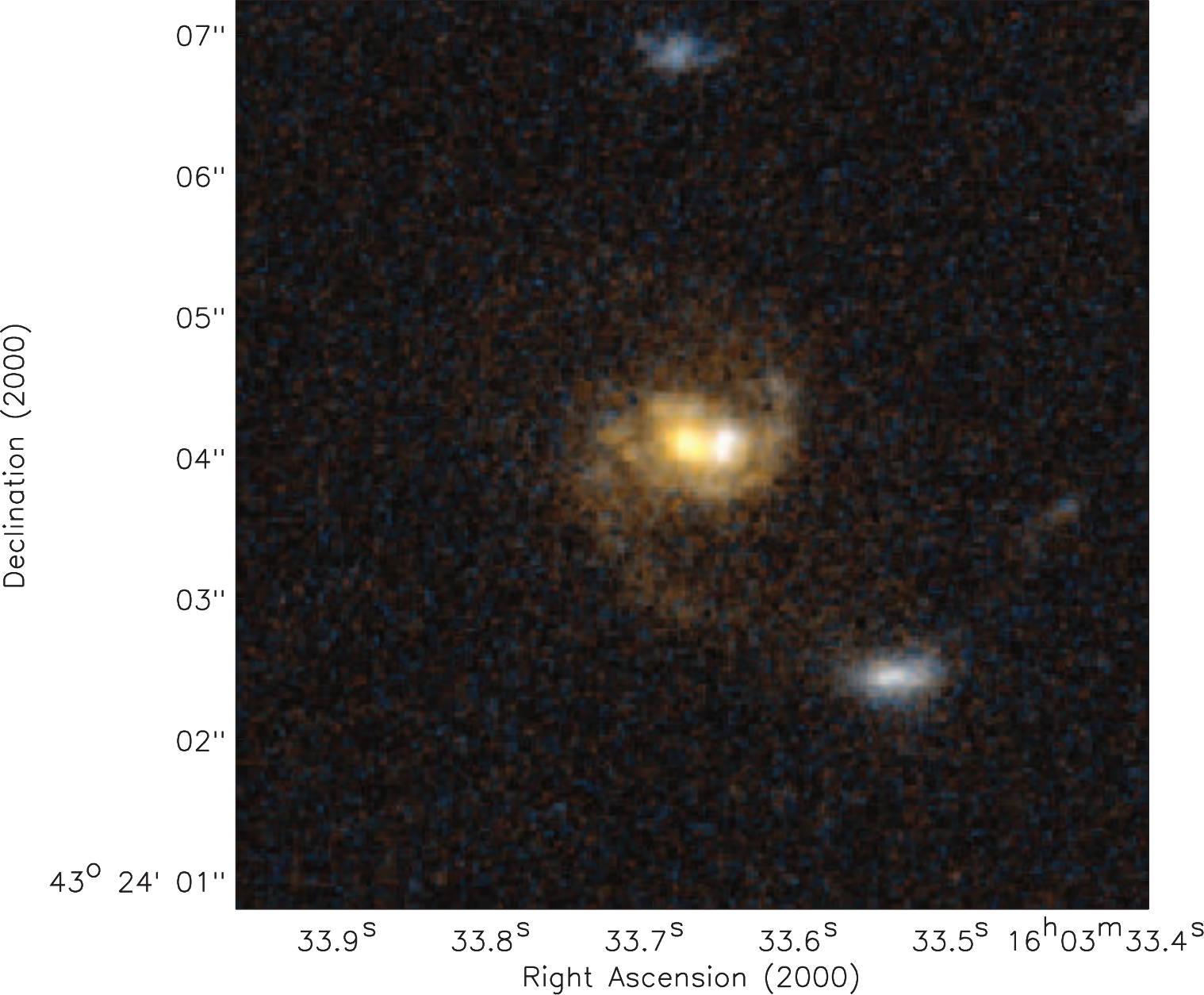} 
\includegraphics[clip,angle=0,width=0.3\hsize]{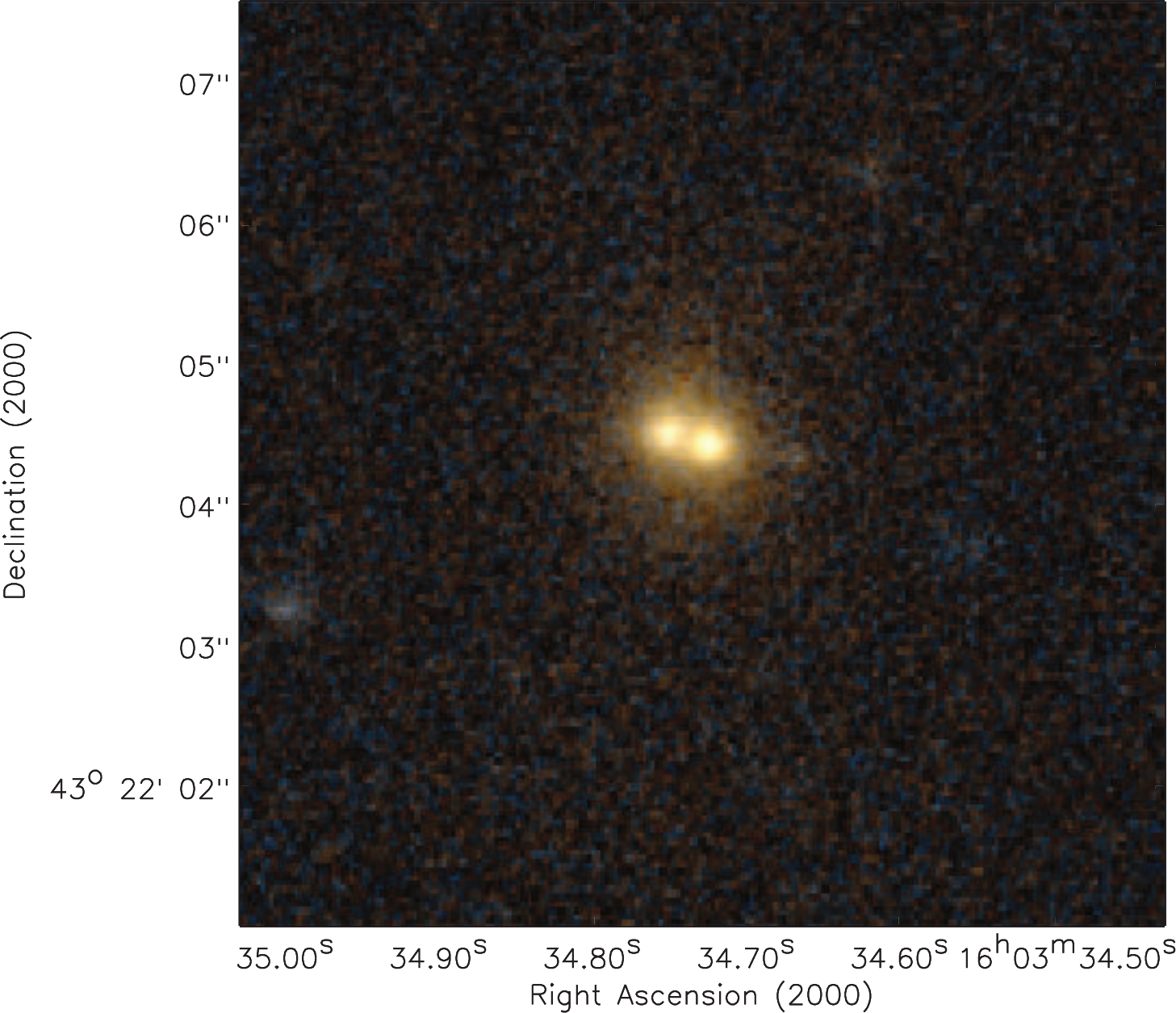} 
\includegraphics[clip,angle=0,width=0.3\hsize]{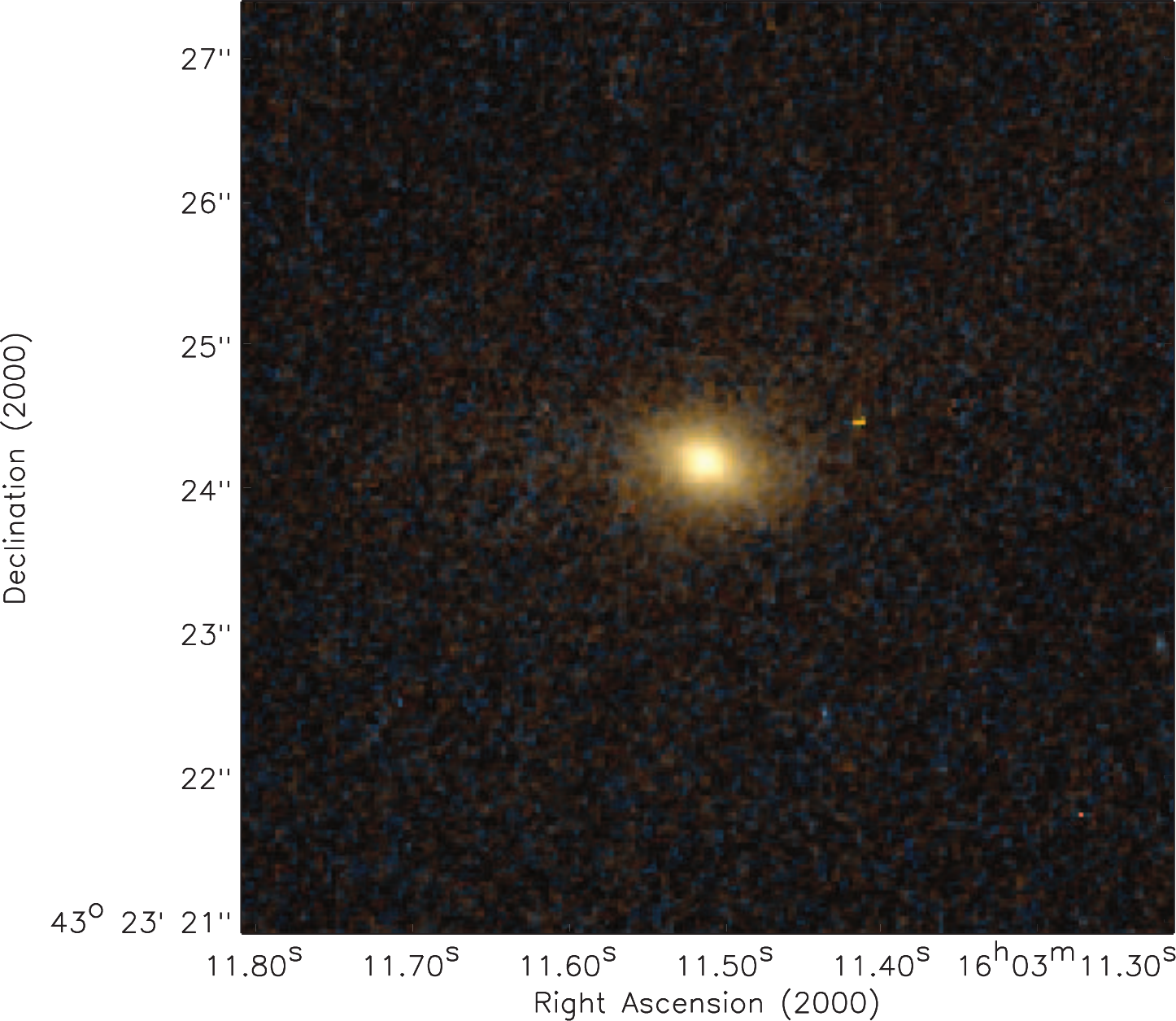} 
\caption{Montage of \emph{HST} ACS color postage stamps of the BCGs of the constituent clusters and groups of the Cl1604 supercluster. Postage stamps were generated using the ACS imaging in the $F606W$ band (blue channel), the $F814W$ band (red channel), and an average of the two bands (green channel). As in Figures \ref{fig:lowzBCG} and \ref{fig:lowzMMCG}, the physical size of the postage stamps is 50$kpc$ on a side. The group/cluster associated with each galaxy increases in letter from left to right followed by top to bottom. The galaxies in this sample show a full range of morphologies, from a pure elliptical with hints of an extended halo (cluster B, top middle panel), to a grand design spiral (cluster A, top left), to a barred ring spiral (cluster D, middle left), to a double-cored elliptical (group H, bottom left).}  
\label{fig:highzBCG}
\end{figure*}

\begin{figure*}
\centering
\includegraphics[clip,angle=0,width=0.3\hsize]{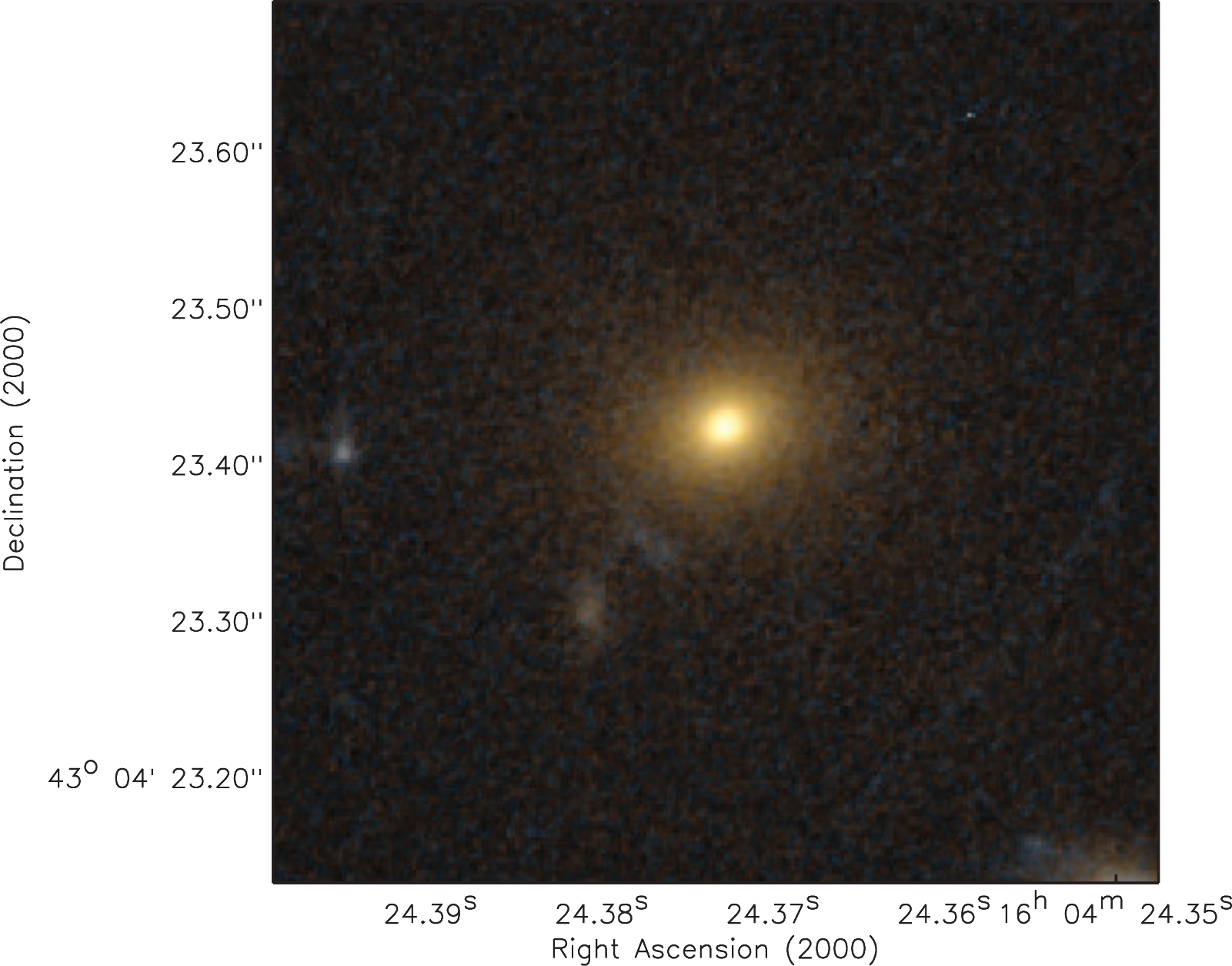} 
\includegraphics[clip,angle=0,width=0.3\hsize]{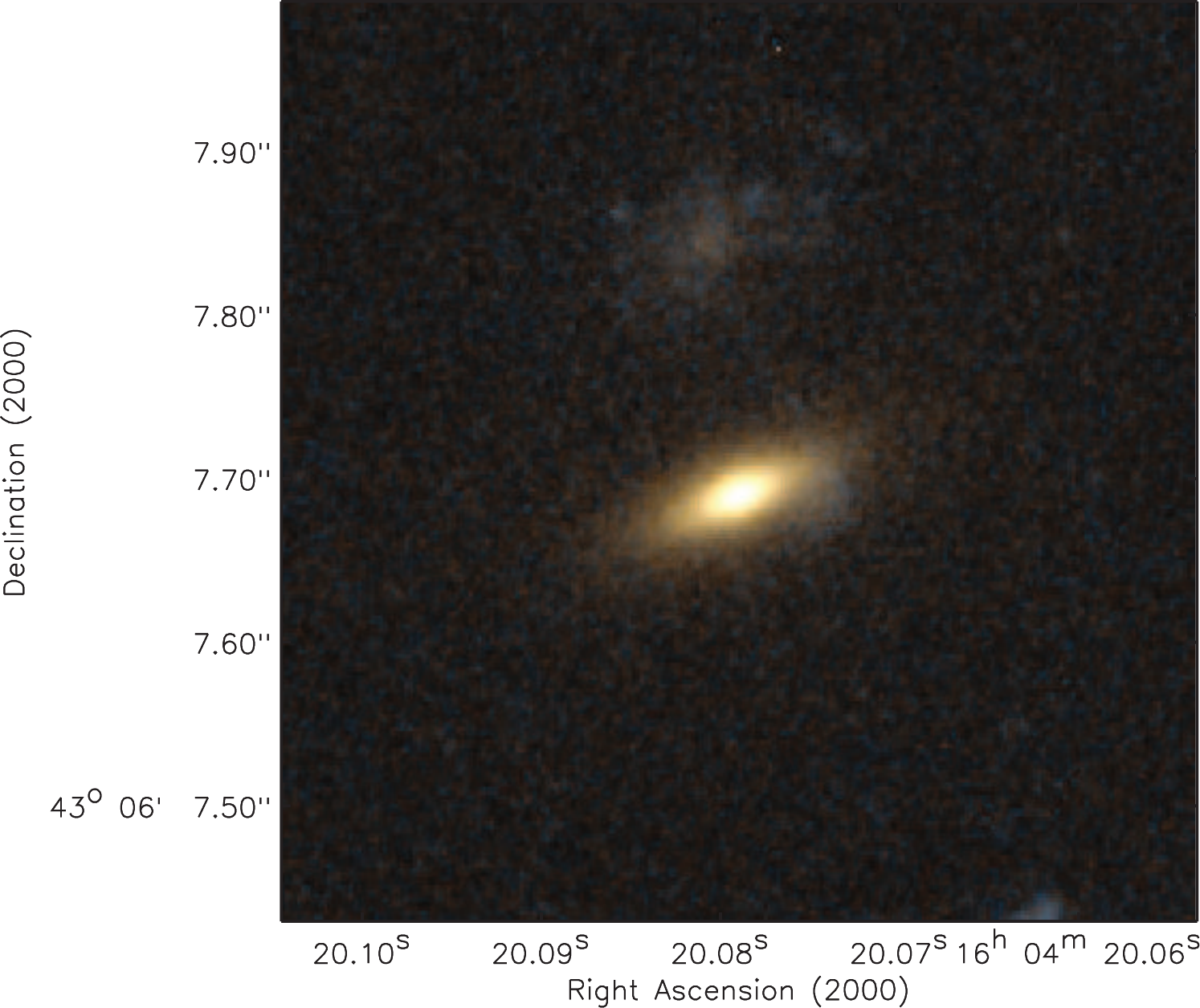} 
\includegraphics[clip,angle=0,width=0.3\hsize]{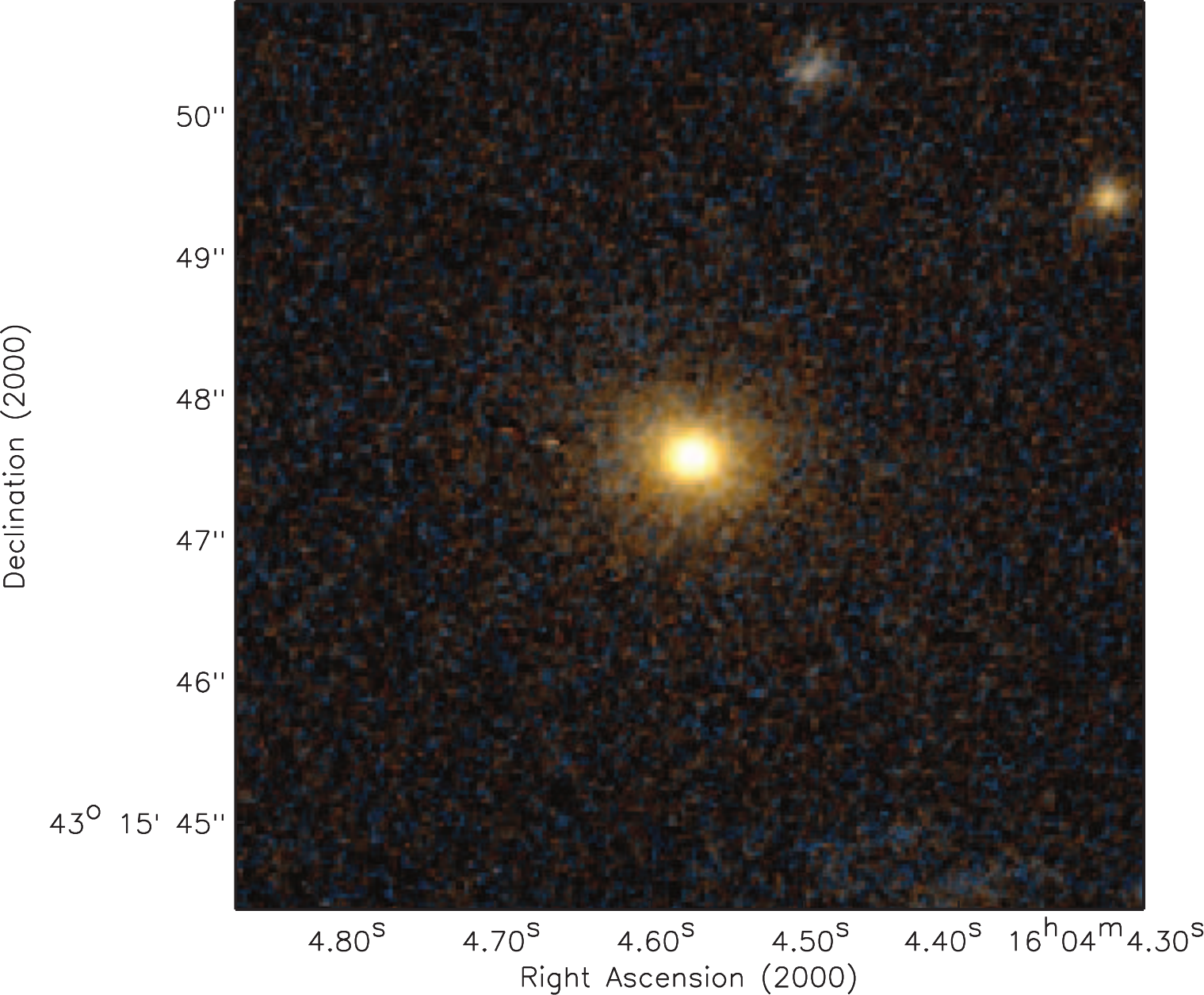} 
\includegraphics[clip,angle=0,width=0.3\hsize]{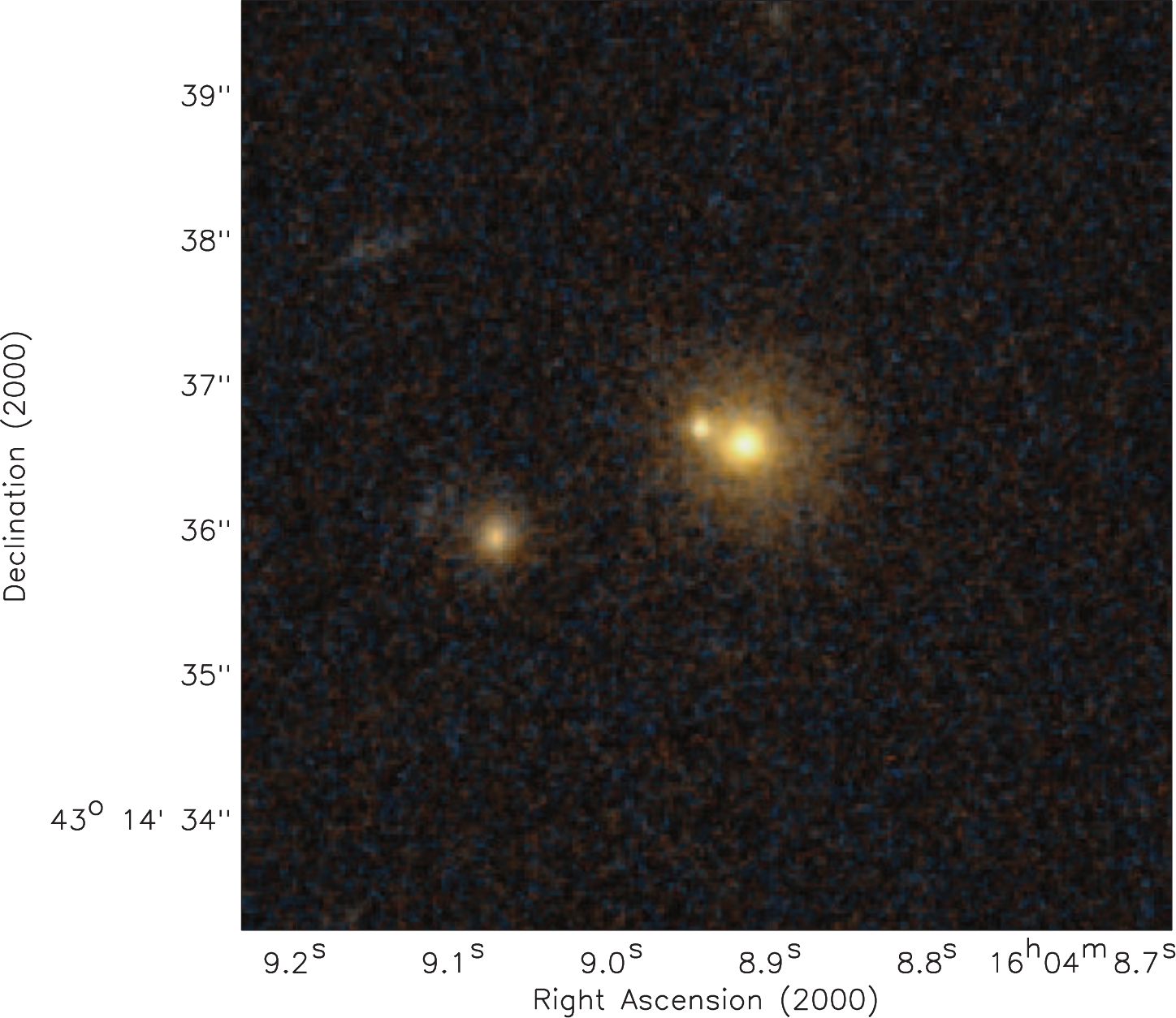} 
\includegraphics[clip,angle=0,width=0.3\hsize]{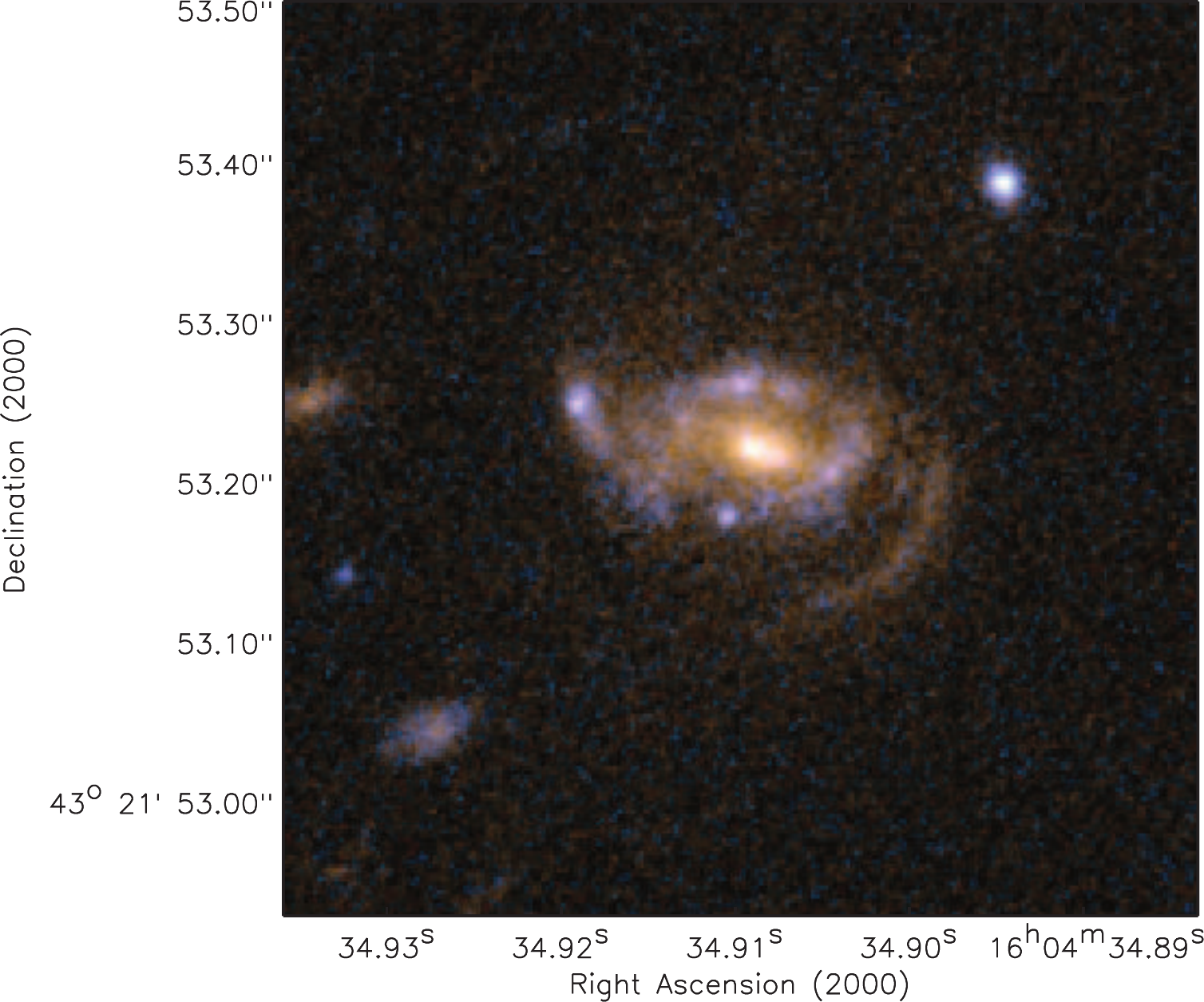} 
\includegraphics[clip,angle=0,width=0.3\hsize]{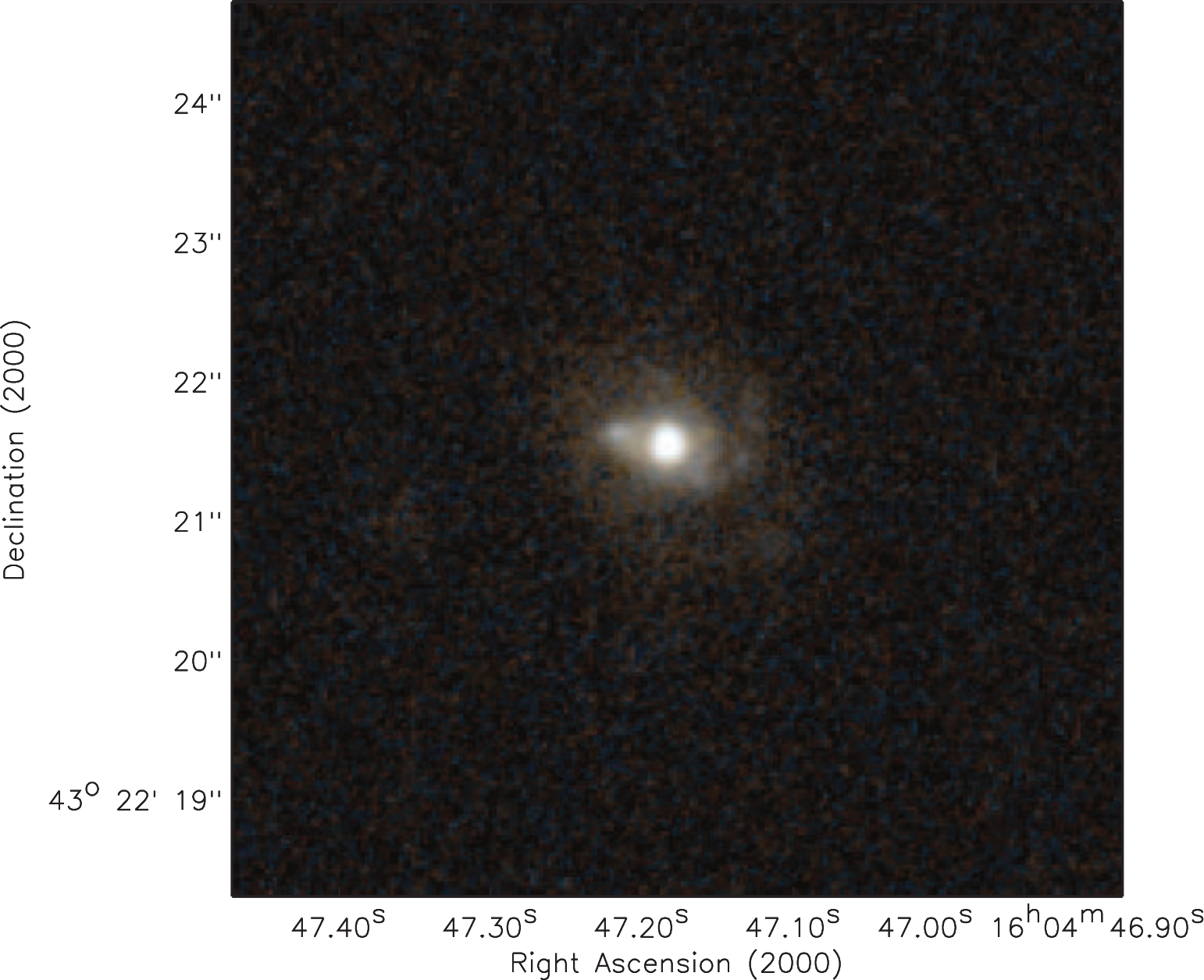} 
\includegraphics[clip,angle=0,width=0.3\hsize]{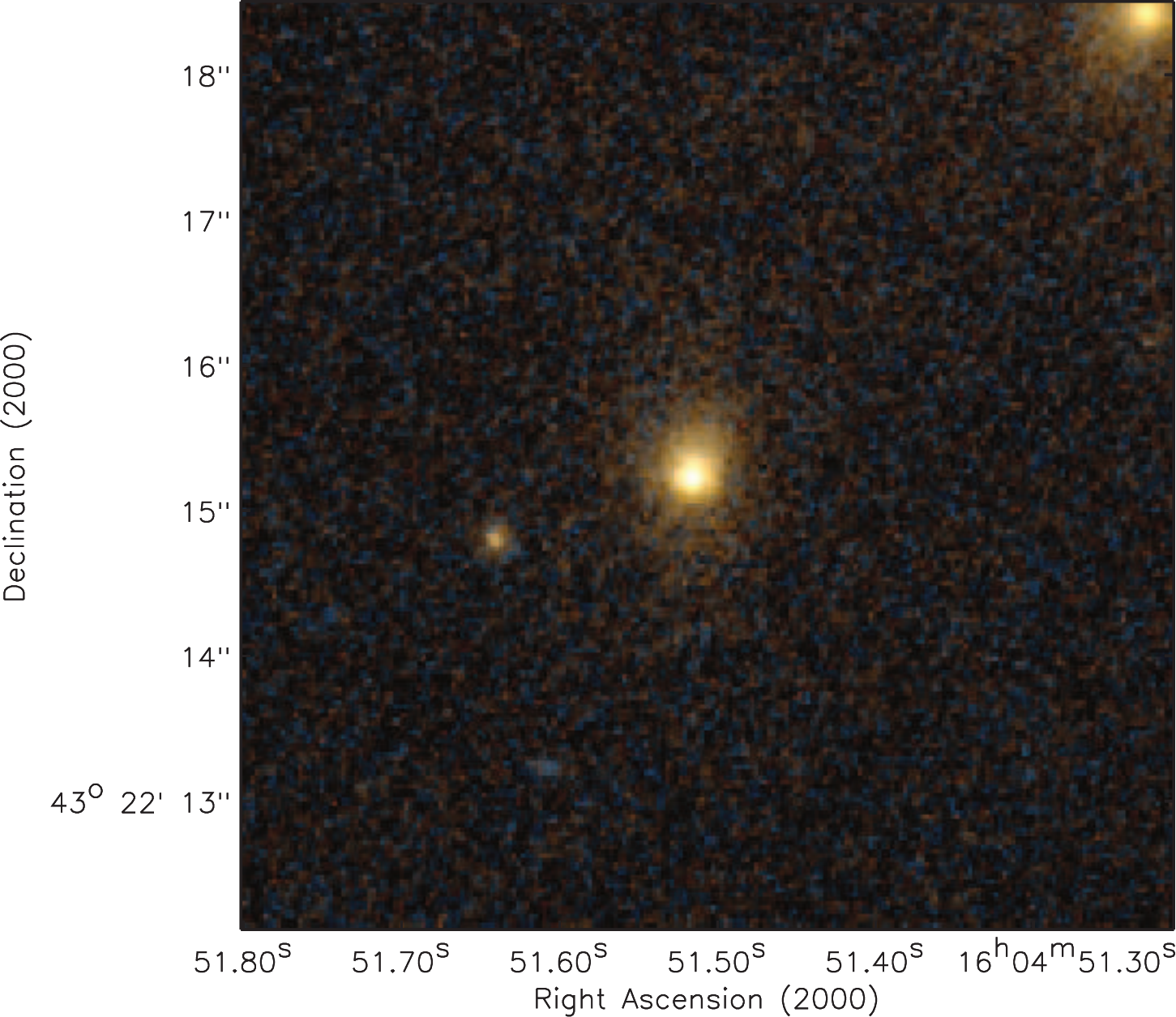} 
\includegraphics[clip,angle=0,width=0.3\hsize]{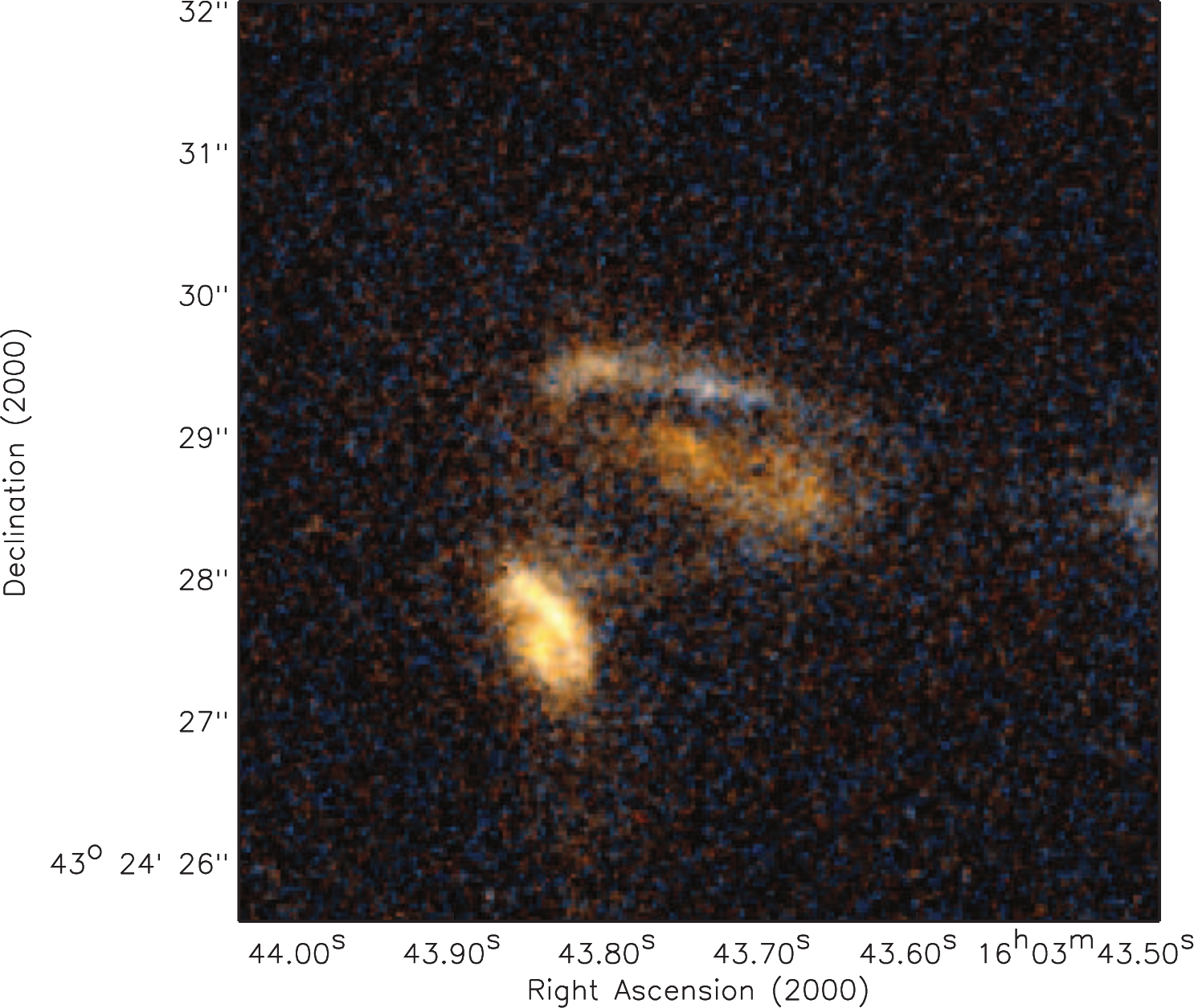} 
\includegraphics[clip,angle=0,width=0.3\hsize]{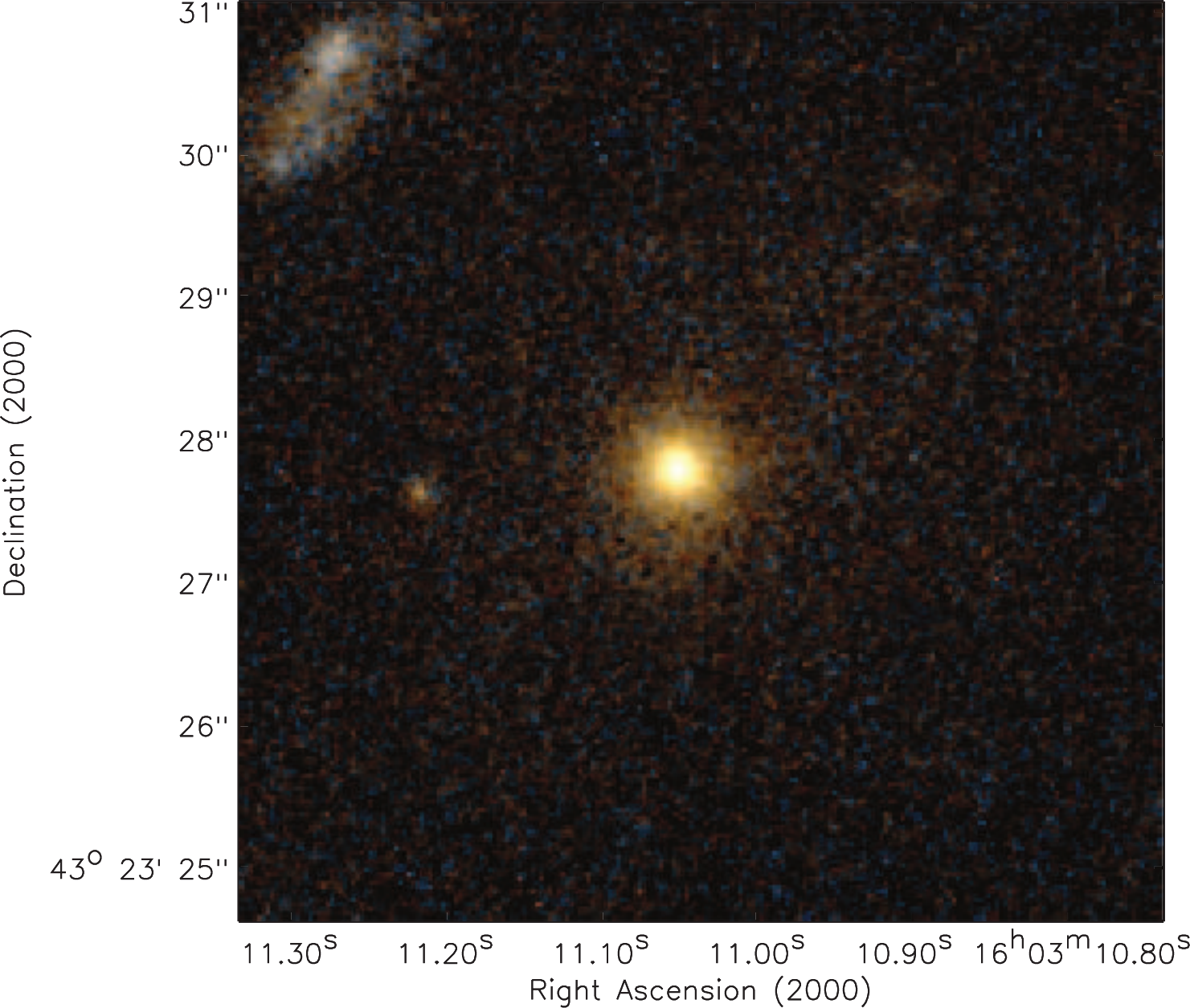} 
\caption{Montage of \emph{HST} ACS color postage stamps of MMCG ``candidates" (see text) of the constituent clusters and groups of the Cl1604 supercluster which were not already selected as BCGs. The physical size and method of generating the postage stamps are the same as Figure \ref{fig:highzBCG}. The range of morphologies spanned by this sample is similar to the one seen in the Cl1604 BCG sample. Notice that many of the Cl1604 MMCG candidates as well as  the Cl1604 BCGs show companions or signs of recent interaction.}
\label{fig:highzMMCG}
\end{figure*}

In Figure \ref{fig:lowzBCG} color postage stamps using the SDSS $u^{\prime} g^{\prime} r^{\prime}$ band images are shown for nine example low-redshift BCGs. The BCGs selected for Figure \ref{fig:lowzBCG} were selected by binning the 81 MCXC clusters with measured velocity dispersions into nine bins equally spaced in $\log(M_{vir})$ and selecting a BCG from a random cluster from each bin. In Figure \ref{fig:lowzMMCG} we show the color postage stamps of nine MCXC MMCGs, generated in the same manner as those of the BCGs. This figure includes only those galaxies from clusters where the MCXC MMCG was different than the BCG. Since there were 13 such galaxies, the nine galaxies shown were drawn randomly in a similar manner to the MCXC BCGs shown in Figure \ref{fig:lowzBCG}. In Figures \ref{fig:highzBCG} and \ref{fig:highzMMCG} we show the \emph{HST}/ACS color postage stamps  for the BCGs and MMCG candidates, respectively, of the Cl1604 supercluster. For the MMCG candidates, we show only those galaxies which were not also selected as BCGs. The SDSS and Cl1604 postage stamps were created in such a way that they span the same physical scale (50 $h_{70}^{-1}$ kpc) in each dimension. 

Visually, the SDSS BCGs and MMCGs comprise an extremely homogenous sample. Nearly all of the galaxies are elliptical and most have an obvious extended halo typical of low-redshift BCGs \citep{seigar07,donzelli11,ascaso11}. Conversely, the BCGs and MMCGs in the Cl1604 have wildly varying appearances. The most massive cluster of the Cl1604 supercluster (A), a cluster whose galaxies are dynamically evolved and whose ICM exhibits features consistent with hydrostatic equilibrium \citep{rumbaugh13} indicative of a older, virialized cluster, has a BCG with a morphology that is, surprisingly, late-type (see the upper left panel of Figure \ref{fig:highzBCG}). This appears to be a common phenomenon amongst the Cl1604 BCGs/MMCGs as a high fraction of BCGs/MMCGs in the Cl1604 supercluster have late-type morphologies. The Cl1604 BCGs/MMCGs also appear to have a high incidence of visual signs of interaction or merging, signs that appear largely absent for the low-redshift MCXC/SDSS sample. Both of these fractions are quantified more rigorously later in this section. Another obvious difference between the two samples is the absence of large luminous haloes around the high-redshift BCGs/MMCGs, a point which we will return to discuss in detail in \S\ref{morphevo}. 

The visual morphology of all MCXC and Cl1604 BCGs/MMCGs was classified by one of us (LML). The process of this visual classification and the ``errors" associated with this process are discussed in detail in \citet{lemaux12}. As will be shown later (see \S\ref{morphevo}), the surface brightness depths of the SDSS and \emph{HST}/ACS imaging are similar enough in physical units (i.e., $L_{\odot}$ pc$^{-2}$) that main morphological parameters can be recovered comparably in the two sets of imaging. Thus, it is unlikely that the results of the visual classification presented here are biased between the two samples, with one minor exception which is  discussed later in this section. For each galaxy a morphological class was assigned  and any visual signatures of interaction were noted (for the various interaction classes, see \citealt{kocevski11a}). As noted earlier, the fraction of Cl1604 BCGs/MMCGs with early-type morphologies (which includes elliptical and S0 galaxies) is dwarfed by the fraction amongst low-redshift clusters. Of the MCXC BCGs and MMCGs, only 2/79 and 3/80, respectively, of the galaxies classified as early-type in each sample are comprised of galaxies that have a lenticular (S0) morphology (2.5\% and 3.8\%, respectively). Conversely, 31\% of BCGs/MMCGs in Cl1604 classified as early-type have an S0 morphology, which, when combined with the large fraction of BCGs/MMCGs observed with late-type morphologies suggests that significant morphological transformation is required between $z\sim0.9$ and $z\sim0.1$. 

It is common to appeal to major merging events as the responsible mechanism to morphologically transform a disc galaxy into an elliptical (see, e.g., \citealt{faber07,hopkins10}). While the likelihood of this scenario is investigated more  rigorously in \S\ref{massrad}, visual inspection of the Cl1604 BCGs/MMCGs broadly support this picture. Though no differentiation is made between minor and major mergers (or interactions) here, the fraction of BCGs/MMCGs with visual signs of interaction in Cl1604 were more than double that of the MCXC/SDSS sample (35.3\% and 16.1\%, respectively)\footnote{Though these numbers may seem inconsistent with those in Table \ref{tab:fractions}, here and elsewhere we consider the combined BCG+MMCG sample without double counting those BCGs also selected as MMCGs. In Table \ref{tab:fractions} we consider the BCG and MMCG samples separately.}. The one exception to the lack of internal biases induced by the differing (physical) surface brightness depths of the SDSS imaging and Cl1604 \emph{HST}/ACS imaging alluded to earlier comes here. The surface brightness depth of the Cl1604 imaging in physical units (i.e., $L_{\odot}$ $pc^{-2}$) in the shallowest regions\footnote{Two deeper pointings of the 17 pointing ACS mosaic, centered on clusters A and D, reach surface brightness limits that are a factor of $\sim2$ shallower than the SDSS $g^{\prime}$ imaging in physical units} is roughly a factor of four shallower than the SDSS imaging (see \S\ref{morphevo}). The effect of this differing depth is to bias us against the detection of faint signatures of interaction in the Cl1604 imaging relative to the SDSS images. Thus, the disparity observed between the fraction of galaxies with noticeable interactions in the two samples is rather a lower limit, i.e., if the Cl1604 BCGs/MMCGs were to be imaged to the same physical depth as the SDSS galaxies it can only serve to increase the observed fraction of Cl1604 BCGs/MMCGs undergoing interactions. If we instead consider only those galaxies undergoing obvious mergers, i.e., those galaxies which definitively appear in some stage of a merging event with one (or more) close companions, a quantity which is essentially unaffected by the differing depths of the two sets of images, the disparity between the two samples increases: 17.6\% of the Cl1604 BCGs/MMCGs are in a state of merging, while only 3.2\% of the low-redshift BCGs/MMCGs are in a similar state. It appears that at least the initial stages of the morphological transformation of the brightest and most  massive galaxies at $z\sim0.9$ appears largely underway. We will return to discuss this point further in \S\ref{massrad}.

\begin{table*}
      \caption{Color, Morphology, and Star-Formation Fractions}
      \[
         \begin{array}{ccccccccc}
		\hline
		\multicolumn{1}{c}{\rm }&
		\multicolumn{1}{c}{\rm ~~~~~~~~~~~f_{Late-Type (1)}}&
		\multicolumn{1}{r}{} &
		\multicolumn{1}{c}{\rm ~~~~~~~~~~~f_{Blue (2)}} &
		\multicolumn{1}{r}{} &
		\multicolumn{1}{c}{\rm ~~~~~~~~~~~f_{SF (3)}} &
		\multicolumn{1}{r}{} & 
		\multicolumn{1}{r}{\rm ~~~~~~~~~~~f_{Inter (4)}} &
		\multicolumn{1}{c}{} \\\hline\hline
		{} & {\rm Low-z} & {\rm High-z} & {\rm Low-z} & {\rm High-z} & {\rm Low-z} & {\rm High-z} & {\rm Low-z} & {\rm High-z}\\\hline
		\multicolumn{1}{l}{\rm BCGs}  & 2.5\% &  50\% &  3.7\% & 37.5\% & 5.2\%  & 50.0\% &  17.2\%  &  37.5\% \\
	        \multicolumn{1}{l}{\rm MMCGs} & 1.2\% &  35.7\% & 1.2\%  & 28.5\% & 3.4\%  & 50.0\%  & 19.8\%   & 28.5\% \\\hline
	\end{array}
      \]
\begin{flushleft} 
 1: Fraction of late-type galaxies includes spirals and mergers with disc features, S0 galaxies are considered as early-type\\
 2: Blue refers to bluer than the lower 3$\sigma$ envelope to the red sequence\\
 3: Galaxies are considered to be star-forming if either $|$EW(H$\alpha$)$|$ or $|$EW([OII])$|$ is $\ge3$\AA\\
 4: Fraction of galaxies undergoing an interaction (see text)\\
\end{flushleft}      
\label{tab:fractions}
   \end{table*}

\subsection{Spectral Properties}
\label{spectral}

One of the main virtues of the datasets presented in this study is the prevalence of high-quality spectroscopic information for both the low-redshift and high-redshift samples. In the next two sections we heavily utilize this information to  draw inferences about the evolution of the BCGs and MMCGs from $z\sim0.9$ to $z\sim0.1$. In this section we focus specifically on the star-forming properties of the BCG/MMCG samples and the plausibility of significant stellar mass build-up resulting from \emph{in situ} star formation. In the subsequent section we will focus on the role of both minor and major merging processes to achieve the same end. 

For the MCXC/SDSS sample, rest-frame equivalent widths (EWs) of common emission and absorption features were measured for all  available spectra as part of DR8\footnote{See http://skyserver.sdss3.org/dr8/en/help/browser/description.asp?n=galSpecLine\&t=U}  using methods described in \citet{tremonti04} and \citet{brinchmann04}. Of these features,  the most pertinent to this work is that of the H$\alpha$ $\lambda6563$\AA\ line, as this line provides a relatively dust-independent measure of the star formation rate (SFR) of galaxies over the last 10 Myr. For Cl1604, the rest-frame coverage of the DEIMOS and LRIS spectra do not afford the opportunity of measuring the EW of the H$\alpha$ feature.  For this sample we instead rely on the [OII] $\lambda3727$\AA\ feature. Though the nature of this line is dubious with respect to star formation properties, especially in the case of red-sequence galaxies (see, e.g., \citealt{yan06,lemaux10}), in the absence of strong AGN or AGN-like phenomenon the [OII] emission feature can closely approximate the function of the $H\alpha$ line. As has been shown already, the samples of BCGs and MMCGs in Cl1604 are not solely comprised of red, elliptical galaxies typically associated with the type of processes that  confound the interpretation of [OII], but rather span a large range in colors and visual morphologies. Thus, we tentatively adopt the [OII] line here as a SFR indicator, relying on other lines of evidence to bolster the conclusions reached solely by interpreting the strength of the [OII] emission feature. The EWs of the [OII] and H$\delta$ feature of the Cl1604 BCGs/MMCGs, as well as the EW(H$\alpha$) feature of the MCXC BCG/MMCG stacked spectrum (see below), were calculated using the bandpass methods described in \citet{lemaux10} and the bandpasses of \citet{fisher98} and \citet{yan06}. Infill corrections to the EW(H$\delta$) were applied using the methodology of \citet{lemaux12}. Measurements of the strength of the continuum break at 4000\AA\ (i.e., $D_n(4000)$) were made using the methodology of \citet{balogh99}. 

We begin by comparing the number of BCGs and MMCGs that have EWs indicative of star-forming galaxies. For the MCXC/SDSS and Cl1604 samples an EW(H$\alpha$)$<-3$\AA\ and EW([OII])$<-3$\AA\ (where a negative EW indicates emission) threshold was adopted as the delineation point between star-forming and quiescent galaxies (for a thorough discussion of various EW thresholds and their consequences see \citealt{yan06} and \citealt{pofeng13}). Under this definition, the two samples show a marked difference in star-forming properties. Of the 58 MCXC BCGs with reliable EW(H$\alpha$) measured from their spectra, only 3 are classified as star forming (5.2\%). This fraction drops to 3.4\% if we consider only the MMCGs and remains largely unchanged if we instead adopt slightly different EW thresholds. These numbers are largely consistent with the fraction of BCGs with ``significant" ongoing star formation found by \citet{liu12} at slightly higher redshifts ($0.1<z<0.4$), though as is noted in that study, this fraction is a strong function of cluster richness. In contrast to the low-$z$ sample, a \emph{majority} (52.9\%)\footnote{Though this number appears to be inconsistent with the numbers given in Table \ref{tab:fractions}, as noted previously we do not double count galaxies which are classified as BCGs and MMCGs in the combined BCG/MMCG sample.}  of the full BCG/MMCG sample in Cl1604 is undergoing significant star formation, a fraction that also remains largely unchanged with differing star formation thresholds.  Such a phenomenon has also observed in an X-Ray selected galaxy group at $z\sim1.1$ \citep{jeltema09}. However, one possible issue in comparing the MCXC and Cl1604 samples in this study arises from the differing physical scales probed by the SDSS fibers of the MCXC/SDSS sample and the DEIMOS/LRIS observations of the Cl1604 galaxies. The 3$\arcsec$ fibers of the SDSS cover, on average, only the central 5 kpc of the MCXC BCGs/MMCGs, while the $1\arcsec$ slits afforded by the DEIMOS/LRIS observations cover the central 8 kpc at the mean redshift of the Cl1604 BCGs/MMCGs. This issue was discussed at length in \citet{oemler13} for a sample of cluster galaxies whose range in redshift reasonably approximate the samples presented here. In this study, the authors concluded that the effect of variable physical apertures had a negligible effect on the derived SFR properties of their cluster galaxies. Thus, we chose to ignore this effect for the remainder of this paper with the caveat that, if significant star formation is occurring on the outskirts of the low-redshift BCGs/MMCGs, it will be missed by the SDSS spectroscopy. 

\begin{figure*}
\centering
\centerline{\includegraphics[clip,angle=0,width=1.0\columnwidth]{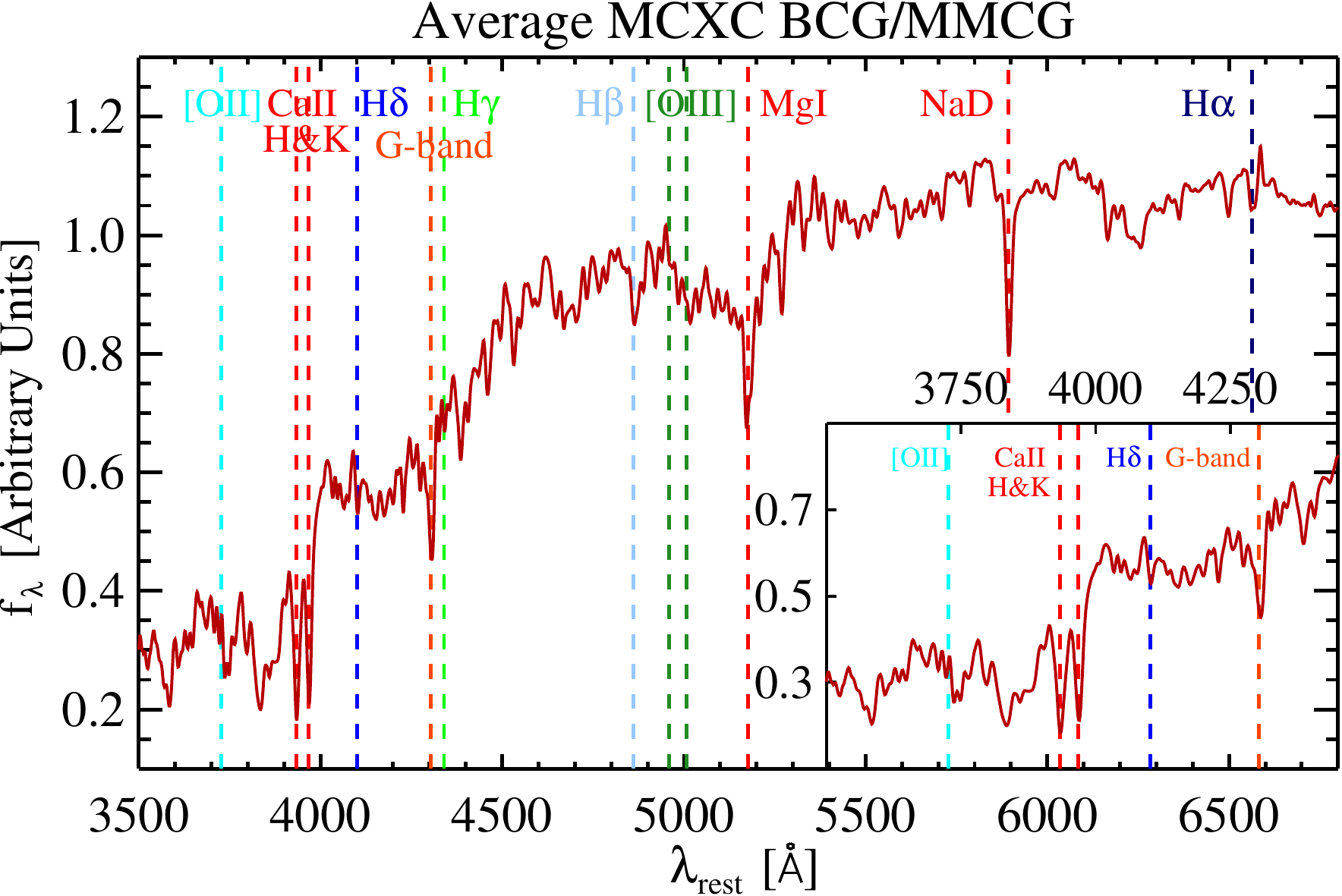}\includegraphics[clip,angle=0,width=1.0\columnwidth]{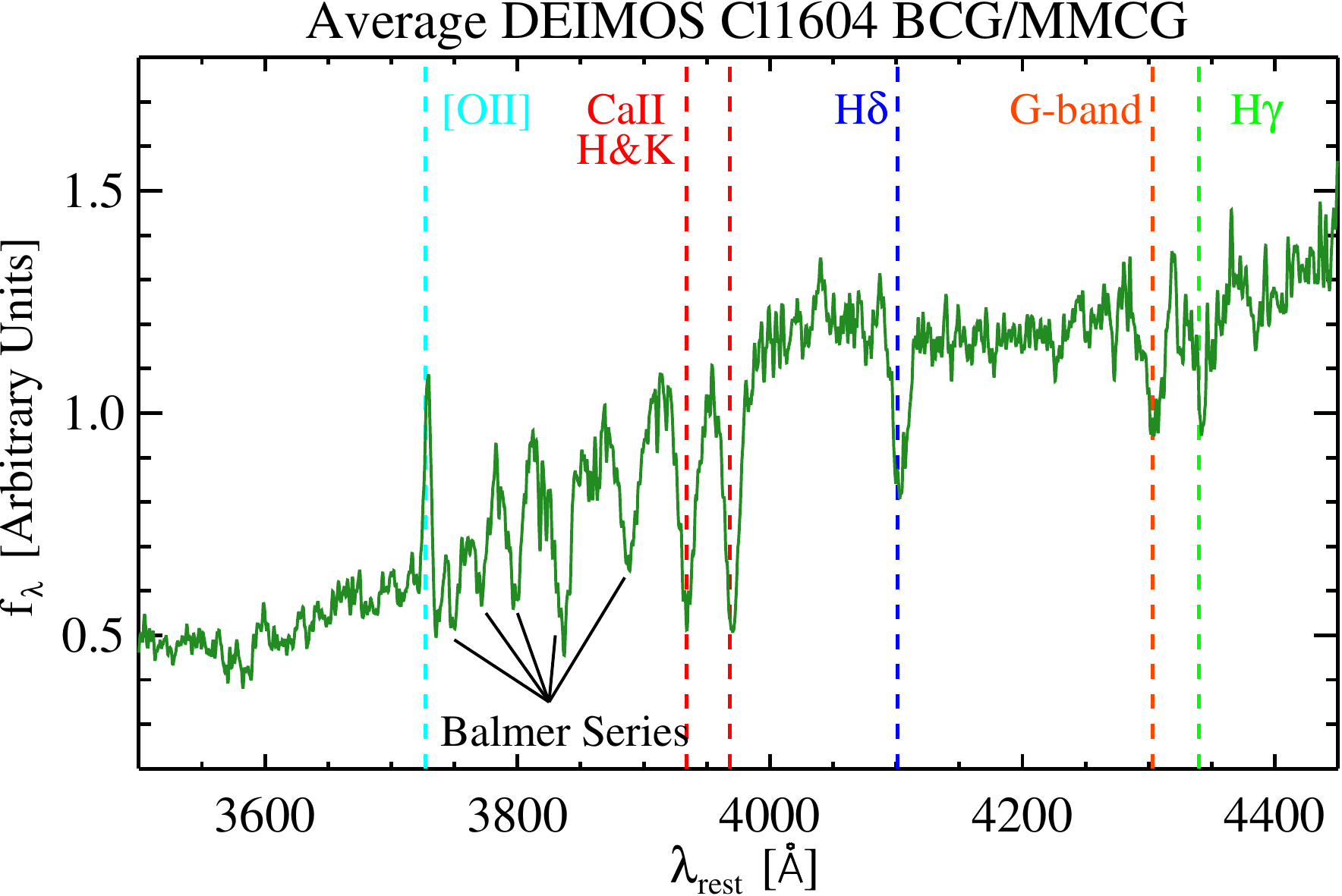}}
\caption{\emph{Left:} Rest-frame unit-weighted ``coadded" spectrum of the 69 BCGs/MMCGs in the MCXC/SDSS sample with usable SDSS spectra. Important spectral features are marked with vertical dashed lines. The inset shows the portion of the coadded SDSS spectrum which spans the same rest-frame wavelength coverage of the Cl1604 coadded spectrum. \emph{Right:} Rest-frame unit-weighted coadded spectrum of the 11 Cl1604 BCGs/MMCGs observed with DEIMOS. In addition to these galaxies, six BCGs/MMCGs were observed with two different epochs of LRIS observations and were coadded separately. These coadds are not shown here, though they appear qualitatively and quantitatively similar to the Cl1604 DEIMOS coadd. As in the left panel, important spectral features are marked with vertical dashed lines. The spectral location of many of the higher order Balmer absorption lines is also marked. While the average MCXC BCG/MMCG shows no signs of either ongoing or recent ($\la 1$ Gyr) star formation, the average Cl1604 BCG/MMCG shows strong signs of both.}
\label{fig:spectralmosaic}
\end{figure*}

To further quantify the differences between the two samples, plotted in Figure \ref{fig:spectralmosaic} are spectral ``co-additions" of the low- and high-redshift BCGs/MMCGs. These co-additions (or ``coadds") were generated in a method identical to that of \citet{lemaux12} and represent unit-weighted mean spectrum of each sample. Because of the differing resolutions of DEIMOS and the various epochs of LRIS observations of the Cl1604 galaxies, co-additions were performed for each subset separately and $EW$ measurements were combined in the manner described in \citet{lemaux12}. Since the majority of the BCGs/MMCGs in Cl1604 were observed with DEIMOS, and because of the similarity of the LRIS co-additions, only the DEIMOS co-addition of the Cl1604 BCGs/MMCGs is shown in the right panel of Figure \ref{fig:spectralmosaic}. In the inset of the left panel of  Figure \ref{fig:spectralmosaic} we plot a zoom in of the coadded SDSS spectrum over the same rest-frame range as that covered by the Cl1604 DEIMOS observations. The SDSS spectra of all MCXC BCGs/MMCGs were inspected and only those free from  reduction artifacts were included in the coaddition. In addition, since a large number of BCGs went untargeted in SDSS due to previously measured spectroscopic redshifts, these galaxies were excluded due to lack of SDSS spectra. The final MCXC/SDSS sample was comprised of 68 BCGs/MMCGs. 

Visually, the average spectrum of the low- and high-redshift BCGs/MMCGs  appear drastically different. While the spectrum of the average MCXC BCGs/MMCGs appears similar to other typical  red, quiescent galaxies observed at low redshift (e.g., \citealt{bernardi03,eisenstein03,dressler04}), the average Cl1604  BCG/MMCG does not appear to be dominated by an old stellar population. Rather, both the appearance of strong Balmer absorption features ($\langle EW(H\delta) \rangle=3.27\pm0.13$) and the relatively weak continuum break at 4000\AA\ ($\langle D_n(4000) \rangle$=1.43$\pm$0.01), a rough proxy of mean stellar age, observed in the Cl1604 BCG/MMCG coadd indicate that significant star formation has occurred in the average $z\sim0.9$ BCG/MMCG within the last Gyr. Additionally, the presence of relatively strong [OII] emission in the average spectrum of the Cl1604 BCGs/MMCGs ($\langle EW([OII]) \rangle_{Cl1604}=-4.53\pm0.15$) suggests that star formation is ongoing in these galaxies. In contrast, the H$\alpha$ feature is not observed significantly in emission in the MCXC coadd ($\langle EW(H\alpha) \rangle_{MCXC}=0.32\pm0.03$), consistent with no ongoing star-formation activity. It is encouraging to observe the lack of significant emission in [OII] ($\langle EW([OII]) \rangle_{MCXC}=0.22\pm0.08$) in concert with the lack of significant H$\alpha$ emission. This suggests that the types of AGN or other processes that mimic star formation via the [OII] line are not prevalent at least amongst the low-redshift BCGs/MMCGs. The rest-frame \emph{EWs} of important spectral features as measured on the coadded spectra of the MCXC and Cl1604 samples is given in Table \ref{tab:spectralproperties}.

\begin{table*}
\caption{Spectral Properties of the Cl1604 and MCXC BCGs and MMCGs}
\begin{tabular}{cccccccc}
\hline
Sample & N & $EW$([OII])$^{1}$ & $EW$(H$\delta$)$^{1,2}$ & $EW$(H$\alpha$)$^{1}$ & $D_{n}(4000)$ & $\langle M_{U}\rangle$ & $\langle$SFR$\rangle$ \\
 & & [\AA] & [\AA] & [\AA] & & [$\mathcal{M}_{\odot}$ yr$^{-1}$] \\ [0.5ex]
\hline \hline
Cl1604 BCGs & 8 & -5.92$\pm$0.25 & 3.45$\pm$0.23 & ---$^{5}$ & 1.440$\pm$0.008 & -20.72$\pm$0.02 & 15.9$\pm$0.9 \\[4pt]
Cl1604 MMCGs & 14 & -4.13$\pm$0.15 & 3.26$\pm$0.17 & ---$^{5}$ & 1.395$\pm$0.008& -20.47$\pm$0.03 & 12.1$\pm$0.6 \\[4pt]
Cl1604 Combined$^{3}$ & 17 & -4.53$\pm$0.15 & 3.27$\pm$0.13 & ---$^{5}$ & 1.410$\pm$0.005 & -20.53$\pm$0.02 & 10.5$\pm$0.5\\[4pt]
Cl1604 Passive$^{4}$ & 11 & -1.43$\pm$0.22 & 2.07$\pm$0.15 & ---$^{5}$ & 1.730$\pm$0.009 & ---$^{6}$ & ---$^{6}$\\[4pt]
MCXC BCGs & 58 & 0.35$\pm$0.08 & -1.32$\pm$0.05 & 0.35$\pm$0.02 & 1.866$\pm$0.003 & ---$^{7}$ & ---$^{7}$ \\[4pt]
MCXC MMCGs & 57 & 0.16$\pm$0.08 & -1.17$\pm$0.05 & 0.32$\pm$0.02 & 1.845$\pm$0.003 & ---$^{7}$ & ---$^{7}$ \\[4pt]
MCXC Combined$^{3}$ & 68 & 0.22$\pm$0.08 & -1.19$\pm$0.05 & 0.32$\pm$0.01 & 1.853$\pm$0.003 & ---$^{7}$ & ---$^{7}$ \\[4pt]
\hline
\end{tabular}
\begin{flushleft}
1: Negative EWs correspond to features observed in emission, positive to those in absorption\\ 
2: Corrected for infill using the method of \citet{lemaux12} \\
3: Combined samples were created such that BCGs that were also MMCGs were not counted twice \\
4: Passive galaxies were defined as those with SSFR $< 1\times10^{11}$ yr$^{-1}$. BCGs and MMCGs were not counted twice \\ 
5: The rest-frame wavelength coverage of DEIMOS and LRIS spectra do not allow for the measurement of H$\alpha$ \\
6: These quantities were not calculated for the passive Cl1604 BCGs/MMCGs \\ 
7: These quantities were not calculated for the MCXC/SDSS sample as the H$\alpha$ feature was consistent with zero  
\end{flushleft}
\label{tab:spectralproperties}
\end{table*}

With this encouragement, we calculate the average SFR of the Cl1604 BCGs/MMCGs in the following manner. First,  the \emph{HST}/ACS $F606W$ \& $F814W$  of each of the Cl1604 BCGs and MMCGs were transformed into an absolute $U-$band magnitude ($M_U$) using the conversion of \citet{homeier06} and the absolute $U-$band flux density was used to calculate a population average $M_U$. This average $M_U$ was then used in conjunction with the average [OII] $EW$ to calculate the average [OII]-derived  SFR using a methodology nearly identical to that of \citet{lemaux13} but with a change of IMFs from \citet{salpeter55} to \citet{chabrier03} to match the IMF used for our stellar mass measurements. An additional slight correction was made to this method to account for the  different filter curves of the SDSS $u^{\prime}$ and the Johnson-Cousins $U$ band: 

\begin{eqnarray}
SFR(L_{[OII]}) = (1.03\pm0.26) \times10^{10}(10^{\frac{-(\langle M_{U} \rangle +48.6)}{2.5}}) \nonumber \\ 
-\langle EW([OII]) \rangle \mathcal{M}_{\odot} \rm{yr}^{-1}
\label{eqn:SFROII}
\end{eqnarray}

\noindent The constant of proportionality is adopted from the [OII] SFR formula of \citet{kewley04} and adapted for the purposes of Equation \ref{eqn:SFROII}. The change of IMFs was performed using the methods of \citet{magnelli13}. This [OII]-derived SFR was corrected for extinction using a \citet{calzetti00} reddening law and adopting $E_s(B-V)=0.25$, the average value of all Cl1604 member galaxies as derived from our SED fitting. While $EW$ values are generally relatively insensitive to internal extinction (see discussion in \citealt{lemaux10}), this correction is necessary to correct the $M_{U}$ values calculated from the observed ACS magnitudes and was applied at the effective wavelength of the Johnson-Cousins $U$ band of 3650 \AA. To this average extinction-corrected [OII]-derived SFR we incorporate 24$\mu$m observations from the Multiband Imaging Photometer for \emph{Spitzer} (MIPS; \citealt{rieke04}) described in detail in \citet{kocevski11a}. Three of the Cl1604 BCGs/MMCGs were detected at a significant level at these wavelengths and their total SFRs were corrected for their infrared luminosities using the method described in \citet{kennicutt09} and subsequently incorporated into the population average. For those galaxies with MIPS detections the [OII]-derived SFR was not corrected for extinction. The resulting average SFR of the Cl1604 galaxies is $\langle SFR\rangle=10.1\pm0.5$ $\mathcal{M}_{\odot}$ yr$^{-1}$. Even at the relatively large stellar masses of the Cl1604 BCGs/MMCGs ($\langle \log(\mathcal{M}_s) \rangle=11.26\pm0.23$), the large SFR observed for the Cl1604 BCGs/MMCGs allows for a substantial percentage of stellar mass to be built up when integrated over several Gyr. In other words, if the average SFR observed at $z\sim0.9$ is a ``typical" value for a BCG/MMCG over the redshift range $z\sim0.9$ to $z\sim0.1$, the time that it takes for the Cl1604 BCGs/MMCGs to double their ensemble stellar content simply through \emph{in situ} star formation (i.e., $\sum\limits_{i=1}^n\mathcal{M}_{S,i}/\sum\limits_{i=1}^n SFR_i$) is 12.9$^{+4.0}_{-2.5}$ Gyr, where the errors on this quantity are derived through a combination of the SFR errors and the adopted errors on the stellar mass described in \S\ref{cl1604}. This value is mildly consistent within the errors of the difference in cosmic time between the two redshifts. This argument does not change if the Cl1604 BCG and MMCG samples are instead coadded separately (see Table \ref{tab:spectralproperties}).

\begin{figure*}
\centering
\centerline{\includegraphics[clip,angle=0,width=1.0\columnwidth]{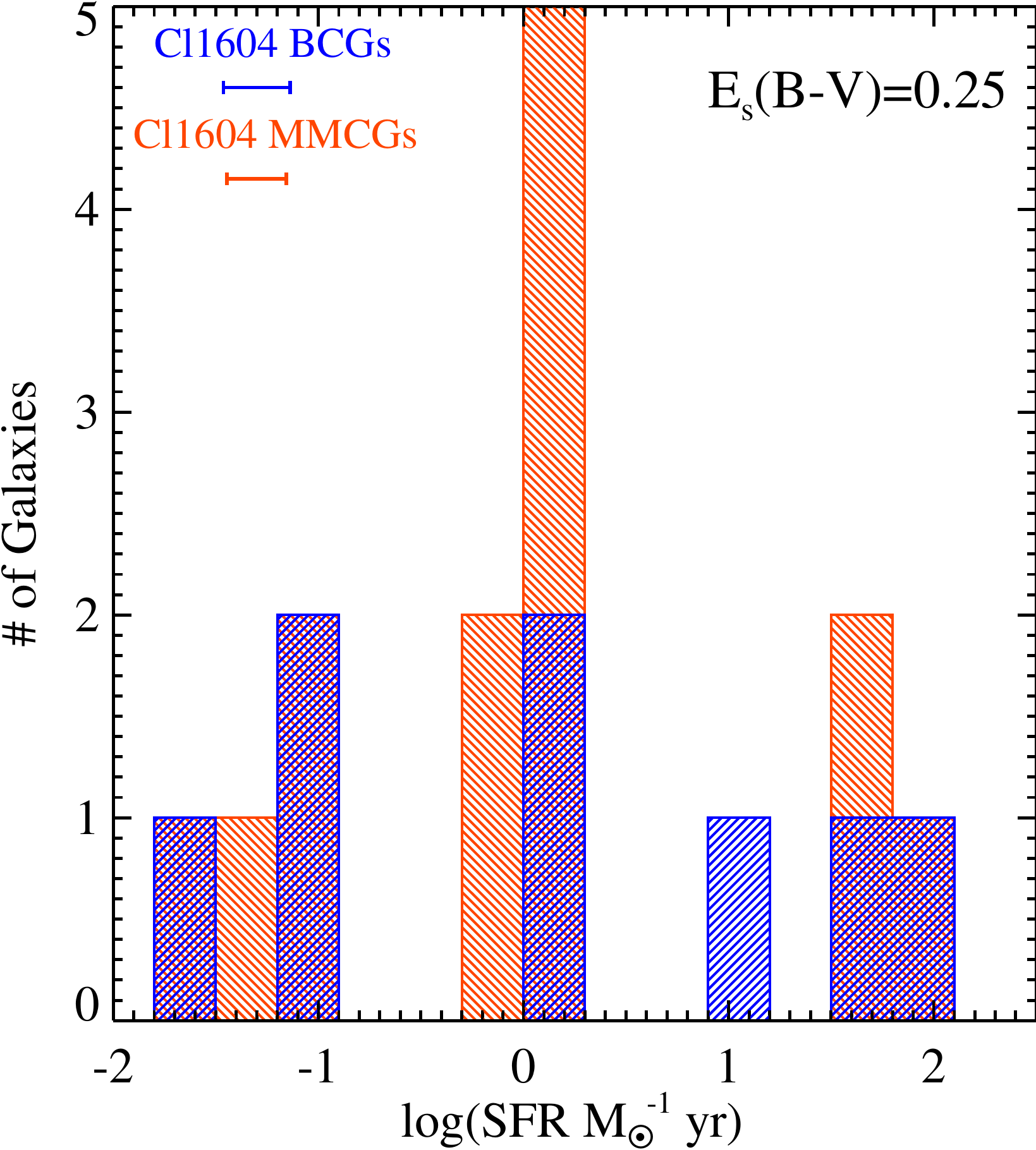}\includegraphics[clip,angle=0,width=1.0\columnwidth]{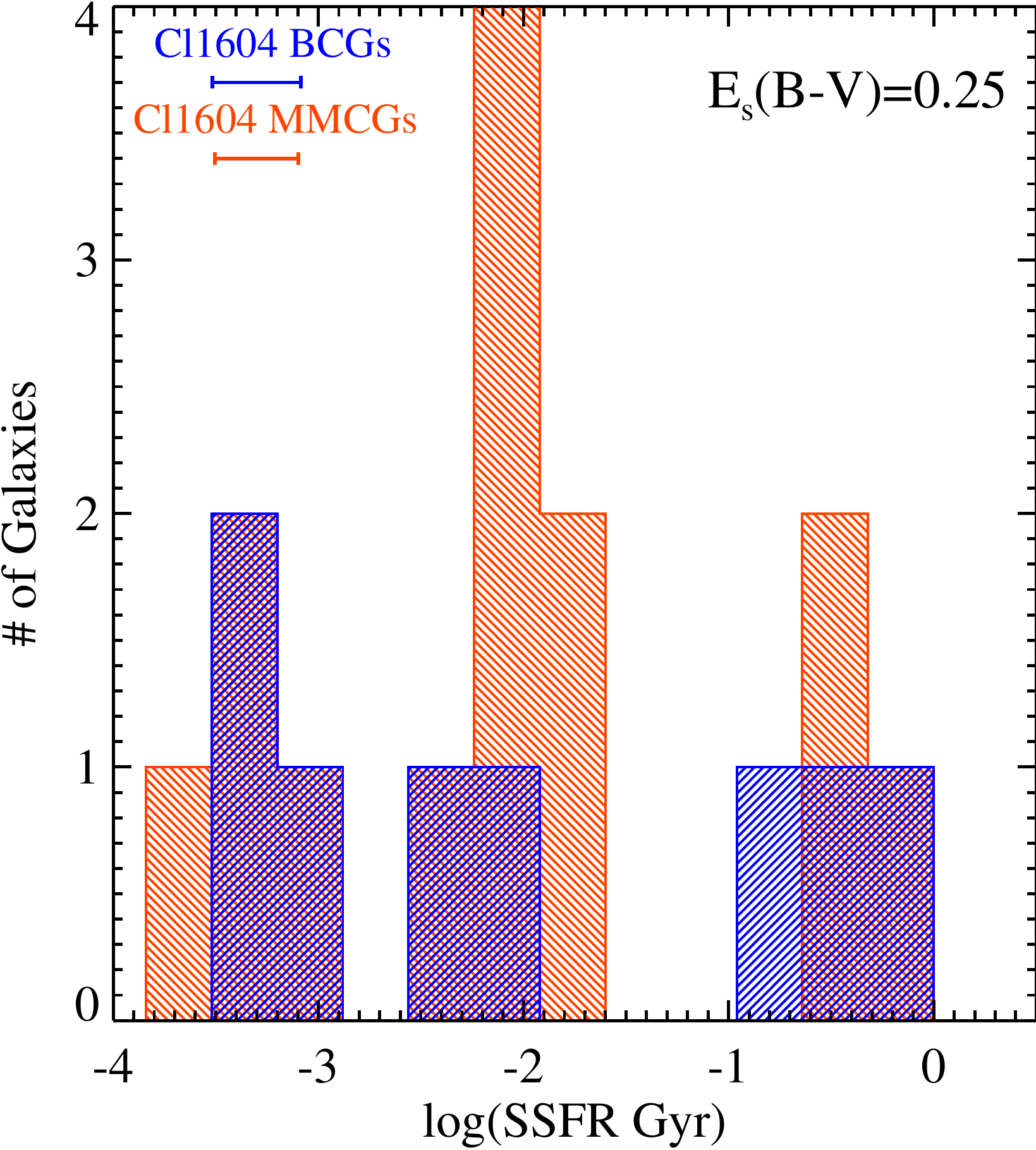}}
\caption{\emph{Left:} Distribution of extinction corrected [OII]-derived SFRs supplemented by MIPS 24$\mu$m observations for the eight Cl1604 BCGs 
(blue hashed histogram) and 14 Cl1604 MMCG candidates (orange hashed histogram). The average $1\sigma$ error bar is shown in the top left corner for
each sample. For those galaxies undetected at 24$\mu$m, extinction corrections were made to the SFRs using an $E_s(B-V)=0.25$, which is the average value for Cl1604 member galaxies as derived from our SED fitting. \emph{Right:} Distribution of specific SFRs (SSFRs) of the Cl1604 BCGs and MMCG candidates. The colors and symbols have identical meanings to those in the left panel. Both the BCG and MMCG samples exhibit a wide range in SFR and SSFR, suggestive of the existence of a stochastic process driving the star formation in these galaxies. The high level of SFR observed in many of the Cl1604 BCGs/MMCGs in addition to the strong starburst features observed in their average spectra indicate that significant stellar mass growth can occur \emph{in situ} in high redshift BCGs/MMCGs.}
\label{fig:SFRnSSFRhist}
\end{figure*}

Instead of considering the samples in the method presented above, it is also possible to calculate the SFR and the specific SFR (SSFR) of each of the galaxies in the Cl1604 sample using the formalism of Equation \ref{eqn:SFROII}. While this results in larger errors, as measurements on individual galaxy spectra are much noiser than those on coadded spectra, it serves the purpose of allowing the observation of the distribution of these quantities amongst the Cl1604  BCGs/MMCGs. This exercise is not performed for the MCXC BCGs and MMCGs as nearly all of the individual spectra have SFRs consistent with zero (see Table \ref{tab:fractions}). Plotted in Figure \ref{fig:SFRnSSFRhist} is the SFR and SSFR distributions  for the Cl1604 BCGs and MMCGs. Both sets of galaxies exhibit a wide range of star-formation properties, spanning nearly four  orders of magnitude in both quantities. Such a spread in properties over a relatively small sample of galaxies is suggestive of a  sporadic process which induces or quenches star formation, though it is impossible to draw definitive conclusions from this number of galaxies. As we will show in the next section, wet or mixed merging between the Cl1604 BCGs/MMCGs and their massive companions provides a plausible candidate for this process. The average doubling time of the Cl1604 BCGs and MMCG candidates calculated from the mean SSFR in Figure \ref{fig:SFRnSSFRhist} is 6.5$^{+1.5}_{-1.0}$ Gyr and 8.6$^{+1.7}_{-1.2}$ Gyr, also consistent within the errors of the difference in cosmic time from $z\sim0.9$ to $z\sim0.1$.

\begin{figure}
\includegraphics[clip,angle=0,width=1.0\columnwidth]{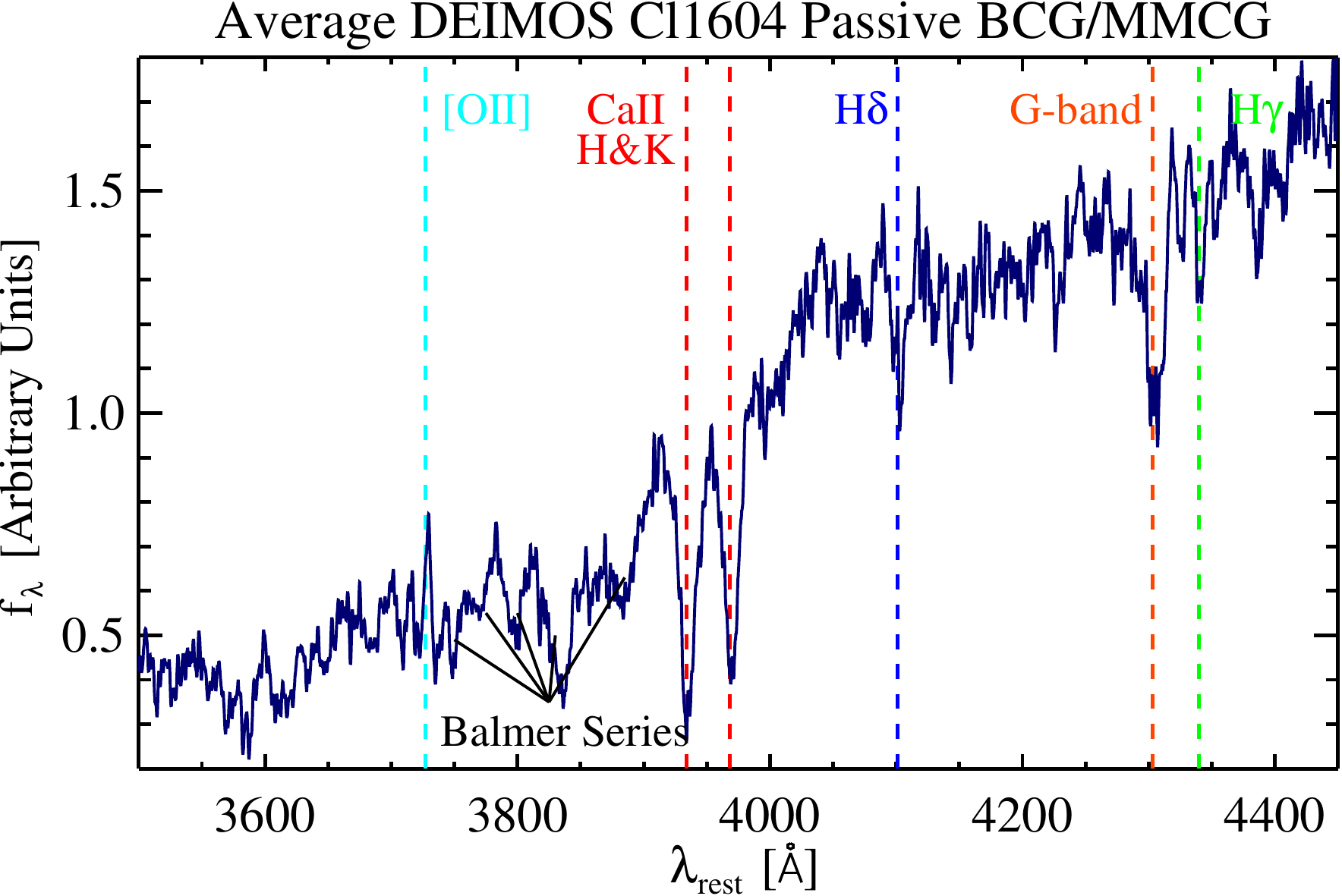}
\caption{Rest-frame unit-weighted coadded spectrum of the six Cl1604 ``passive" BCGs/MMCGs observed with DEIMOS. Passive BCGs/MMCGs are defined as those galaxies with a SSFR $< 1\times10^{11}$ yr$^{-1}$ as measured by the methods described in the text. In addition to these galaxies, five passive BCGs/MMCGs were observed with two different epochs of LRIS observations and were coadded separately. These coadds are not shown here, though they appear qualitatively and quantitatively similar to the Cl1604 DEIMOS coadd. Important spectral features are marked. The average spectral properties of passive Cl1604 BCGs/MMCGs depart considerably from the ensemble average shown in the right panel of Figure \ref{fig:spectralmosaic}, with the average passive BCG/MMCG exhibiting less [OII] emission and a redder color than the average Cl1604 BCG/MMCG. However, relatively strong Balmer features are still observed in the average spectrum of the passive Cl1604 BCGs/MMCGs, which indicates the presence of a component of moderately young stars.} 
\label{fig:spectralmosaic2}
\end{figure}

Given the inhomogeneity observed in Figure \ref{fig:SFRnSSFRhist}, ensemble-averaged SFR and SSFR properties of the Cl1604 BCG/MMCG sample provides only part of the picture. While there are several Cl1604 BCGs/MMCGs which appear to be forming stars prodigiously (i.e., $\ga10$ $\mathcal{M}_{\odot}$ yr$^{-1}$),  the majority of the Cl1604 BCG/MMCG sample is estimated to be forming stars at rates of 1 $\mathcal{M}_{\odot}$ yr$^{-1}$ or lower. Such galaxies do not necessarily fit into the picture derived from the average properties of the Cl1604 BCGs/MMCGs. However, if the process driving star formation in these galaxies is a stochastic one, as suggested above, one would still expect to see signs of recent star formation in such galaxies. To differentiate BCGs/MMCGs that were actively forming stars from those passive with respect to star formation process (hereafter ``passive" BCGs/MMCGs), we adopted the SSFR threshold of SSFR $< 1\times10^{11}$ yr$^{-1}$ (see \citealt{ilbert13} and references therein). In total, 11 of the 17 combined Cl1604 BCG/MMCG sample fell below this threshold. Plotted in Figure \ref{fig:spectralmosaic2} is the coadded spectrum, generated in the same way as in Figure \ref{fig:spectralmosaic}, of the six passive BCGs/MMCGs in Cl1604 observed by DEIMOS. As before, the five remaining galaxies observed with LRIS were coadded separately and their measurements combined with the DEIMOS coaddition. Given in Table \ref{tab:spectralproperties} are the combined $EW($[OII]$)$, $EW($H$\delta)$, and $D_{n}(4000)$ measurements from these coadds. While it is clear that an older stellar population dominates the average passive Cl1604 BCG/MMCG, as evidenced by the strong CaII and G-band features and the strong 4000\AA\ break, there remains significant Balmer absorption. While this absorption is qualitatively and quantitatively weaker than that in the spectrum shown in the right panel of Figure \ref{fig:spectralmosaic}, it is strong enough to imply either moderate levels of star formation which ended in the recent past or higher levels of star formation within the last $\sim$2 Gyr in the \emph{average} passive Cl1604 BCG/MMCG.

The implied significant buildup of stellar mass of the Cl1604 BCGs/MMCGs through star formation is in broad agreement with the results of \citet{kaviraj08}, in which rest-frame UV-optical colors were used to reveal the prevalence of low- to moderate- star formation events in elliptical galaxies over the redshift $0.5<z<1$. Of course, given the plethora of physical mechanisms hostile to star formation in cluster environments, the average SFR in the Cl1604 BCGs/MMCGs is likely to decline with decreasing redshift rather than plateau (though the presence of cooling flows into the BCG could result in increased and possibly sustained levels of high star formation, see \citealt{pipino09}). Additionally, as we will show in the next section, even a doubling of the stellar mass of the Cl1604 BCGs/MMCGs cannot account for the observed difference in stellar masses between the MCXC and Cl1604 samples. Still, the fact remains that \emph{in situ} star formation appears to contribute substantially to the buildup of mass in BCGs and MMCGs from $z\sim0.9$ to the present day. 

\subsection{Stellar mass radial distribution}
\label{massrad}

In this section we compare the concentration of stellar baryonic mass contained in galaxies around the MCXC and Cl1604 BCG/MMCG samples in an attempt to determine how much stellar mass these galaxies could accrete through merging processes over the last 7 Gyr. In order to make valid comparisons between the galaxy populations of MCXC clusters and those of the Cl1604 clusters an important consideration is necessary which has been eluded to in previous sections. It has been observed that correlations between the stellar mass of the BCG and the total halo mass of the cluster in which it 
resides (e.g., \citealt{edge91}) exist across a wide variety of clusters and redshifts. The reason it is necessary to account for this correlation here is, unlike other sections, the quantities being compared in this section, stellar mass of the BCG/MMCG and the total stellar mass of the clusters, are inherently linked to mass of the parent cluster in a way that the previous properties are not. In addition, the large homogeneity of the properties investigated in the previous sections of the MCXC BCGs/MMCGs resulted in invariance in our results with respect to the sample being chosen. In this section we are now studying the relationship between the BCG/MMCG and the cluster population as a whole and, as such, we no longer have the luxury of this invariance.

Corrections for the correlation between the stellar mass of a BCG and the total mass of the cluster in which it resides were explored in great detail by \citet{lidman12} in a study of 160 BCGs observed over a large redshift range (0.03$<z<1.63$). The general approach used in \citet{lidman12} was to evolve the total mass of each of the high-redshift clusters in their sample to the epoch of their low-redshift samples using the mass evolution of clusters observed by \citet{fakhouri10} in the Millennium \citep{springel05} and Millennium-II \citep{boylan-kolchin09} $N$-body simulations. The evolved total masses of the high-$z$ clusters were then compared to the observed total masses of the low-$z$ clusters and a variety of methods were utilized to account for the observed disparity. Here we rely heavily on the philosophy of \citet{lidman12}. All Cl1604 clusters and groups were evolved to the mean redshift of the MCXC clusters in a method identical to that of \citet{lidman12} using the virial mass estimated from $\sigma_v$ as the starting point of the evolutionary model. The resulting  average increase in total (virial) cluster mass was found to be a factor of $\sim 2.5$ between $z\sim0.9$ and $z\sim0.1$. The average evolved virial cluster mass of the Cl1604 clusters and groups ($\langle \log(\mathcal{M}_{s}) \rangle = 14.67$) is, however, slightly higher than the mean observed virial mass for the MCXC/SDSS sample. To correct for this, we began cutting the MCXC/SDSS sample starting with the lowest virial mass cluster until the mean observed MCXC cluster virial mass equaled that of the evolved Cl1604 clusters and groups. The final MCXC comparison sample comprised a large fraction of the original sample; of the sample of 53 MCXC clusters defined for this analysis in \S\ref{clustermass}, the 48 most massive  were retained as the MCXC comparison sample (hereafter simply ``comparison sample"). In addition to the high level of similarity between the average mass of the MCXC comparison sample and the evolved Cl1604 clusters and groups, it is important to note that the evolved virial mass range of the Cl1604 clusters and groups essentially spans the entire virial mass range of the observed MCXC clusters (i.e., those in Figure \ref{fig:Mvirdist} with $\log(\mathcal{M}_{s})\ga13.6$), thereby sampling the sample variance inherent in the MCXC cluster sample. The minimal difference in the observed virial masses of the $z\sim0.1$ clusters and the evolved virial masses of the $z\sim0.9$ Cl1604 clusters and groups will also be important when we discuss the morphological evolution of the Cl1604 BCGs/MMCGs in the next section using a sample of all 81 MCXC clusters.

\begin{figure*}
\centering
\centerline{\includegraphics[clip,angle=0,width=1.0\columnwidth]{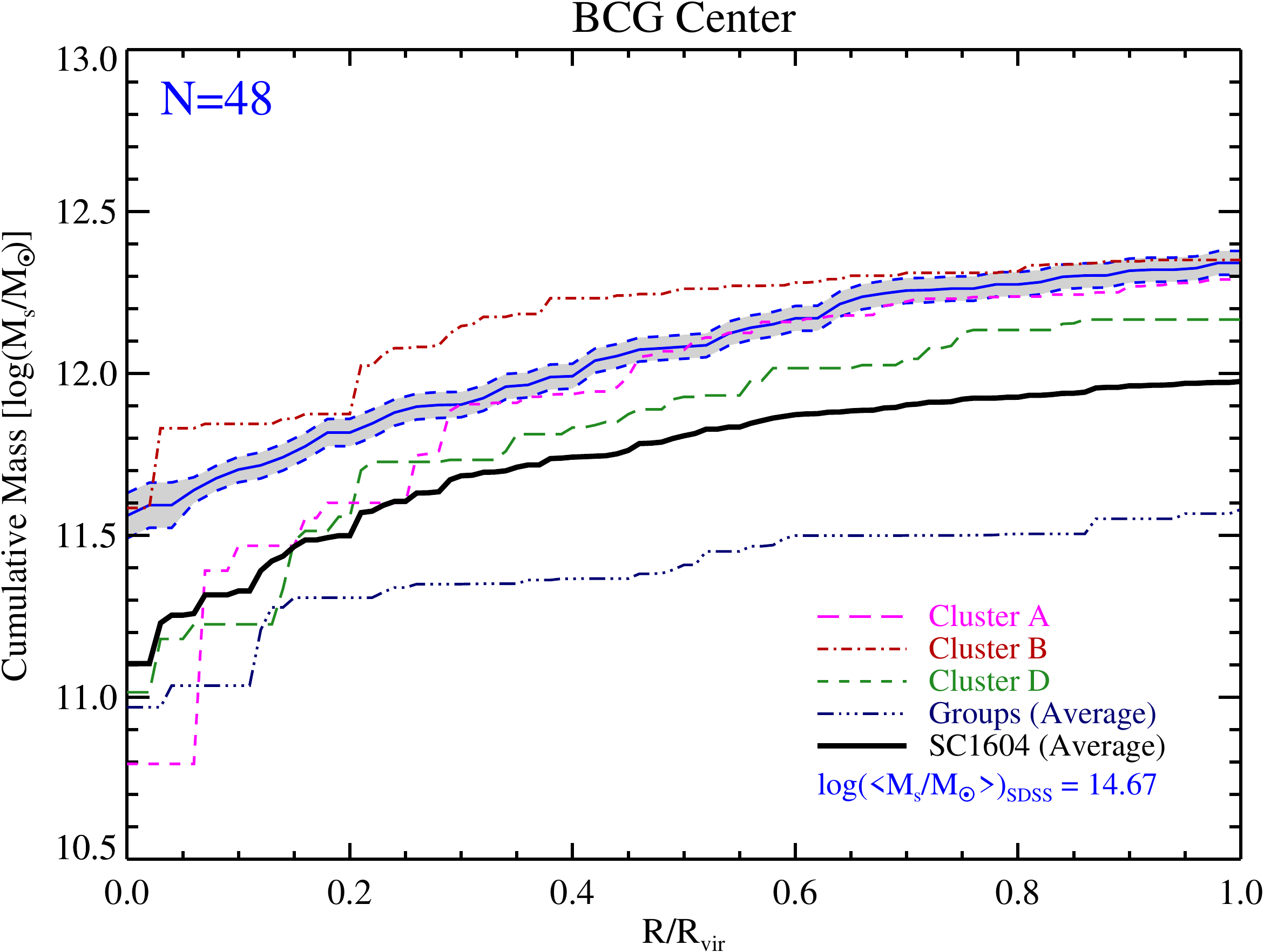}\includegraphics[clip,angle=0,width=1.0\columnwidth]{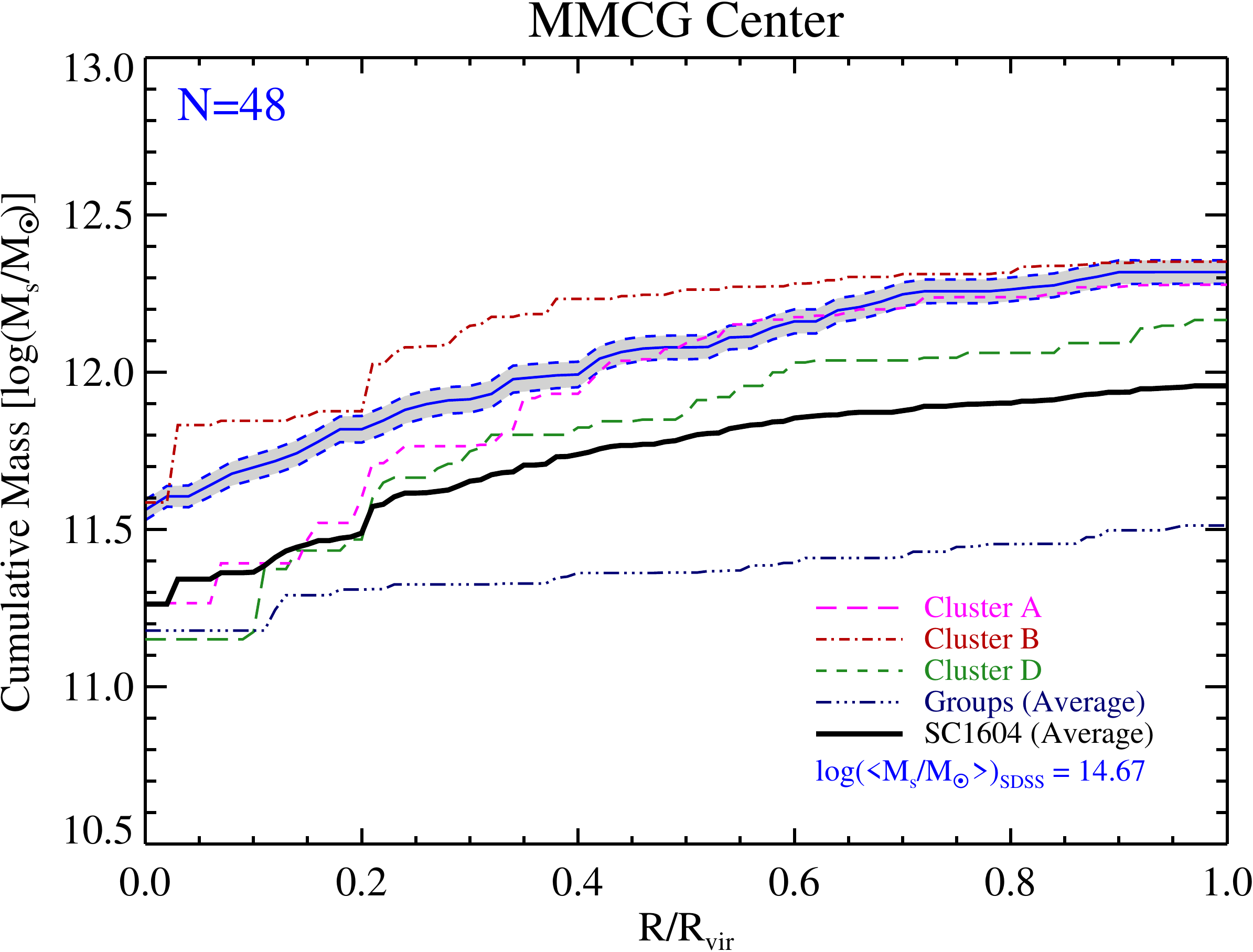}}
\caption{Cumulative stellar mass distribution as a function of viral radius centered on the BCG (left panel) and the MMCG (right panel) for the MCXC/SDSS comparison sample and for the Cl1604 clusters and groups. The blue solid line indicates the average cumulative stellar mass distribution for the 48 most massive MCXC clusters with well-measured velocity dispersions. This number was chosen such that the average $M_{vir}$ of the MCXC/SDSS sample used was matched to the average evolved virial mass of the Cl1604 groups and clusters (shown in the bottom right of each panel, see text). The shaded region surrounding the blue  solid line denotes the $1\sigma$ sample variance of the MCXC clusters. The cumulative mass distribution of the members of the individual Cl1604 clusters, a combined Cl1604 group sample, and a combined Cl1604 sample is also plotted. On average, there is a rapid increase in cumulative stellar mass surrounding the $z\sim0.9$ Cl1604 BCGs/MMCGs at low (virial) radii that is not observed in the MCXC/SDSS sample at $z\sim0.1$. This indicates the presence of massive companions surrounding these galaxies. The amount of stellar mass in all galaxies within $R<0.2R_{vir}$ of the average Cl1604 BCG/MMCG is sufficient to equal the stellar mass of the average MCXC BCG/MMCG.}
\label{fig:stellarmasscum}
\end{figure*}

\begin{figure*}
\centering
\centerline{\includegraphics[clip,angle=0,width=1.0\columnwidth]{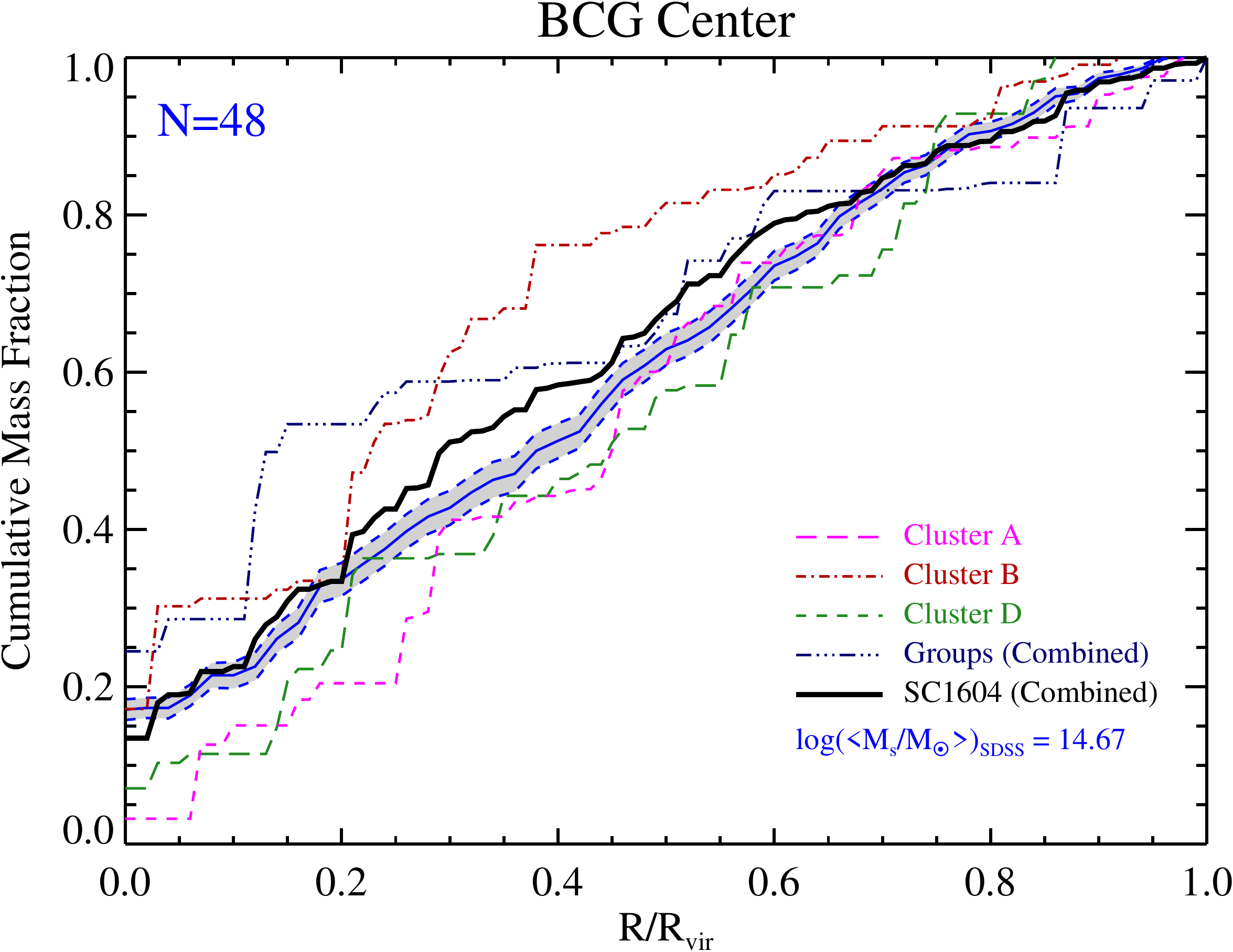}\includegraphics[clip,angle=0,width=1.0\columnwidth]{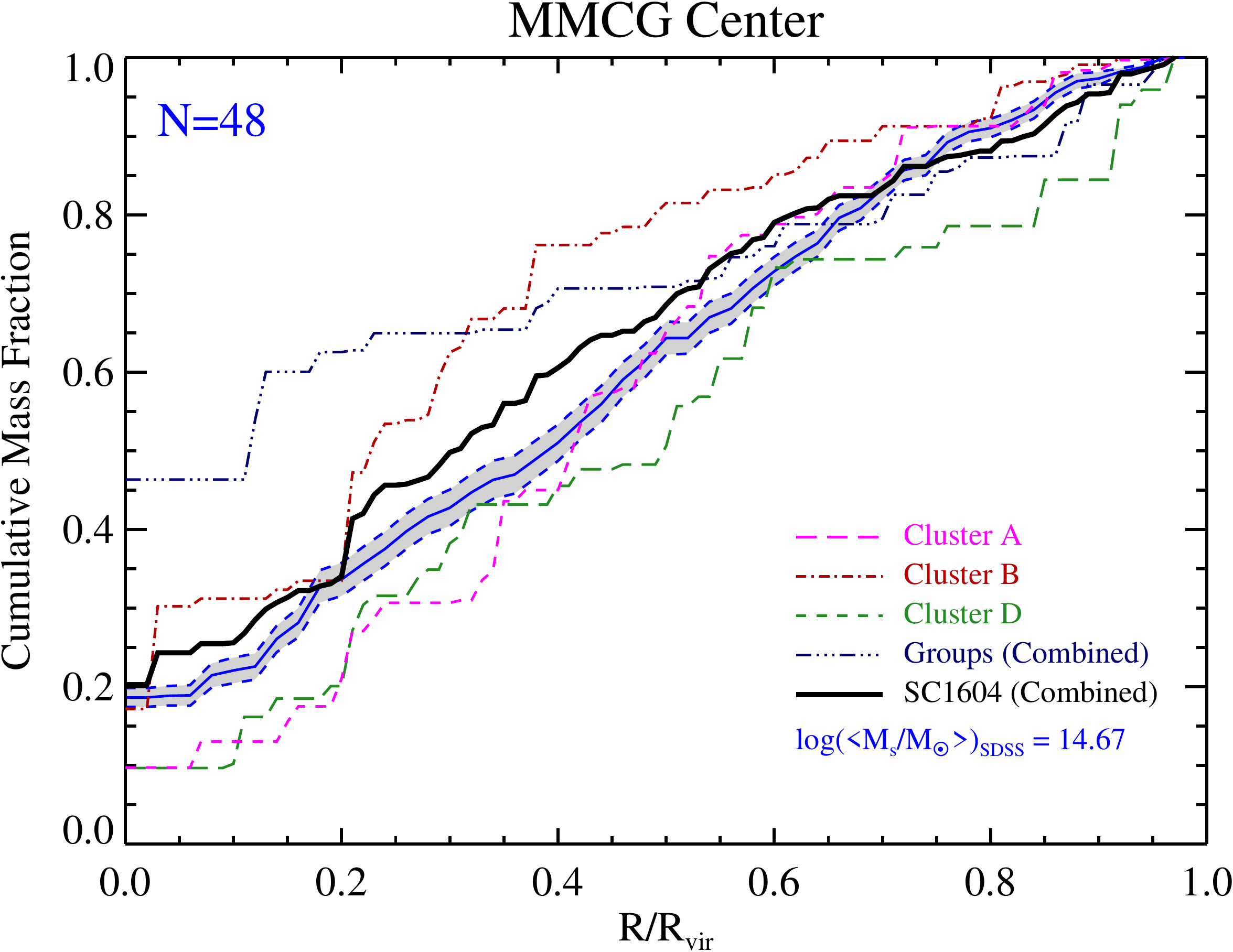}}
\caption{Normalized radial cumulative stellar mass distribution centered on the BCG (left panel) and the MMCG (right panel) for the MCXC/SDSS comparison samples and for the Cl1604 clusters and groups. The normalization in all cases is by the total amount of stellar mass contained within 1 $R_{vir}$ of the BCG (left panel) and MMCG (right panel). The samples plotted and the meaning of the lines and shaded region are identical to Figure \ref{fig:stellarmasscum}. Notice that the fraction of stellar mass contained in the average high-redshift Cl1604 BCG/MMCG relative to the total amount in the cluster at $\le1$ $R_{vir}$ is identical to that of the average BCGs/MMCG at $z\sim0.1$. The presence of massive companions to the Cl1604 BCGs/MMCGs is still apparent in this representation.} 
\label{fig:stellarmassnorm}
\end{figure*}

Plotted in Figure \ref{fig:stellarmasscum} is the radial cumulative stellar mass distribution surrounding Cl1604 BCGs/MMCGs and those of the comparison sample. For each cluster/group in each sample, the radial distribution was normalized to the virial radius of that cluster/group in order to average the samples at a common scaled physical scale rather than an absolute (projected) radial distance. As will be true for the remainder of this section, only those galaxies which are \emph{spectroscopically confirmed members} of each cluster/group are used in this analysis. The high level of spectroscopic completeness of the Cl1604 and MCXC clusters ensures that a large fraction of the true members, including all of the brightest, and, therefore, most massive members in each cluster are represented in the radial distributions. For the Cl1604 groups, the level of spectroscopic completeness is not as high as that of the Cl1604 clusters and varies considerably from group to group. While it is still true that nearly all of the most massive member galaxies in the groups will be represented in this plot, the groups were combined into an average sample to mitigate the effects of differing completeness. This sample is further combined into a full Cl1604 average in Figure \ref{fig:stellarmasscum} (black line) whose spectroscopic completeness (by construction) is identical to that of the comparison sample.

Several observations can be made from this plot. The first is the steep inner slope of the radial cumulative stellar mass distribution for the average Cl1604 clusters/groups, a phenomenon not observed for the average BCG/MMCG in the comparison sample. This point will be returned to later in this section. From this plot it is also possible to directly determine the average growth in stellar mass of the average BCG/MMCG from $z\sim0.9$ to $z\sim0.1$. The BCGs in Cl1604 are deficient in mass by an average factor of 2.51$\pm$0.71 relative to the comparison sample, where the errors on this quotient are calculated by a combination of the Cl1604 stellar mass errors given in \S\ref{cl1604} and the sample variance of the comparison sample. This result is in  direct contradiction to previous results that utilized a large sample high-redshift clusters from the European Distant Cluster Survey (EDisCS; \citealt{white05}) to probe the stellar mass growth of BCGs over the same redshift range \citep{whiley08}. In that study, no stellar mass growth of BCGs was observed between redshift $z\sim1$ and $z\sim0$. It is unclear whether this contradiction arises due simply to cosmic variance or to some differential bias induced by different methods of comparison between BCGs high- and low-redshift. However, a more recent study by the same group of people \citep{valentinuzzi10} in which a factor of two growth in BCG stellar mass is observed using a similar sample over the same redshift range strongly suggests that it is the latter. The factor of stellar mass growth derived for the Cl1604 BCGs is similar to that derived in other recent works that combine data from a large number of clusters over similar redshift ranges \citep{lidman12,lin13}. In addition, the Cl1604 MMCGs also appear to grow in stellar mass appreciably over this redshift, suggesting that, at the very least, band-induced selection effects are largely negligible with regard to the binary question of whether or not BCGs/MMCGs have grown in stellar mass over this redshift range. However, the stellar mass growth of MMCGs is milder, with the Cl1604 MMCGs being deficient on average by a factor of 1.78$\pm$0.45 relative to their low-redshift counterparts. The main difference between this quotient and the one determined for the BCGs is the large difference in stellar mass of the Cl1604 MMCGs with respect to the BCGs, as the MCXC BCG and MMCG samples are largely  the same. Regardless, both growth factors are (to varying degrees) consistent with the mass growth of the cluster as a whole over the same redshift range as derived from $N$-body simulations (a factor of $\sim$ 2.5 in total mass), which suggests that the stellar mass growth of BCGs/MMCGs occurs in lockstep and with the growth of the total mass of the cluster.  

Exploring this relationship in more detail, the dependence between the stellar mass growth within $R_{vir}$ and the total mass of the cluster growth is parameterized with $\gamma_{R_{vir}}$ through: 

\begin{equation}
\frac{SM_{MCXC}(R<R_{vir})}{SM_{Cl1604}(R<R_{vir})} = (\frac{M_{vir,MCXC}}{M_{vir,Cl1604}})^{\gamma_{R_{vir}}}
\label{eqn:gammarvir}
\end{equation}
and the BCG stellar mass growth and the total mass of the cluster growth with  $\gamma_{BCG}$:

\begin{equation}
\frac{SM_{MCXC}(BCG)}{SM_{Cl1604}(BCG)} = (\frac{M_{vir,MCXC}}{M_{vir,Cl1604}})^{\gamma_{BCG}}
\label{eqn:gammabcg}
\end{equation}

\noindent In this work, we obtain a value of $\gamma_{R_{vir}} =0.87 \pm 1.27$ and $\gamma_{BCG} =1.05 \pm 1.46$, These values can be directly compared to the slope of the stellar halo mass relation estimated  from simulations and optical data ($\sim 0.2-0.5$; \citealt{moster10,behroozi10}) or weak lensing ($\sim 0.2$; \citealt{leauthaud12}). While our results are compatible within the errors, we notice that our average value is significantly higher than those found by other works. Several effects can justify this difference. First, we are assuming no evolution in the stellar halo mass relation. Although different works \citep{conroy09,moster10} support little evolution with redshift at the massive end, small evolutionary corrections might vary those quantities. Furthermore, theoretical works (e.g.: \citealt{behroozi10}) have extensively discussed the wide range of uncertainties in modeling this relation which can lead to  0.25 dex or even more. In addition, Fig. 10 in  \cite{leauthaud12} shows that, to date, there exist few samples able to constrain the high mass regime, in particular the range of halo (virial) masses spanned by the Cl1604 and MCXC structures (i.e., $13.6<\log M_{vir}<14.5$). We note that the sample by \cite{leauthaud12} in the halo mass range $13<\log M <14$ appears to agree well with both the $\gamma_{R_{vir}}$ and $\gamma_{BCG}$ derived from our observations in that the average halo-to-stellar mass ratio of the MCXC and Cl1604 structures appears to simply extend the trend observed in these structures. Higher than this threshold there exists a significant departure from this trend. However, for such clusters halo masses are measured via weak lensing and, consequently, the estimation of the dark matter haloes and their errors are not directly comparable to the kinematically-derived halo masses in this work. This exercise, however, serves to illustrate the large uncertainty in the high end of this relation. Unfortunately, it is observationally difficult to improve these results at present. The main contributor to the errors in $\gamma_{R_{vir}}$ and $\gamma_{BCG}$ are the errors on the high-z virial masses ($\sim 60\%$ of the errors) and the systematic uncertainties in the high-z stellar mass estimates ($\sim 25 \%$ of the error). The first source of error depends directly on the spectroscopic sampling of the cluster as well as the overall number of clusters to be averaged over, while the second depends on the depth of the rest-frame NIR imaging. Given that the high-$z$ sample in this work is one of the most complete dataset in terms of spectroscopy coverage available at $z\sim 0.9$, a larger sample of clusters spectroscopic covered down to at least the same limiting magnitude as the observations of Cl1604 are necessary to constrain these slopes more accurately through observations. This will perhaps be possible through an inclusion of the full ORELSE dataset. 

\begin{figure*}
\centering
\centerline{\includegraphics[clip,angle=0,width=1.0\columnwidth]{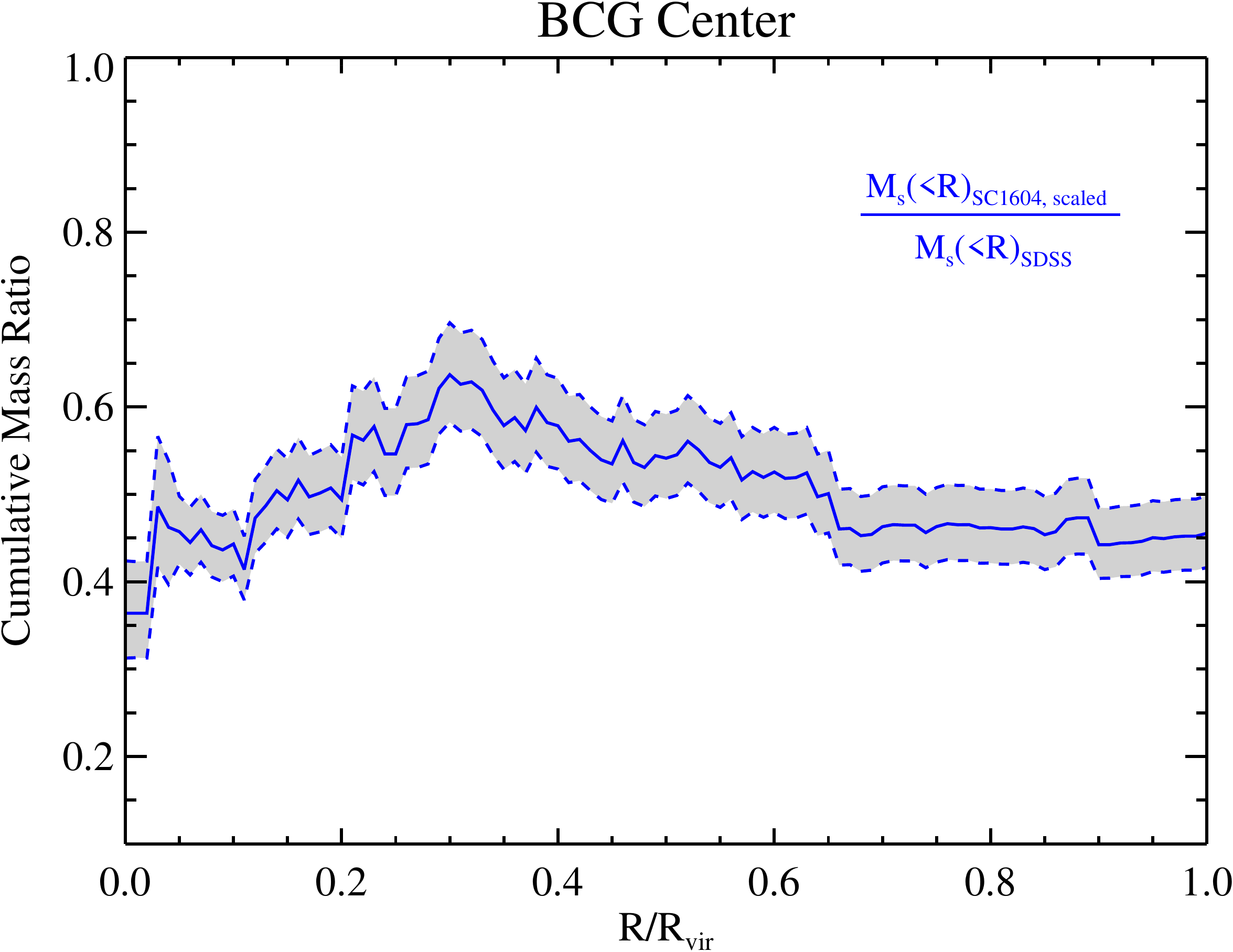}\includegraphics[clip,angle=0,width=1.0\columnwidth]{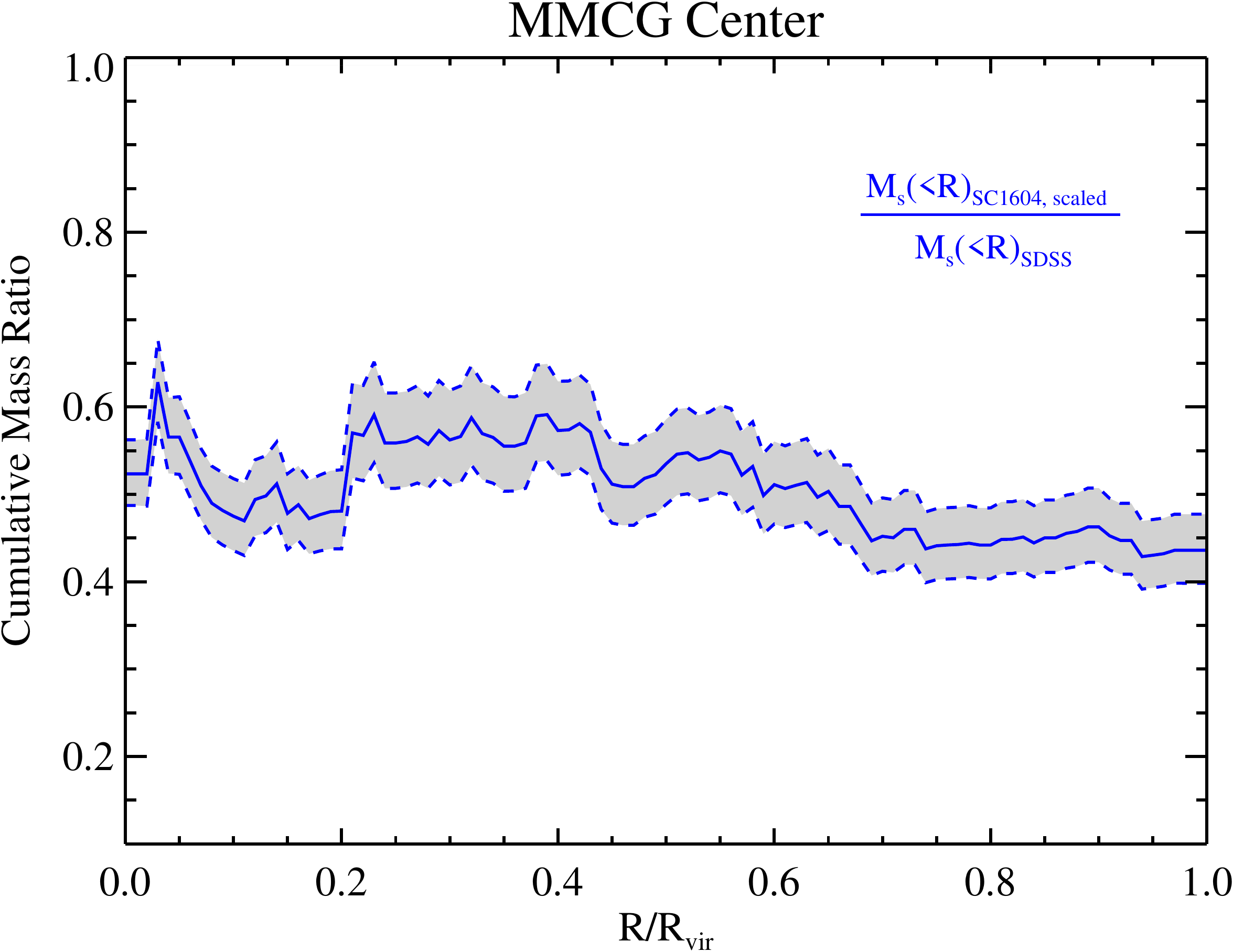}}
\caption{The ratio of the Cl1604 to the MCXC radial stellar mass cumulative distribution using the BCG (left panel) and the MMCG (right panel) as centers. The blue line shows the average ratio between the the cumulative stellar mass distribution in the average Cl1604 cluster/group and that of the average MCXC cluster in the comparison sample. The gray shaded region denotes the sample variance at each radius. The projected radial distance of members of all clusters and groups are normalized by the virial radii of their parent structures. An increase in the ratio in either panel indicates an excess in stellar mass surrounding the average BCG/MMCG in the high-redshift clusters and groups relative to that surrounding the average low-redshift BCG/MMCG. The average BCG at high redshift shows a decided excess of stellar mass at a relatively low (normalized) radius increasing steadily out to $R\sim0.3R_{vir}$. The average MMCG at high redshift also shows a sharp excess of stellar mass in its immediate surroundings relative to the average low-redshift MMCG, but only at very small (normalized) radius ($R<0.05R_{vir}$). This excess indicates the presence of extremely massive companions of the high-redshift MMCGs that are not present at low redshift.} 
\label{fig:stellarmassratio}
\end{figure*}

In Figure \ref{fig:stellarmassnorm} we again plot the radial cumulative stellar mass distribution of galaxies surrounding the Cl1604 and comparison sample BCGs/MMCGs. This time, however, the distribution is normalized by the total stellar mass in each cluster. It is apparent from the nearly identical fraction  of total stellar mass observed at low radii in the combined Cl1604 and comparison samples that not only does the stellar mass of BCG/MMCG grow in lockstep with the mass of the cluster as a whole, but it also evolves in tandem with the total stellar mass content of the cluster. It appears that \emph{the stellar mass of the BCG is fundamentally linked with the growth of both the baryonic and dark matter mass of the cluster as a whole}, though, as discussed previously, there exists a large uncertainty on these growth factors. In both the case of the average Cl1604 BCG and MMCG, it can again be seen that the inner profile of the normalized radial cumulative stellar  mass distribution is steeper than that of the comparison sample. This is recast in Figure \ref{fig:stellarmassratio} in which we plot the ratio of the Cl1604 and comparison sample (un-normalized) radial cumulative stellar mass distributions (hereafter ``cumulative mass ratio"). The value of the cumulative mass ratio increases  when there is more stellar mass within a given radius (normalized by the virial radius) surrounding the Cl1604 BCGs/MMCGs relative to the low-$z$ comparison sample and drops in the reverse case. In the left panel  of Figure \ref{fig:stellarmassratio} we plot the cumulative mass ratio centered on the BCGs samples. In this light, the steep inner slope of the Cl1604 radial cumulative stellar mass distribution relative to the  comparison sample becomes obvious. At very low projected radii ($R_{\rm{proj}}<0.05R_{vir}$) there is a sharp increase in the cumulative mass ratio, which implies a extremely nearby, massive companion or companions surrounding the Cl1604 BCGs that are largely absent in the low-$z$ comparison sample. This ratio continues to increase nearly monotonically from $R_{\rm{proj}}=0$ to $R_{\rm{proj}}\sim0.3R_{vir}$, meaning the number or stellar mass (or both) of companions surrounding the \emph{average} Cl1604 BCG at low projected radii far exceeds those of the low-$z$ comparison sample. The large number of massive companions surrounding the BCGs of the Cl1604 clusters and groups is largely consistent with observations of  ``BCGs" in a large sample X-Ray selected groups that span a large redshift range ($0.4<z<1.1$; \citealt{jeltema06,jeltema08,jeltema09}). In the right panel of Figure \ref{fig:stellarmassratio} the same exercise is performed for the MMCGs. Though the same signature of extremely nearby massive companion or companions is seen surrounding the \emph{average} Cl1604 MMCG, the profile of the cumulative mass ratio remains largely flat beyond these radii. 

\begin{figure*}
\centering
\centerline{\includegraphics[clip,angle=0,width=0.97\columnwidth]{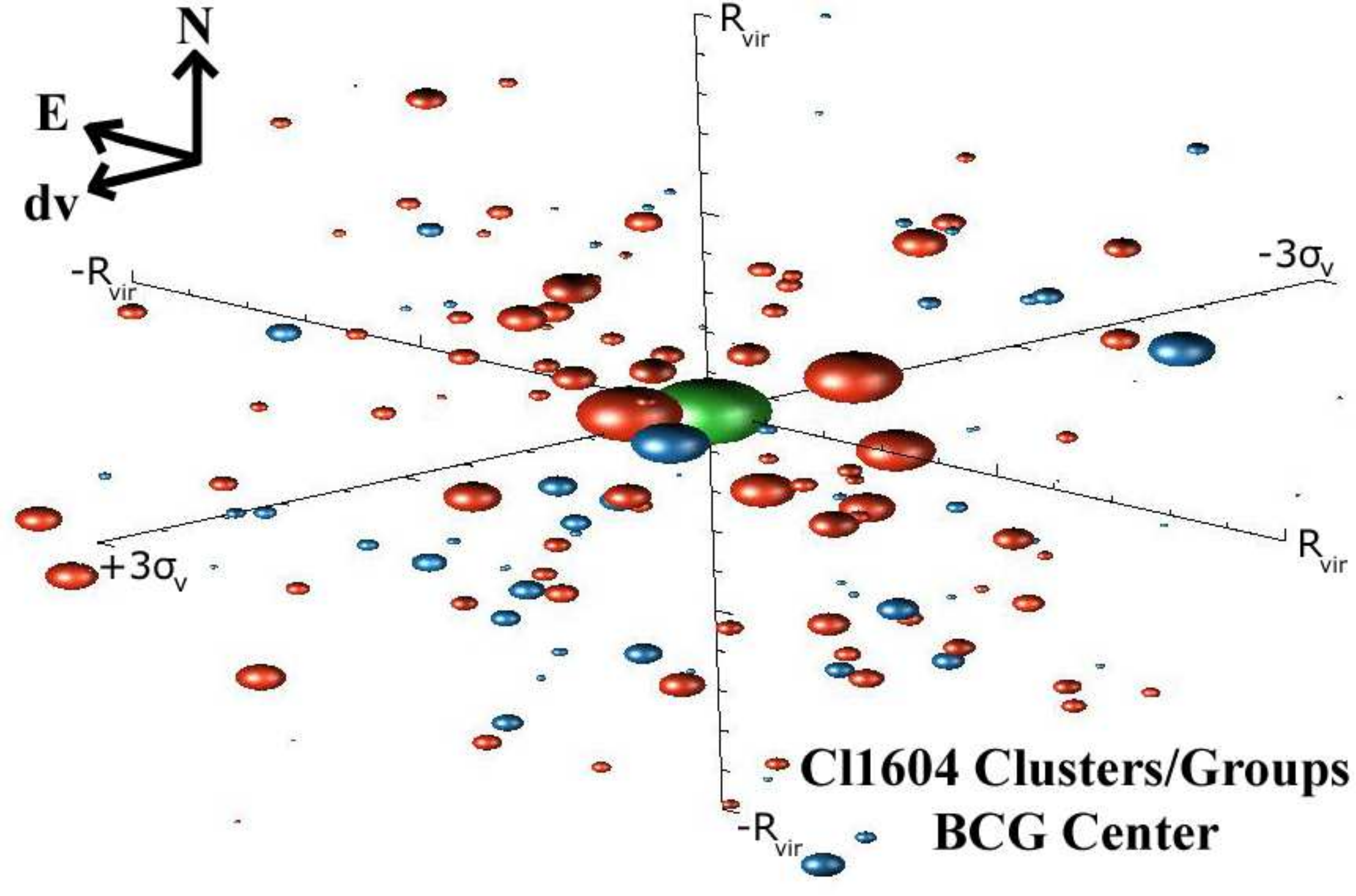}\includegraphics[clip,angle=0,width=0.97\columnwidth]{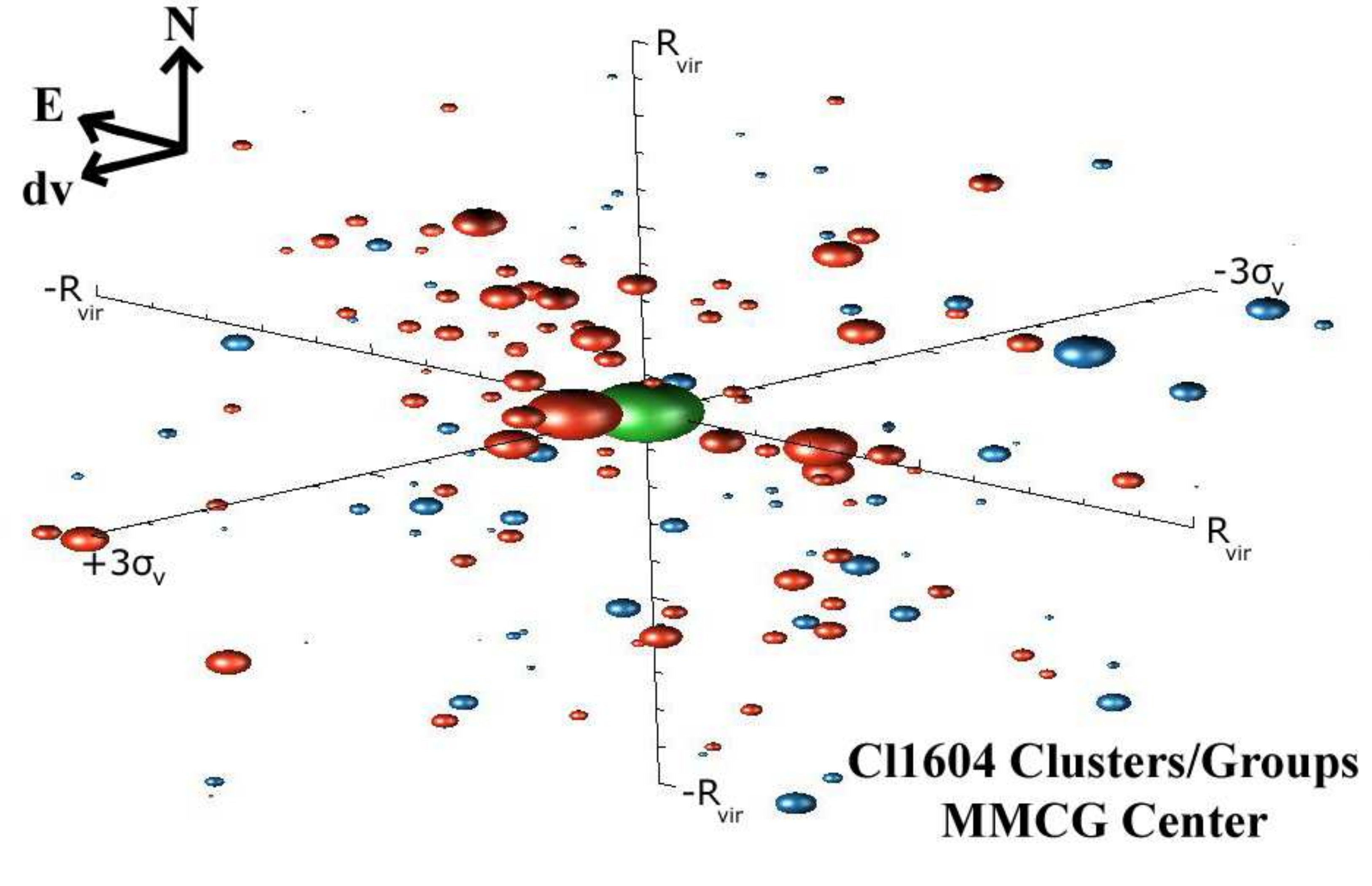}}
\caption{\emph{Left:} Three-dimensional plot of the (projected) spatial and differential velocity distribution of galaxies surrounding the Cl1604 BCGs. All Cl1604 clusters and groups are combined into a single plot, with the spatial axes normalized by the virial radius of each structure and the velocity axis normalized by $3\sigma_{v}$ for each cluster/group. The color of each sphere denotes whether a galaxy is in the blue cloud or the red sequence for a particular cluster or group and the size of each sphere is scaled (logarithmically) with the stellar mass of each galaxy. The BCGs, comprised of both blue-cloud and red-sequence galaxies, are shown as a green sphere, with the size of the sphere set by the most massive BCG (that of cluster B). In this plot, BCGs are placed at the center of the spatial and velocity dimensions, with the differential velocity and spatial offsets of all galaxies calculated with respect to the BCGs. \emph{Right:} Identical to the left panel except that the spatial and differential velocity distribution is show with respect to the Cl1604 MMCGs. The Cl1604 MMCG sample is denoted by a green sphere with the size of that sphere again set by the mass of the most massive MMCG (that of cluster B). Notice the large numbers of massive galaxies surrounding both the BCGs and MMCGs and the relative paucity of lower mass galaxies. Though many of the more massive galaxies surrounding the BCGs/MMCGs belong to the red sequence, the high incidence of dusty starbursts, spiral morphologies, and high star formation rates of both the BCG/MMCG samples as well as the galaxies surrounding the BCGs/MMCGs suggest that a large fraction of the eventual mergers will be wet (i.e., between two gas-rich galaxies) or mixed (i.e., between a gas-rich and gas-depleted galaxy).}
\label{fig:spherez}
\end{figure*}

Now that we have firmly established the close proximity of a substantial amount of stellar mass surrounding both the high-$z$ BCGs and MMCGs, the plausibility of significant stellar mass buildup through merging processes increases considerably. In order to quantify this possibility further, the merging timescale of each galaxy within $\Delta R_{\rm{proj}}<0.3R_{vir}$ of each Cl1604 BCG and MMCG was calculated using Equation 3 of \citet{burke13}, 
which is based on the method presented in \citet{binney87}. Under this formalism, the merging timescale increases as $\Delta R_{\rm{proj}}^2$ and linearly with the difference in line-of-sight velocities ($dv$) and decreases linearly with the mass of the companion galaxy. Here we define major merging candidates as those galaxies which have a stellar mass ratio with respect to the BCG or MMCG of 4:1 or lower and minor merging candidates as those galaxies with stellar mass ratios between 4:1 and 25:1 (the upper limit being set by the rough stellar mass completeness limit of the Cl1604 spectroscopy, see \citealt{lemaux12}), definitions which are similar to those of other studies (e.g., \citealt{lidman13}). Within this projected radius, a total of 47 and 41 merging candidates surround the Cl1604 BCGs and MMCGs, respectively. These galaxies are represented in Figure \ref{fig:spherez}, where $\Delta R_{\rm{proj}}$ and $dv$ is plotted for all spectroscopically confirmed member galaxies within $\Delta R_{\rm{proj}}<R_{vir}$ surrounding the BCGs and MMCGs of all Cl1604 clusters and groups. Of the merging candidates within $R_{\rm{proj}}<0.3R_{vir}$, 51.0\% and 31.7\% for the BCG and MMCG samples, respectively, are major merging candidates. The high incidence of major merging candidates at surrounding the Cl1604 BCGs/MMCGs can be seen in Figure \ref{fig:spherez}, as many of the galaxies at low $\Delta R_{\rm{proj}}$ and low $dv$ are of reasonably large stellar mass. However, because of the way we have chosen to represent the galaxies in Figure \ref{fig:spherez}, with the size of the sphere of all BCGs/MMCGs represented by the most massive BCG/MMCG, and because the definition of major merger is dependent on stellar mass ratios rather than the absolute magnitude of the stellar mass of the BCG/MMCG or its companion, it is not immediately obvious which companion galaxies are of the major merging candidate flavor. Regardless, it is sufficient to say that a large fraction of the galaxies at close (projected) spatial distances and small differential velocities with respect to the Cl1604 BCGs/MMCGs are massive and are comprised of nearly all the most massive galaxies in the entire Cl1604 clusters/groups  sample.

For the 47 and 41 merging candidates surrounding the Cl1604 BCGs and MMCGs, respectively, we now make the assumption that all galaxies which have merging timescales lower than the difference in cosmic time between the average redshift of the Cl1604 and MCXC/SDSS samples (e.g., $\la6.5$ Gyr) merge with their BCG/MMCG companion. This assumption leads to an average increase in BCG stellar mass of a factor of 2.23$\pm$0.73 and  an average increase in MMCG stellar mass of a factor of 1.35$\pm$0.31. Of the 15 merging candidates with a small enough timescale to have merged by $z\sim0.1$, all but one has a mass ratio lower than 4:1 (i.e., 14/15 are major mergers). This high incidence of major merging candidates is similar to that recently found in amongst the BCGs of ten $z\sim1$ clusters \citet{lidman13}, which suggests this phenomenon is not unique to the BCGs/MMCGs of the Cl1604 supercluster, but is rather common at these redshifts. It is important to note, however, that the merging events that result from the companion galaxies of the Cl1604 BCGs/MMCGs are typically not dry; a majority of the major merging candidates  have morphologies or colors or 24$\mu m$ detections that imply, along with the properties of the BCG/MMCG, a mixed or wet merging event (see, e.g., \citealt{lin08} for the definitions of dry, mixed, and wet merging as they are applied here). Such events subsequently lead to increased star formation activity (see, e.g., \citealt{hopkins08} for a theoretical view and \citealt{kocevski11a} for an observation of this phenomenon in the Cl1604 supercluster), which will in turn increase the resultant stellar mass beyond the simple addition of the stellar masses of the two galaxies. In conjunction with the analysis presented in the previous section, the combination of \emph{in situ} star 
formation, major merging, and merging induced star formation appears more than sufficient to buildup the stellar mass in BCGs/MMCGs from $z\sim0.9$ to $z\sim0.1$. As will be discussed in the subsequent section, however, these results do not preclude the possibility of additional minor merging, as substantial stellar mass of the BCG can be lost to the cluster medium in major merging events. In the next section we attempt to contextualize these results in terms of parametric morphological evolution between the BCGs/MMCGs of the two samples. 

\subsection{Evolution of Structural Parameters}
\label{morphevo}

\begin{figure*}
\centering
\includegraphics[clip,angle=0,width=0.475\hsize]{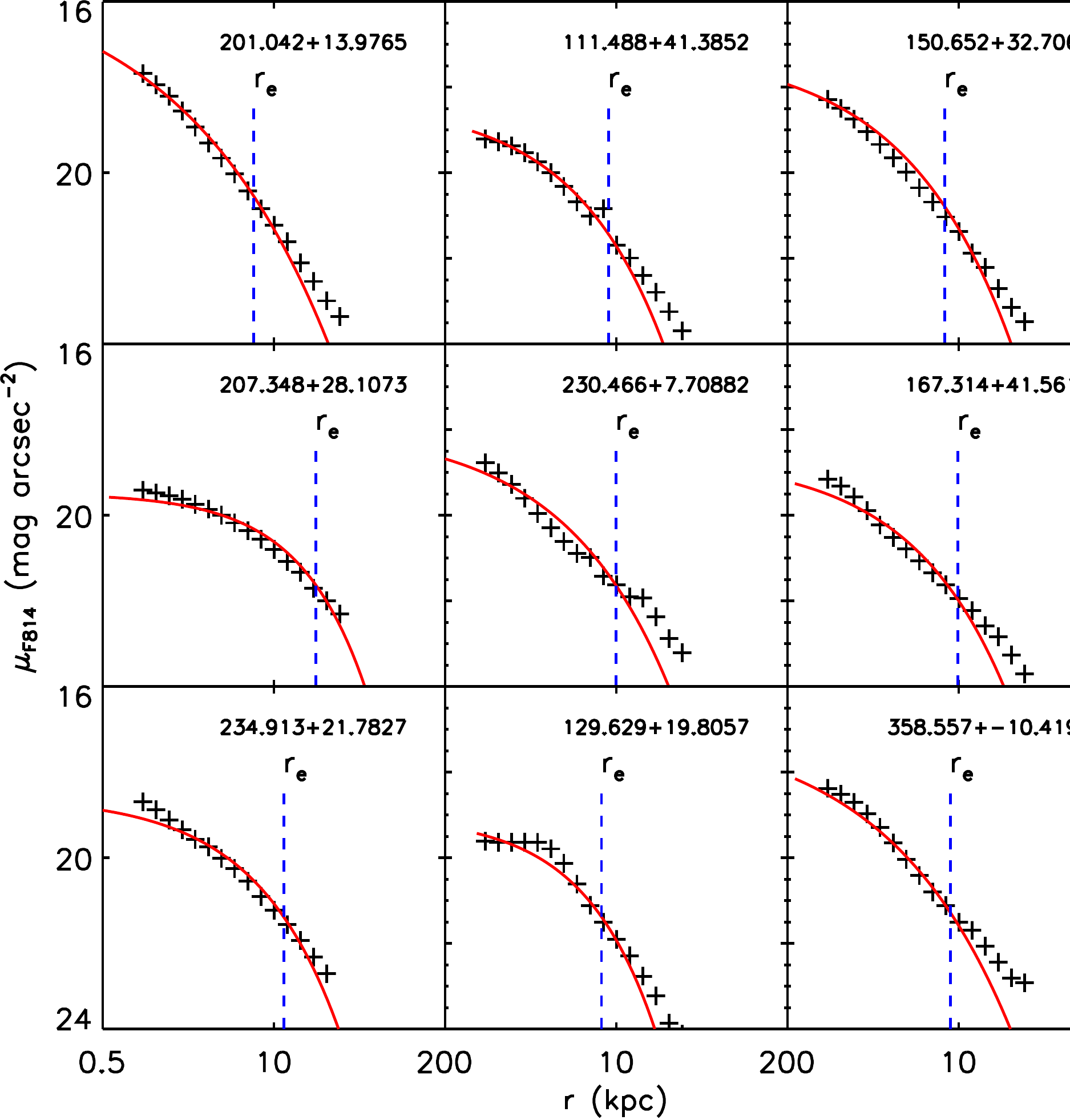}
\includegraphics[clip,angle=0,width=0.475\hsize]{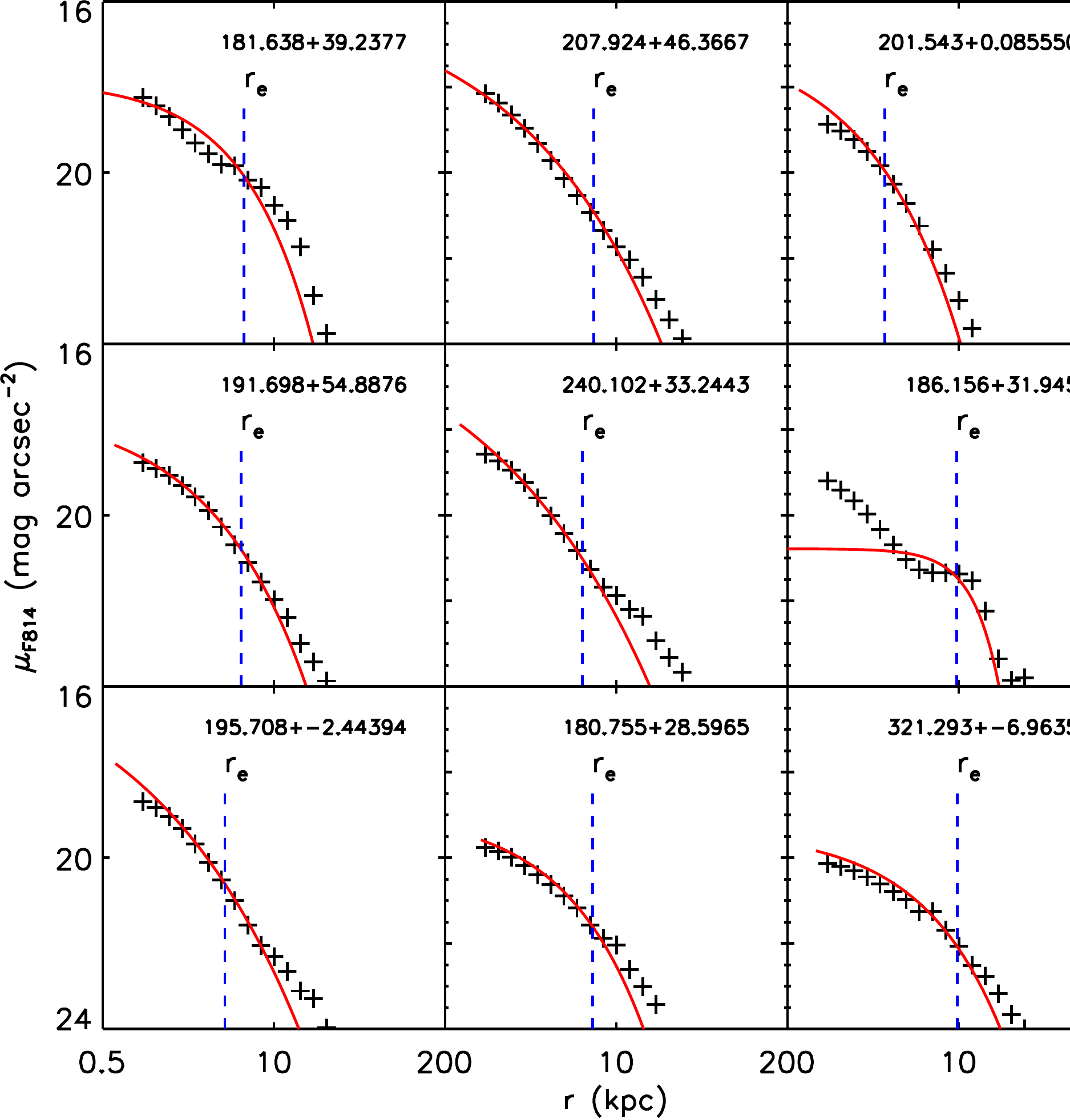}
\caption{Nine examples of the 1D-surface brightness profile for the MCXC/SDSS sample. The left panels show fits to MCXC BCGs and the right panels to MMCGs. Only those MMCGs which were not selected as BCGs are shown. The plus signs indicate the measured, azimuthally-averaged surface brightness radial profile of the galaxy as measured from the SDSS $g^{\prime}$ imaging and the curved solid red line shows the best-fit single S\'ersic profile. The blue vertical dashed line indicates the effective radius for each galaxy.}
\label{fig:SBfitlow}
\end{figure*}

\begin{figure*}
\centering
\includegraphics[clip,angle=0,width=0.475\hsize]{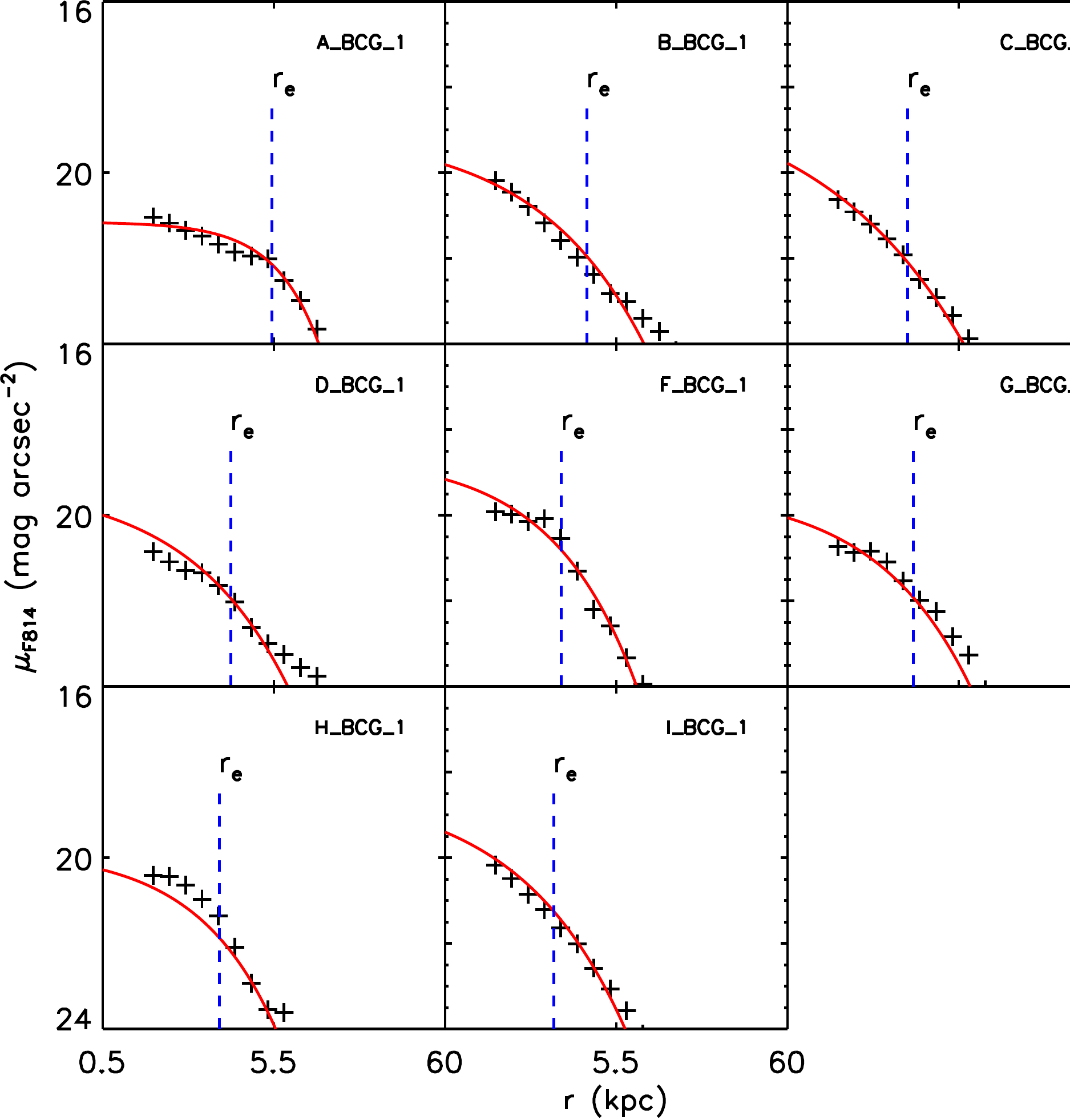}
\includegraphics[clip,angle=0,width=0.475\hsize]{Fig18b.pdf}
\caption{The same as Figure \ref{fig:SBfitlow} but for the Cl1604 BCGs/MMCGs. The meanings of the symbols and lines are identical to Figure \ref{fig:SBfitlow}. In the right panels, only those galaxies selected as (potential) MMCGs which were not already selected as BCGs are shown. Note that the features associated with late-type morphologies or disturbances coming from interactions are common amongst the Cl1604 BCGs/MMCGs,  which results into very different profiles with respect the MCXC/SDSS sample.}
\label{fig:SBfithigh}
\end{figure*}

Comparing BCG/MMCG structural parameters over cosmic time can, in conjunction with stellar mass and spectral evolution, constrain evolutionary scenarios and differentiate scenarios which solely involve merging from those involving solely adiabatic expansion (\citealt{hopkins10, ascaso11}). 
To this end, we fit the two dimensional surface brightness (SB) profiles of all the BCGs/MMCGs in both the Cl1604 and MCXC/SDSS samples using GALFIT 3.0 \citep{peng02,peng10}, with a single S\'ersic model \citep{sersic68}. A single S\'ersic profile is usually sufficient to describe the profiles of ``typical" elliptical galaxies \citep{gonzalez05,seigar07,bernardi07,donzelli11,ascaso11} though BCGs with large haloes, also called dominant (cD) galaxies, sometimes require an extra component to describe its light  \citep{bernardi07,donzelli11}. Many of the BCGs/MMCGs in our sample exhibit large haloes, primarily in the MCXC/SDSS sample. While a second component can, in principle, be used to model some portion of the halo light, modeling in this manner introduces degeneracies between the different parameters \citep{peng02,haussler07,barden08,peng10}. Furthermore, most of the halo light likely falls below even the SDSS surface brightness limit, confounding the interpretation of this second component in terms of a physical picture (e.g., \citealt{gonzalez05}).  Thus, in the rest of this subsection, structural parameters are derived utilizing only a single S\'ersic profile model for the fitting.

The fitting procedure was as follows. First, the postage stamps of BCGs and MMCGs from the full sample of the 81 MCXC clusters which defined our full sample were created from the SDSS $g^{\prime}$ imaging. The same process was followed for the Cl1604 BCGs/MMCGs using the ACS $F814W$ imaging. The $g^{\prime}$-band SDSS imaging was used for the generation of the MCXC postage stamps because this band roughly samples the same rest-frame wavelength range at $z\sim0.1$ as the ACS $F814W$ band at $z\sim0.9$. The size of the generated postage stamps were 200 and 100 $h_{70}^{-1}$ kpc on a side for the MCXC and Cl1604 samples, respectively, considerably larger than those shown in \S\ref{vismorph}. These sizes were chosen in order to allow for the measurement of the total profile of the BCG/MMCG in addition to enough ``blank" sky coverage to make a proper determination of the noise properties of each image. Following the generation of the postage stamps, point spread function (PSF) images were created following the procedure explained in the Postage Stamp Pipeline (PSP; \citealt{stoughton02}) for the SDSS images and \cite{jee07} for the ACS images. We then ran SExtractor \citep{bertin96} on the postage stamps with a twofold objective: to create masks of the surrounding objects and to obtain initial positions, orientation angle, and sizes of the central galaxy to be used as initial conditions for GALFIT. Finally, GALFIT was run on the all the masked SDSS and ACS postage stamps.

As alluded to earlier, the SDSS sample is approximately four times deeper than the ACS in limiting surface brightness in terms physical units (i.e., 23.5 $L_{\odot}/pc^2$ vs. 94.0 $L_{\odot}/pc^2$, respectively). In order to determine the effect that this differing surface brightness limit might have on our fitting results we performed the following test. A random subsample of five MCXC BCGs was selected and noise was added to the masked SDSS postage stamps to create 7 ``noise-added" SDSS postage stamps for each MCXC BCG. The noise was added such a way that the surface brightness limit of the noise-added SDSS images, in physical units, ranged between the nominal SDSS limit and a limit shallower than that of the ACS imaging in equally spaced steps. At each step, the fitting process was repeated. Over the relatively large range of surface brightnesses tested here we found essentially no difference in the best-fit parameters; the variation of the best-fit effective radius and S\'ersic index was $<2$\% and $\sim$6\%,  respectively. This is also seen in other works using similar tests (e.g., \citealt{haussler07,barden08}). The results presented in this paper are robust to a change of this level on either parameter and we, therefore, take no account of the difference in physical surface brightness depths between the two samples for the remainder of the paper.  For illustration, we plot in Figures \ref{fig:SBfitlow} and \ref{fig:SBfithigh} the 1D-surface brightness profiles of the MCXC and Cl1604 BCGs/MMCGs, respectively, with the results of the single S\'ersic model overplotted. Note that the fits performed in this work are made in two dimensions, while we represent in these Figures the one dimensional azimuthally-averaged surface brightness radial profiles including possible satellite galaxies, which are excluded from the fit. The one component S\'ersic profile sufficiently describes the low- and high-redshift samples in almost all cases.

\begin{figure}
\centering
\includegraphics[clip,angle=0,width=1.0\hsize]{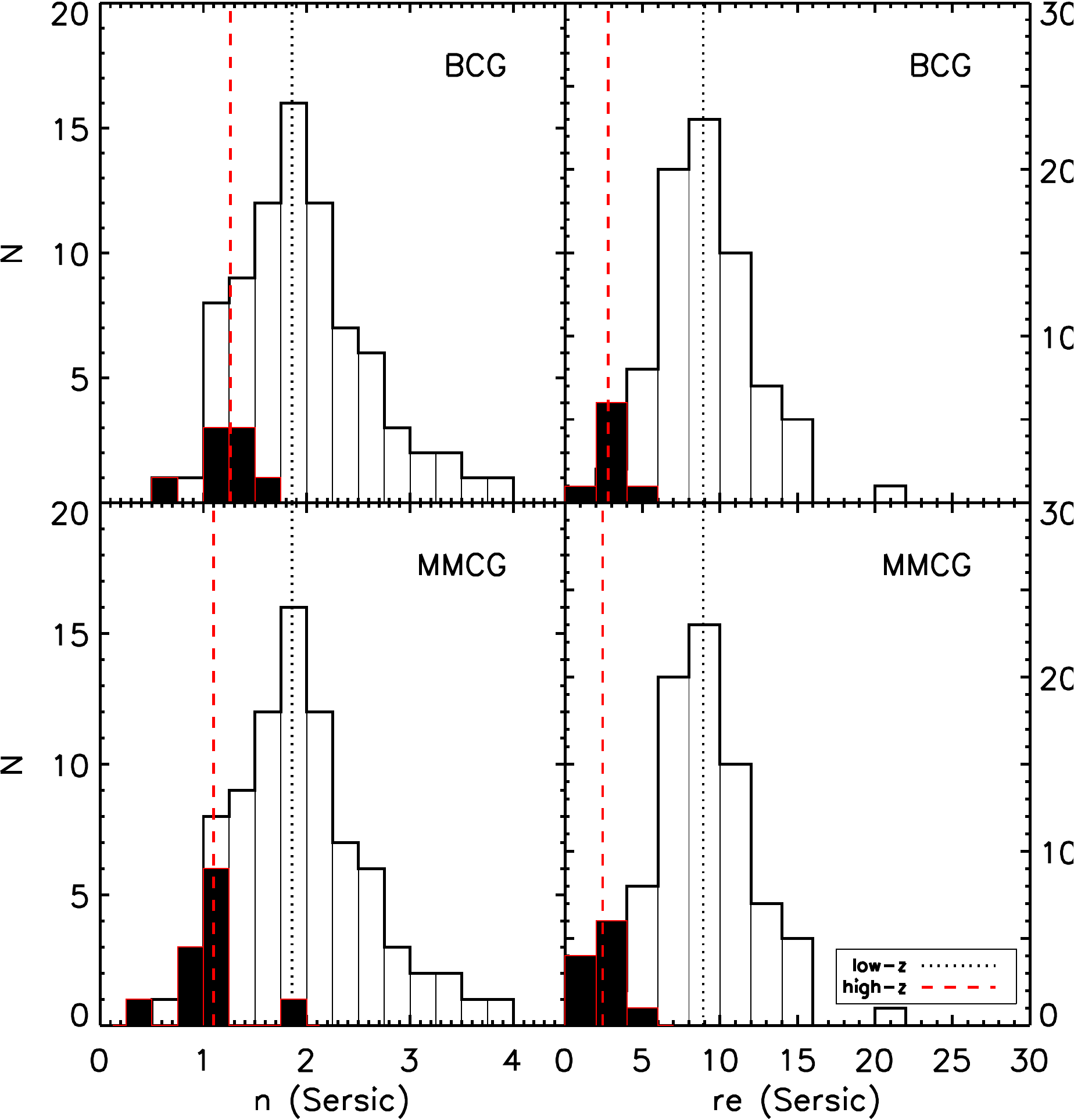}
\caption{Distribution of the BCG S\'ersic indices (top left panel), BCG effective radius (top right panel), MMCG S\'ersic indices (bottom left panel) and MMCG effective radius (bottom right panel) for the one S\'ersic component fit. The low$-z$ sample is shown in white and the high$-z$ sample in black. The vertical solid dotted and red dashed line refers to the median value for each sample. The low$-z$ BCGs/MMCGs have both a higher effective radius and S\'ersic index, on average, than the high$-z$ BCGs/MMCGs. These results are quantified in the text.}
\label{fig:histo1comp}
\end{figure}

In Figure \ref{fig:histo1comp}, the distributions of the best-fit main structural parameters, effective radius ($r_e$) and S\'ersic index ($n$), derived from the S\'ersic fit are plotted for both the BCGs and MMCGs of the low- and high-redshift samples. It is apparent from this figure that the BCGs/MMCGs in the low-redshift MCXC/SDSS sample are both larger and have increased S\'ersic indices relative to their high-redshift counterparts. In Table \ref{tab:resultfitsS} the average values, along with their errors, of the main structural parameters are given for all samples of BCGs/MMCGs. A factor of $3.07 \pm 0.01$ growth in $r_{e}$ is observed from $z\sim0.9$ to $z\sim0.1$ in the average BCG and an evolution of the average S\'ersic index of $n(z=0)-n(z)=0.45 \pm 0.03$ is observed over the same redshift range. For the MMCG sample the observed evolution in both parameters is similar, with an increase in $r_{e}$ of a factor of $3.33 \pm 0.01$ and an increase of the average S\'ersic index of $n(z=0)-n(z)=0.63 \pm 0.03$. As mentioned earlier in this section, these numbers are derived from the full sample of 81 MCXC clusters defined in \S\ref{mcxc}. It was shown in \S\ref{clustermass} that there was not an appreciable difference in cluster properties between the 81 MCXC clusters we selected, the full MCXC cluster sample and the 53 MCXC clusters used as a basis for the sample in \S\ref{massrad}. Regardless, the average values of the S\'ersic index and size of the BCGs/MMCGs were also calculated from the comparison sample defined in \S\ref{massrad}. No significant difference in either of the median structural parameters relative to those derived from our full sample of 81 MCXC clusters were found. Thus, the results presented in this section are broadly applicable to the results presented in other sections.

The factor of $\sim3$ evolution in size for both the BCGs and MMCGs between $z\sim0.9$ and $z\sim0.1$ is consistent with evolutionary factors found in other cluster studies \citep{nelson02,bernardi09,ascaso11} over a variety of redshift ranges. Conversely, in the field ``typical" massive elliptical galaxies have been observed to increase their size by only a factor of $\sim 2$ from $z\sim$1 to the present day \citep{daddi05,trujillo06,ryan12,huertas-company13}. Given our previous results, it is likely that the Cl1604 BCGs/MMCGs will accrete stellar mass faster than similar galaxies in the field (primarily through major mergers), which would in turn cause a differential increase in size relative to comparable field ellipticals 

However, the evolution in the S\'ersic index from $z\sim0.9$ to $z\sim0.1$ is moderate ($\sim 0.45$). This evolution is smaller than that expected from many similar measurements for massive field ellipticals \citep{vandokkum10,lopez-sanjuan12}. The magnitude of this evolution, however, still remains an open question, and the evolution observed amongst the samples presented here are in broad agreement with a measurement of the evolution amongst field and group ellipticals from $z\sim1$ to the present day (see Fig. 19 in \citealt{huertas-company13}). Environmental effects may play a large role in this evolution and, indeed, in a study of BCGs across a redshift baseline of $z\sim0.5$ to $z\sim0$, \citet{ascaso11} found no redshift evolution of the S\'ersic index. Thus, in order to attempt to understand the moderate evolution of the S\'ersic index of the BCG/MMCG in our sample, we compared our results with numerical simulations by \citet{hopkins10}, who considered different scenarios, such as adiabatic expansion and major or minor merging, to explain the evolution of structural parameters in massive galaxies. Through these simulations, the authors of this study claimed that major and minor mergers will cause both the size and the S\'ersic index of ellipticals to grow in time, while adiabatic expansion could serve to soften the change in the S\'ersic index induced by merging events. Hence, a pure analysis of the structural parameters makes it plausible that the Cl1604 BCGs/MMCGs evolve through a mixture of different mechanisms, with major/minor merging and adiabatic expansion contributing at different levels and at different times than for massive field ellipticals. In the remainder of the paper we continue this discussion, using all of our results to contextualize and constrain the prevalence and significance of such mechanisms.

\begin{table}
      \caption{S\'ersic indices and Sizes from 1-component S\'ersic fit}
      \[
         \begin{array}{lllll}
            \hline\noalign{\smallskip}
\multicolumn{1}{c}{\rm }&
\multicolumn{2}{c}{\rm Low-z}&
\multicolumn{2}{c}{\rm High-z}\\
\multicolumn{1}{c}{\rm }&
\multicolumn{1}{c}{\rm <n> }&
\multicolumn{1}{c}{\rm <r_e (kpc)>}&
\multicolumn{1}{c}{\rm <n> }&
\multicolumn{1}{c}{\rm <r_e (kpc)>}\\
\hline\noalign{\smallskip}
{\rm BCGs}  &  1.86\pm         0.02 &         8.92 \pm         0.06 &  1.41\pm         0.02 &         2.90 \pm         0.02    \\ 
{\rm MMCGs} &  1.86\pm         0.02 &         8.92 \pm         0.06 &      1.23\pm         0.02 &         2.68 \pm         0.02   \\ 
\hline
         \end{array}
      \]
\label{tab:resultfitsS}
   \end{table}
   
\section{Discussion}
\label{connecting}

\subsection{Major/minor mergers}

Throughout this work we have shown different lines of evidence, based on the spectral properties, morphological properties, and the radial stellar mass
distribution around high- and low-redshift BCGs/MMCGs, that wet or mixed major merging events in tandem with merger induced star formation appears more than sufficient to explain the buildup of the stellar mass from $z\sim0.9$ to $z\sim0.1$ (see \S\ref{massrad}). We now attempt to contextualize these
results in terms of semi-analytic and hydrodynamical simulations. 

In a seminal study, \citet{delucia07} used the Millenium simulation to study the formation and evolution of BCGs from a hierarchical point of view. Here we compare our results with their merger tree predictions. In section \S\ref{massrad}, we determined that the mean mass quotient of the BCGs is $z\sim$0.9 and those at $z\sim0.1$ was a factor of $\sim$ 2.51$\pm$0.71. Also in that section it was determined that the stellar mass growth of the BCGs over the same redshift range through major mergers alone was 2.23$\pm$0.73 which is perhaps the more fundamental quantity when comparing to a hierarchical $N$-body simulation. This value is similar to the stellar mass growth found in recent observational works over similar redshift ranges (e.g. \citealt{lidman12,lin13} find a factor of $\sim1.8$ since $z\sim 0.9$). In the simulations of \citet{delucia07} a large fraction ($\sim70$\%) of the stellar mass buildup in BCGs is a result of major merging events, which is in broad agreement with our observational results. The expected accretion rate estimated from hierarchical merging in \citet{delucia07} over redshift range of our sample is $\sim$3.3. This value is higher than either of the values determined from our observational sample and other observational works \citep{lidman12,lin13}, though the difference is only marginally significant.

However, many works have presented evidence of large intracluster light (ICL) fractions in local clusters (e.g., \citealt{gonzalez05,krick06,gonzalez07}). 
Predictions from $N$-body simulations (e.g., \citealt{conroy07,murante07,rudick11}) show that anywhere from 25$-$80\% of the stellar matter accreted onto BCGs from major merging events ends up lost to the ICL. In the previous sections this effect was not taken into account. Thus, it possible that additional stellar mass buildup is necessary in excess of what is added to the BCGs/MMCGs through the major merging events discussed in \S\ref{massrad}. As is also discussed in that section and in \S\ref{spectral}, the large number of mixed or wet major merger candidates around the Cl1604 BCGs/MMCGs will likely increase or at least serve to maintain the already high level of \emph{in situ} star formation observed in the average high-redshift BCG/MMCG (e.g., \citealt{hopkins08,kocevski11a}). Because of the large potential of these galaxies, such stellar mass buildup would likely not be lost to the intracluster medium through supernovae winds or other processes associated strictly with star formation. However, even with the relatively large star formation rate of the average Cl1604 BCG/MMCG and assuming a continuous rate from $z\sim0.9$ to $z\sim0.1$, a large stellar mass loss  (i.e., $>25$\%) during major merging processes still leaves some room for additional stellar mass buildup through other means. This is true both when comparing to the observed quotient in BCG stellar mass from the MCXC and Cl1604 samples (2.51$\pm$0.71) and to the predictions of \citet{delucia07}. 

In the case of significant mass loss of satellite galaxies during major merging processes, it is possible that minor merging could, at least in small part, contribute to the stellar mass buildup of the Cl1604 BCGs/MMCGs over cosmic time. The introduction of subsequent minor mergers to major mergers  events could also help reconcile the observed evolution in the structural parameters, as minor mergers inflate $r_{e}$ more quickly for a given evolution in the S\'ersic index than do major mergers \citep{hopkins10}. However, such minor merging candidates are not detected in our Cl1604 spectroscopy, meaning candidate galaxies would have to have extreme mass ratios relative to the BCGs/MMCGs, i.e., $>25:1$. Depending on the amount of stellar mass lost to the ICL in each major and minor merging event (but requiring that at least 25\% is lost) and assuming a constant SFR of the Cl1604 BCGs/MMCGs, anywhere from $\sim$10 to $\sim100$ 25:1 minor mergers are allowed by the difference in mass observed between the MCXC and Cl1604 samples. Given that many galaxies in the Cl1604 supercluster have high-quality redshifts and stellar mass measurements at magnitudes fainter (and stellar masses lower) than our formal completeness limit, the complete lack of such galaxies observed in the data rules out a large number of minor merging events of this magnitude. This result is supported by other works, which also find evidence for a very high incidence of major merging candidates in $z\sim1$ clusters \citep{lidman13}. Still it is not inconsistent with our data that a few minor merging events with extreme mass ratios could help to build up the stellar mass in high-redshift BCGs. Given the wealth of lower mass blue, gas-rich galaxies still remaining in the cluster and group environments of Cl1604, such events could also aid stellar mass buildup through starbursting events subsequent to cannibalization. 

\subsection{Adiabatic expansion}

A different mechanism arises when merging events ignite AGN activity. Such merging events can be the mixed or wet major merging events, events which we have shown that the Cl1604 BCGs/MMCGs are fated to undergo, or the cannibalism of low-mass, gas-rich galaxies discussed in the previous section. Such processes are typically characterized by a galaxy losing mass from its central regions in an adiabatic manner producing a response of the stars and dark matter to ``puff up". The effects of this mechanism, known as adiabatic expansion, have been observed and characterized by a variety of different observational studies \citep{fan08,collins09,stott11,ascaso11}. As mentioned earlier, this mechanism is also studied in detail by \citet{hopkins10}, in which numerical simulations were used to study the effect of this mechanism on measured structural parameters.  

In this work, we have found strong evidence of ongoing or impending major merging amongst the Cl1604 BCGs/MMCGs. This observation should be accompanied by a strong evolution of the S\'ersic index between the Cl1604 and MCXC/SDSS samples. An evolution of the S\'ersic index is seen from $z\sim0.9$ to $z\sim0.1$ in our samples, but that evolution is more moderate, only approximately 50\% of that expected from simulations. However, observational work on BCGs \citep{ascaso11} in two separate samples spanning from $z\sim0$ to $z\sim0.5$ found no evolution in the S\'ersic index between the two samples, with an increase in $r_{e}$ similar to what is found in the study presented here. According to \cite{hopkins10}, a small decrease of S\'ersic index is expected under adiabatic expansion. Thus, a significant amount of adiabatic expansion occurring in conjunction with major (and possibly minor) merging events amongst the Cl1604 BCGs/MMCGs could provide a means to mitigate the increase in the S\'ersic index from merging alone. In other words, if the Cl1604 BCGs were to evolve only through some combination of major and minor merging events from $z\sim0.9$ to $z\sim0.1$ an increase in S\'ersic parameter of a factor between two and four is expected. However, as we have shown, the major merging candidates surrounding the Cl1604 BCGs/MMCGs will predominantly result in mixed or wet merging events. The merging events that are observed to be already underway amongst the Cl1604 BCGs/MMCGs also appear predominantly in late-type hosts. Such events are commonly thought to ignite AGN activity (e.g., \citealt{hopkins08}), which can lead to subsequent adiabatic expansion and a subsequent decrease in the S\'ersic index as measured by the light profile of the galaxy. The difference between the expected and observed evolution of the S\'ersic index, a difference that is statistically relevant at $\gg3\sigma$ \emph{for both subsamples}, strongly suggests that at least some level of adiabatic expansion must occur in BCGs/MMCGs over the last 7 Gyr.

\section{Conclusions}

In this study we have analyzed the properties of brightest and most massive cluster and group galaxies at high ($z\sim0.9$) and low ($z\sim0.1$) redshift. At high redshift, the properties of the BCGs/MMCGs of the eight groups and clusters of the Cl1604 supercluster were considered. These were compared to a sample of BCGs/MMCGs of 81 low-redshift clusters drawn from a cross-correlation of the MCXC catalog of X-Ray clusters and SDSS. The constituent structures of the MCXC and Cl1604 samples were matched in total virial mass and dynamical states using a variety of techniques and comparisons between the BCGs/MMCGs of the two samples were primarily made in a common rest-frame wavelength range. Using enormous spectroscopic and imaging databases provided at low redshift by the SDSS and at high redshift by ORELSE, we investigated the evolution of color, morphological, stellar mass, and spectral properties of BCGs/MMCGs over the past $\sim7$ Gyr. Our main conclusions are as follows.

\begin{itemize}

\item \emph{Color and luminosity evolution:} A large fraction ($\sim35$\%) of the combined Cl1604 BCG/MMCG sample were observed with colors blueward of the red sequence for its parent cluster or group. In contrast, only a small fraction ($\sim2$\%) of the BCGs/MMCGs at low redshift were observed to be similarly offset of the red sequence. The gap in magnitude between the BCG and the next brightest cluster/group galaxy in the average Cl1604 cluster/group was found to be less than half that of the average MCXC cluster.

\item \emph{Morphology evolution:} Exactly half of the Cl1604 BCG sample were galaxies classified with late-type morphologies and a large fraction ($\sim40$\%) of had signs of interaction. In contrast, only 2.5\% the BCGs at low redshift were classified as late-type and the fraction of those undergoing interactions was less than half of that of the Cl1604 sample. In addition, the Cl1604 BCGs were observed in some stage of a merging event greater than five times more frequently than the low-$z$ BCGs. These numbers did not change appreciably when the MMCGs of the two samples were considered.

\item \emph{Spectral evolution:} A majority ($\sim$53\%) of the combined Cl1604 BCG/MMCG sample show significant ongoing star formation. In contrast, only a small fraction $\sim4$\% of the MCXC BCGs/MMCGs were observed with ongoing star formation. From a stacked spectrum of galaxies, the average star formation rate of the Cl1604 BCGs/MMCGs was found to be $\langle SFR\rangle=10.5\pm0.5$ $\mathcal{M}_{\odot}$ yr$^{-1}$. This value is in stark contrast with the average SFR of the MCXC-SDSS BCG/MMCG, which was consistent with zero. In addition, strong Balmer absorption features and weak features associated with older stellar populations were observed in the Cl1604 BCG/MMCG stacked spectrum, which indicated a considerably younger mean luminosity-weighted stellar age as compared to the MCXC/SDSS sample. Even those Cl1604 BCGs/MMCGs considered passive (SSFR $<10^{11} yr^{-1}$) showed signs of a moderately young stellar population.

\item \emph{Stellar mass evolution:} The average $z\sim0.9$ Cl1604 BCG was observed to be deficient in stellar mass by a factor of 2.51$\pm$0.71 relative to a (cluster total mass) matched sample of $z\sim0.1$ MCXC BCGs. The average MMCG in Cl1604 was found to be deficient in stellar mass by a factor of 1.78$\pm$0.45 relative to the MMCGs same matched low-redshift sample. Surprisingly, this growth factor is consistent with both the increase in total mass and increase of total stellar baryonic mass of the clusters over the same redshift interval. This result strongly suggested that the growth of the stellar mass of a BCG/MMCG is intimately linked with both the total stellar (contained in galaxies) and dark matter growth of the clusters.

\item \emph{Radial distribution of stellar mass:} A comparison was made between the stellar mass surrounding the BCGs/MMCGs at low and high redshift in the form of companion galaxies. A marked increase of stellar mass at low (projected) radii $R_{\rm{proj}}<0.3R_{vir}$ was observed surrounding the Cl1604 BCGs/MMCGs relative to the MCXC BCGs/MMCGs. Merging timescales were calculated for all companion galaxies to the Cl1604 BCGs/MMCGs that had the possibility of merging within $\la$7 Gyr. Of the 15 merger candidates surrounding the Cl1604 BCGs/MMCGs with small enough merging timescales, 14 would result in a major merging event ($\le$4:1 mass ratio). These potential merging events are primarily comprised of the mixed or wet variety. From these merging events alone, the average Cl1604 BCG will increase in stellar mass by a factor of 2.23$\pm$0.73 and the average Cl1604 MMCG by a factor of 1.35$\pm$0.31 under the assumption of 100\% retention of stellar matter.

\item \emph{Structural parameter evolution}: By fitting the surface brightness profiles of all BCGs/MMCGs in both the high- and low-redshift samples to a single S\'ersic profile, we found an increase of a factor of $\sim 3$ of the size ($r_{e}$) of BCGs/MMCGs over the past $\sim7$ Gyr. The factor of this size increase was invariant with respect to which low-redshift sample we chose to compare the Cl1604 BCGs/MMCGs to. An increase in the average S\'ersic index was also measured over the same redshift range, though its evolution was milder, with an observed increase of $n(z=0)-n(z)\sim0.5$. 

\end{itemize}

From our observational data alone, we strongly favored a scenario in which BCGs/MMCGs grow through a combination of \emph{in situ} star formation and major merging events, the latter likely causing subsequent increases in star formation activity. Though we could not completely rule out the involvement of minor mergers in building up at least a small fraction of the stellar mass of the BCGs/MMCGs over the past $\sim7$ Gyr, the aforementioned scenario is wholly consistent with all of the results in this study. These observational results were then compared to a variety of hydrodynamical simulations and semi-analytic models chosen from the literature. Through these comparisons we found that the observed prevalence of (potentially) impending major merging events amongst the Cl1604 BCGs/MMCGs was sufficient to explain the evolution in the size of BCGs/MMCGs from $z\sim0.9$ to $z\sim0.1$. However, the observed evolution in the average S\'ersic index was not as dramatic as that predicted from a large number of major merging events. In order to explain this mild evolution, we appealed to adiabatic expansion, a process which will serve to soften the evolution of S\'ersic indices of galaxies and a process which naturally follows from the ignition of an AGN as a result of a wet or mixed merging event. 

This study represents one of the most comprehensive studies of the evolution of BCGs/MMCGs over cosmic time to date in terms of the sheer amount of spectroscopic and imaging data utilized for the galaxy populations of clusters at both high and low redshift. From this study we were able to draw a definitive picture of the evolution of BCGs/MMCGs in the Cl1604 supercluster. However, the sample of BCGs/MMCGs used here, especially at high redshift, remains somewhat small, and given the large amount of intrinsic variance amongst galaxy cluster populations observed at all redshifts, it is not entirely clear how applicable this picture is to the evolution of an ``average" BCG over the past $\sim7$ Gyr. It is encouraging that our results, or at least those which can be directly compared, show broad agreement with similar samples of BCGs taken from other surveys. However, future work remains to utilize data from the $\sim50$ high-redshift clusters and groups of the full ORELSE sample, as well as datasets available from other high-redshift cluster surveys (e.g., EDisCS, GCLASS), to determine if the mode of evolution observed here amongst the Cl1604 BCGs/MMCGs is fully representative of that
of typical BCGs observed across the universe.

\section*{Acknowledgments}

We are grateful to the anonymous referee for his/her interesting comments that made us improved the paper significantly. We thank Jeff Newman and Michael Cooper for guidance with the \emph{spec2d} reduction pipeline and for the many useful suggestions and modifications necessary to reduce our DEIMOS data. We also thank the Keck II support astronomers for their dedication, knowledge, and ability to impart that knowledge to us at even the most unreasonable of hours. B.A.\ thanks Mamen Argudo for her help with the SDSS CasJob, Chien Peng for useful suggestions on the surface brightness analysis, Txitxo Ben\'itez for support and suggestions and the Laboratoire d'Astrophysique de Marseille (LAM) for kindly hosting her during her visit. B.C.L.\ thanks Carl Sagan and  C.S.\ Lewis for helping to start this journey of mere observations. B.A. acknowledges the support from Junta de Andaluc\'ia, through the Excellence Project P08-TIC-3531 and the Spanish Ministry for Science and Innovation, through grants  AYA2010-22111-C03-01 and CSD2007-00060. Part of this work was supported by funding from the European Research Council Advanced Grant ERC-2010-AdG-268107-EARLY. In addition, B.C.L.\ acknowledges the support from the National Science Foundation under grant AST-0907858 and by NASA through a grant from the Space Telescope Science Institute (HST-GO-11003), which is operated by the Association of Universities for Research in Astronomy, Incorporated, under NASA contract NAS5-26555.  Portions of this work were based on observations made with the \emph{Spitzer Space Telescope}, which is operated by the Jet Propulsion Laboratory, California Institute of Technology under a contract with NASA. Support for this work was provided by NASA through an award issued by JPL/Caltech.  A part of this work was based on data obtained by the Sloan Digital Sky Survey, which is managed by the Astrophysical Research Consortium for the Participating Institutions. Funding for the SDSS has been provided by the Alfred P.\ Sloan Foundation, the Participating Institutions, the National Science Foundation, the U.S. Department of Energy, the National Aeronautics and Space Administration, the Japanese Monbukagakusho, the Max Planck Society, and the Higher Education Funding Council for England. The SDSS Web Site is http://www.sdss.org/. A portion of the spectrographic data presented herein were obtained at the W.M.\ Keck Observatory, which is operated as a scientific partnership among the California Institute of Technology, the University of  California, and the National Aeronautics and Space Administration. The Observatory was made possible by the generous financial support of the W.M.\ Keck Foundation. We wish to thank the indigenous Hawaiian community for allowing us to be guests on their sacred mountain; we are most fortunate to be able to conduct observations from this site.

\appendix

\section{\normalsize{Dynamical States of the MCXC and Cl1604 Clusters and Groups}}

The nature of the dynamical states of the Cl1604 clusters is somewhat ambiguous. From various lines of evidence, the second most massive cluster in Cl1604 (A) appears largely relaxed \citep{rumbaugh13}, with a galaxy population reflective of a cluster that has fully virialized \citep{lemaux12}. The most massive cluster in Cl1604 (B) shows significant departures from its less massive counterpart, exhibiting both a younger and more active galaxy population and ICM properties which appear inconsistent with a high degree of  virialization, a trend that is continued for the least massive cluster in Cl1604 (D). The dynamical state of the Cl1604 groups is even more ambiguous. With 50ks of ACIS-I observations over the entire supercluster, the \emph{Chandra} observations are simply not deep enough to detect the presence of a hot intragroup  medium (IGM), nor deep enough to put meaningful limits. While we are able to probe the dynamics of the group galaxies, and do not observe any evidence for significant velocity substructure (with the possible exception of Group C), without deep X-Ray observations it is impossible to determine definitively the level of virialization of the Cl1604 groups. 

It has been shown in simulations that massive galaxy clusters which virialize tend to remain in that state, with only minor stochastic fluctuations that cause short-lived departures from virialization (e.g., \citealt{cohn05,arayamelo08,ludlow12}). The situation is, however, much more complicated for lower mass  objects (i.e., the Cl1604 groups), which show a larger spread in dynamical states at every epoch and are easier to significantly disturb due to their shallower potential wells. Though it is likely that the Cl1604 clusters and groups will be largely virialized by $z\sim0.08$ (the mean MCXC cluster redshift), given the inhomogeneity of the Cl1604 sample and our ignorance of their current dynamical states and in the evolution of these states to the redshifts of the MCXC clusters, we required that a fraction of the clusters in the low-redshift sample depart in some way from virialization. In such a way, we can allow for the possibility of at least some fraction of the consitituent clusters and groups of Cl1604 to depart from from virialization as they evolve to the present day. 

Because of the large number of inhomogeneous samples that were drawn upon to create the MCXC catalog, the authors of this catalog chose not to list $T_{X}$ values. In the absence of X-Ray temperatures, we are limited to one point of comparison to infer the level of virialization of the MCXC clusters: the  relationship between the X-Ray luminosity of the ICM and the dynamics of the cluster galaxies. From large samples of local groups and clusters there exists a broad swath of literature discussing the relationship between these two quantities at low redshift, though there has been a dearth of  systematic studies of such structures within the last decade. This relationship is nearly always measured using X-Ray luminosities that are bolometric and are aperture corrected to infinity using a model-dependent correction (typically a $\beta$-model \citealt{cavaliere76,arnaud99}). The X-Ray luminosities available in this study through the MCXC catalog are measured (or modeled) in  a fixed aperture of $R_{500}$ and are measured (or modeled) over an energy range of 0.1-2.4 keV.  While we later adopt the cluster ``self-similarity" relationship of \citet{xue00}, we first discuss the possible implications of the differences in X-Ray luminosity measurements.

Using \emph{XMM} observations of a representative sample of low-redshift clusters from the REXCESS \citep{pratt09} survey, \citet{croston08} measured ICM density profiles for each of the clusters in their sample. In that study, the authors found that the ICM density measured at $R_{500}$ was lower by a factor of $\sim500-1000$ from the value in the inner regions of their clusters, suggesting that the integrated X-Ray flux originating from the ICM at $R<R_{500}$ is an accurate approximation of the total integrated X-Ray output of the ICM. This claim was further supported by \citet{piffaretti11}, in which only minor differences ($\sim5-10$\%) were observed between $L_{X,500}$ and $L_{X}$ measured at much larger radii. Thus, the difference in spatial measurements between the MCXC catalog and $L_{X}$ values used to derive ``self-similar" relations between clusters is negligible. Bolometric corrections to observed X-Ray luminosities are typically realized through the models of \citet{raymond77}. In these models, a large fraction of the luminosity of the ICM ($\ga90$\%) is observed in the rest-frame 0.1-2.4 keV band, essentially independent of temperature for a reasonable range of values. Since the MCXC clusters are at all at reasonably low redshifts, the $k$-correction on $L_{X}$ for any given cluster is small, meaning that a large fraction of the bolometric luminosity is recovered by the MCXC $L_{X}$. Taking both the spatial and energy range differences into account, the difference between the $L_{X,500,0.1-2.4keV}$ measured in the MCXC catalog and the $L_{X,\infty,bol}$ measured in  \citet{xue00} is, at maximum, 20\%, and likely significantly less for most of clusters in the MCXC/SDSS sample. Because we cannot constrain this value for any individual cluster in our sample, we chose to ignore this effect and adopt the self-similar relationship of \citet{xue00} without correction. 

\begin{figure}
\includegraphics[clip,angle=90,width=1.0\columnwidth]{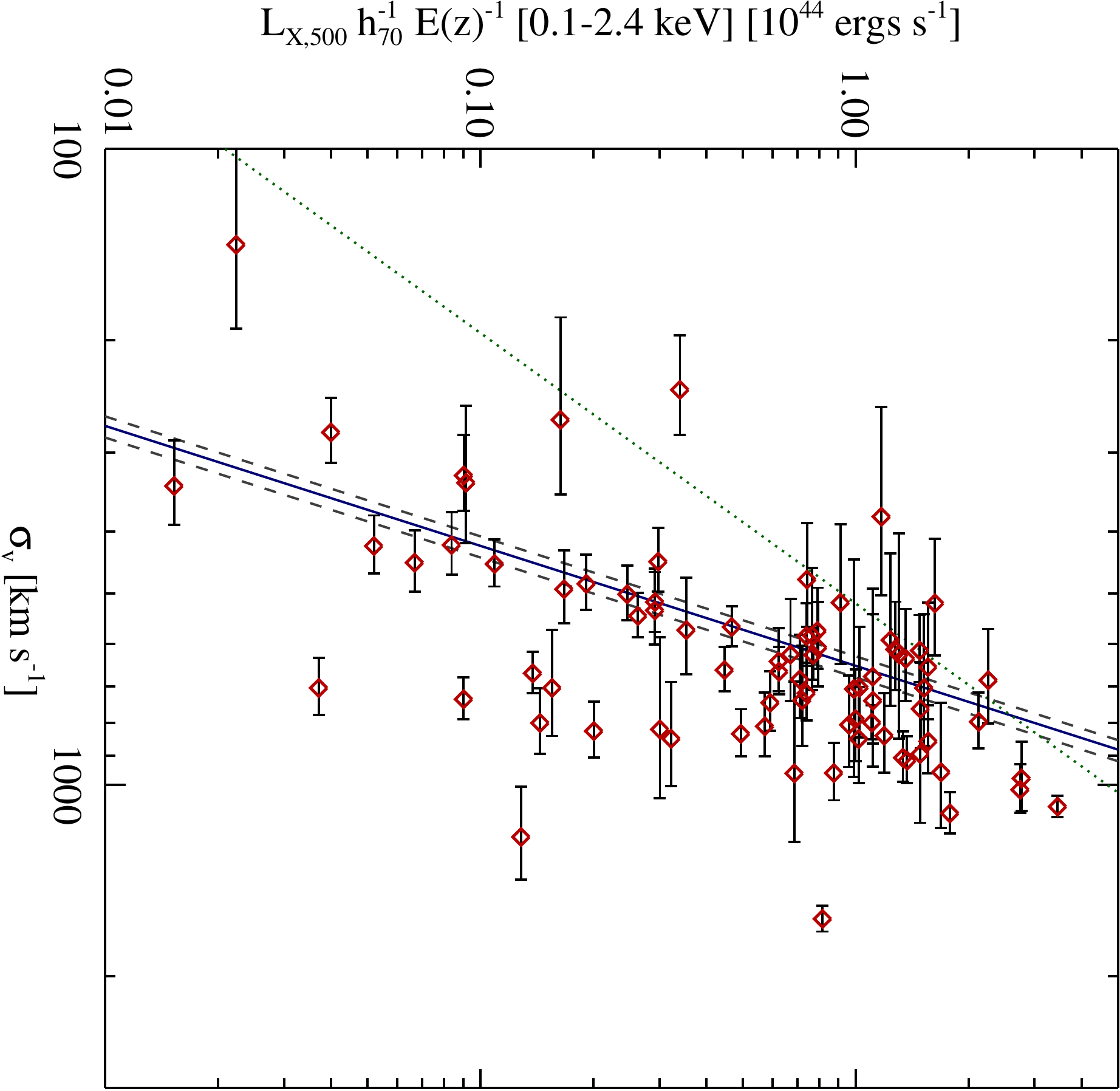}
\caption{Cluster galaxy line-of-sight velocity dispersions for the 81 cluster in the MCXC/SDSS sample with well-measured dispersions plotted against $L_{X,500,0.1-2.4keV}$ drawn from the MCXC catalog. The method used to calculated velocity dispersions and their errors is given in \S\ref{clustermass}. Because of the original inhomogeneity of the X-Ray data used for the MCXC clusters and the complicated method of transformation, errors on $L_{X,500,0.1-2.4keV}$ are not given for each object, though they are roughly $\sim15-20$\% on average. The blue solid line is adapted from \citet{xue00} and shows the expected relation between the two quantities for virialized clusters. The gray dashed lines represent a $\pm$20\% departure from this relationship, which is likely the maximum possible correction to the relationship resulting from the differing methodology used to derive the X-Ray luminosity values of the clusters in \citet{xue00} and those of the MCXC catalog. The green dotted line shows the same relation for groups of galaxies adapted from \citet{xue00}. While a large majority of the MCXC clusters are consistent with virialization, a non-negligible fraction appear to be still in the process of relaxing.}
\label{fig:LXsig}
\end{figure}

In Figure \ref{fig:LXsig} we plot the relationship between $L_{X,500,0.1-2.4keV}$ as given in the MCXC catalog and the galaxy velocity dispersions for the 81 low-redshift clusters with well-measured velocity dispersions (see \S\ref{clustermass}). This is plotted against the backdrop of a best-fit relation for low-redshift virialized clusters \citep{xue00}. Also plotted are lines which show a $\pm$20\% deviation from the best-fit relation, which encompasses the maximum possible fluctuation of an individual MCXC cluster simply resulting from the different methods of calculating $L_{X}$ in \citet{xue00} and \citet{piffaretti11}. In addition, as mentioned in \S\ref{clustermass}, some of the MCXC ``clusters" technically fall under the auspices of the term ``groups". As such, also plotted in Figure \ref{fig:LXsig} is the best-fit relationship determined for groups of galaxies as derived by \citet{xue00}. However, the two relationships (clusters vs. groups) have a high level of similarity in the area of this phase space where the majority of our sample lies. Because of this we do not comment on which of these two relationships  is more appropriate to adopt for each individual MCXC structure other than to say that the group self-similar relationship may be used to explain some of the MCXC structures that deviate significantly from the cluster relationship, structures whose presence in our sample does not significantly affect our analysis regardless  (see, e.g., \S\ref{colormag}). As such, we have simply refered to the MCXC structures as ``clusters" throughout the study and continue to do so here. 

The 81 galaxy clusters plotted in Figure \ref{fig:LXsig} are broadly representative of the 53 MCXC galaxy cluster adopted for use in some analyses in this paper (see \ref{massrad}); a KS test of the X-Ray luminosities of the two sampled show the samples to be statistically indistinguishable. In addition, the fractions presented here are invariant with respect to which sample is chosen. A large fraction of the selected MCXC clusters appear consistent with the \citet{xue00} self-similar relationship. However, a small fraction ($\sim$17\%) of the MCXC clusters deviate significantly ($>3\sigma$) from this relationship, implying that at least some fraction of the sample is still undergoing virialization at $z\sim0.1$. This fraction remains essentially unchanged if we apply the 20\% correction discussed above. The measured fraction is roughly equivalent to the fraction of virialized clusters at $z\sim0$ in halo masses greater than the evolved virial mass of the least massive Cl1604 structure (group I) in simulations (e.g., \citealt{arayamelo08}) and observations of local Abell clusters \citep{popesso07}. In the case of the latter, the data derived from that study were re-analyzed to calculate an equivalent quantity to the one explored here resulting in a fraction of 22.7\%. Thus, it appears that the MCXC/SDSS sample loosely exhibits a representative distribution of dynamical states potential descendants of the Cl1604 clusters and groups. Without further knowledge of the true dynamical states of the Cl1604 clusters and groups and their evolution, it is sufficient for the purposes of this study that the potential descendants of these clusters which serve as the points of comparison for this study span dynamical states which are broadly representative of galaxy clusters in the local universe.

\begin{thebibliography}{}
\bibitem[Araya-Melo(2008)]{arayamelo08} Araya-Melo, P.~A.\ 2008, Ph.D.~Thesis, \url{http://irs.ub.rug.nl/ppn/310289882}
\bibitem[Arnaud \& Evrard(1999)]{arnaud99} Arnaud, M., \& Evrard, A.~E.\ 1999, \mnras, 305, 631 
\bibitem[Ascaso et al.(2011)]{ascaso11} Ascaso, B., Aguerri, J.~A.~L., Varela, J., Cava, A., Bettoni, D., Moles, M., \& D'Onofrio, M.\ 2011, \apj, 726, 69
\bibitem[Balogh et al.(1999)]{balogh99} Balogh, M.~L., Morris, S.~L., Yee, H.~K.~C., Carlberg, R.~G., \& Ellingson, E.\ 1999, \apj, 527, 54
\bibitem[Barden et al.(2008)]{barden08} Barden, M., Jahnke, K.,  {\ H{\"a}uml}u{\ss}ler, B.\ 2008, \apjs, 175, 105 
\bibitem[Behroozi et al.(2010)]{behroozi10} Behroozi, P.~S., Conroy, C., \& Wechsler, R.~H.\ 2010, \apj, 717, 379 
\bibitem[Bernardi et al.(2003)]{bernardi03} Bernardi, M., Sheth, R.~K., Annis, J., et al.\ 2003, \aj, 125, 1817 
\bibitem[Bernardi et al.(2007)]{bernardi07} Bernardi, M., Hyde, J.~B., Sheth, R.~K., Miller, C.~J., \& Nichol, R.~C.\ 2007, \aj, 133, 1741
\bibitem[Bernardi(2009)]{bernardi09} Bernardi, M.\ 2009, \mnras, 395, 1491
\bibitem[Bertin \& Arnouts(1996)]{bertin96} Bertin, E., \& Arnouts, S.\ 1996, \aaps, 117, 393 
\bibitem[Binney \& Tremaine(1987)]{binney87} Binney J., Tremaine S., 1987, Princeton University Press, 747 p.
\bibitem[Biviano et al.(2006)]{biviano06} Biviano, A., Murante, G., Borgani, S., et al.\ 2006, \aap, 456, 23 
\bibitem[Boylan-Kolchin et al.(2009)]{boylan-kolchin09} Boylan-Kolchin, M., Springel, V., White, S.~D.~M., Jenkins, A., \& Lemson, G.\ 2009, \mnras, 398, 1150 
\bibitem[Brinchmann et al.(2004)]{brinchmann04} Brinchmann, J., Charlot, S., White, S.~D.~M., et al.\ 2004, \mnras, 351, 1151 
\bibitem[Brown et al.(2008)]{brown08} Brown, M.~J.~I., Zheng, Z., White, M., et al.\ 2008, \apj, 682, 937
\bibitem[Bruzual \& Charlot(2003)]{bc03} Bruzual, G., \& Charlot, S.\ 2003, \mnras, 344, 1000
\bibitem[Buitrago et al.(2008)]{buitrago08} Buitrago, F., Trujillo, I., Conselice, C.~J., et al.\ 2008, \apjl, 687, L61
\bibitem[Burke \& Collins(2013)]{burke13} Burke, C., \& Collins, C.~A.\ 2013, arXiv:1307.1702 
\bibitem[Calzetti et al.(2000)]{calzetti00} Calzetti, D., Armus, L., Bohlin, R.~C., Kinney, A.~L., Koornneef, J., \& Storchi-Bergmann, T.\ 2000, \apj, 533, 682
\bibitem[Carlberg et al.(1997)]{carlberg97} Carlberg, R.~G., Yee, H.~K.~C., \& Ellingson, E.\ 1997, \apj, 478, 462
\bibitem[Cavaliere \& Fusco-Femiano(1976)]{cavaliere76} Cavaliere, A., \& Fusco-Femiano, R.\ 1976, \aap, 49, 137 
\bibitem[Chabrier(2003)]{chabrier03} Chabrier, G.\ 2003, \pasp, 115, 763
\bibitem[Cohn \& White(2005)]{cohn05} Cohn, J.~D., \& White, M.\ 2005, Astroparticle Physics, 24, 316 
\bibitem[Collins et al.(2009)]{collins09} Collins, C.~A., Stott, J. P., Hilton, M.  et al.\ 2009, \nat, 458, 603 
\bibitem[Conroy et al.(2007)]{conroy07} Conroy, C., Wechsler, R.~H., \& Kravtsov, A.~V.\ 2007, \apj, 668, 826 
\bibitem[Conroy \& Wechsler(2009)]{conroy09} Conroy, C., \& Wechsler, R.~H.\ 2009, \apj, 696, 620
\bibitem[Croston et al.(2008)]{croston08} Croston, J.~H., Pratt, G.~W., B{\"o}hringer, H., et al.\ 2008, \aap, 487, 431 
\bibitem[Daddi et al.(2005)]{daddi05} Daddi, E., Renzini, A., Pirzkal, N., et al.\ 2005, \apj, 626, 680 
\bibitem[Dariush et al.(2007)]{dariush07} Dariush, A., Khosroshahi, H.~G., Ponman, T.~J., et al.\ 2007, \mnras, 382, 433 
\bibitem[Dariush et al.(2010)]{dariush10} Dariush, A.~A., Raychaudhury, S., Ponman, T.~J., et al.\ 2010, \mnras, 405, 1873 
\bibitem[De Lucia \& Blaizot(2007)]{delucia07} De Lucia, G., \& Blaizot, J.\ 2007, \mnras, 375, 2 
\bibitem[Donzelli et al.(2011)]{donzelli11} Donzelli, C.~J., Muriel, H., \& Madrid, J.~P.\ 2011, \apjs, 195, 15
\bibitem[Dressler et al.(2004)]{dressler04} Dressler, A., Oemler, A.~J., Poggianti, B.~M., Smail, I., Trager, S., Shectman, S.~A., Couch, W.~J., \& Ellis, R.~S.\ 2004, \apj, 617, 867
\bibitem[Dubois et al.(2013)]{dubois13} Dubois, Y., Pichon, C., Devriendt, J., et al.\ 2013, \mnras, 428, 2885 
\bibitem[Edge(1991)]{edge91} Edge, A.~C.\ 1991, \mnras, 250, 103 
\bibitem[Edwards \& Patton(2012)]{edwards12} Edwards, L.~O.~V., \& Patton, D.~R.\ 2012, \mnras, 425, 287 
\bibitem[Eisenstein et al.(2003)]{eisenstein03} Eisenstein, D.~J., Hogg, D.~W., Fukugita, M., et al.\ 2003, \apj, 585, 694 
\bibitem[Faber et al.(2003)]{Faber03} Faber, S.~M., et al.\ 2003, \procspie, 4841, 1657
\bibitem[Faber et al.(2007)]{faber07} Faber, S.~M., et al.\ 2007, \apj, 665, 265
\bibitem[Fabian(1994)]{fabian94} Fabian, A.~C.\ 1994, \araa, 32, 277
\bibitem[Fakhouri et al.(2010)]{fakhouri10} Fakhouri, O., Ma, C.-P., \& Boylan-Kolchin, M.\ 2010, \mnras, 406, 2267 
\bibitem[Fan et al.(2008)]{fan08} Fan, L., Lapi, A., De  Zotti, G., \& Danese, L.\ 2008, \apjl, 689, L101 
\bibitem[Fassbender et al.(2011)]{fassbender11} Fassbender, R., B{\"o}hringer, H., Nastasi, A., et al.\ 2011, New Journal of Physics, 13, 125014 
\bibitem[Fazio et al.(2004)]{fazio04} Fazio, G.~G., Hora, J.~L., Allen, L.~E., et al.\ 2004, \apjs, 154, 10 
\bibitem[Fisher et al.(1998)]{fisher98} Fisher, D., Fabricant, D., Franx, M., \& van Dokkum, P.\ 1998, \apj, 498, 195
\bibitem[Ford et al.(1998)]{ford98} Ford, H.~C., et al.\ 1998, \procspie, 3356, 234
\bibitem[Fukugita et al.(1996)]{fukugita96} Fukugita, M., Ichikawa, T., Gunn, J.~E., Doi, M., Shimasaku, K., \& Schneider, D.~P.\ 1996, \aj, 111, 1748
\bibitem[Gal \& Lubin(2004)]{gal04} Gal, R.~R., \& Lubin, L.~M.\ 2004, \apjl, 607, L1 
\bibitem[Gal et al.(2008)]{gal08} Gal, R.~R., Lemaux, B.~C., Lubin, L.~M., Kocevski, D., \& Squires, G.~K.\ 2008, \apj, 684, 933
\bibitem[Gallagher \& Ostriker(1972)]{gallagher72} Gallagher, J.~S., III, \& Ostriker, J.~P.\ 1972, \aj, 77, 288 
\bibitem[Girardi et al.(1993)]{girardi93} Girardi, M., Biviano, A., Giuricin, G., Mardirossian, F., \& Mezzetti, M.\ 1993, \apj, 404, 38 
\bibitem[Gonzalez et al.(2005)]{gonzalez05} Gonzalez, A.~H., Zabludoff, A.~I., \& Zaritsky, D.\ 2005, \apj, 618, 195 
\bibitem[Gonzalez et al.(2007)]{gonzalez07} Gonzalez, A.~H., Zaritsky, D., \& Zabludoff, A.~I.\ 2007, \apj, 666, 147
\bibitem[H{\"a}ussler et al.(2007)]{haussler07} H{\"a}ussler, B., McIntosh, D.~H., Barden, M., et al.\ 2007, \apjs, 172, 615 
\bibitem[Homeier et al.(2006)]{homeier06} Homeier, N.~L., et al.\ 2006, \aj, 131, 143
\bibitem[Hopkins et al.(2008)]{hopkins08} Hopkins, P.~F., Hernquist, L., Cox, T.~J., \& Kere{\v s}, D.\ 2008, \apjs, 175, 356 
\bibitem[Hopkins et al.(2010)]{hopkins10} Hopkins, P.~F., Bundy, K., Hernquist, L. et al.\ 2010, \mnras, 401, 1099 
\bibitem[Huertas-Company et al.(2013)]{huertas-company13} Huertas-Company, M., Mei, S., Shankar, F., et al.\ 2013, \mnras, 428, 1715
\bibitem[Ilbert et al.(2013)]{ilbert13} Ilbert, O., McCracken, H.~J., Le F{\`e}vre, O., et al.\ 2013, \aap, 556, A55 
\bibitem[Jee et al.(2007)]{jee07} Jee, M.~J., Blakeslee, J.~P., Sirianni, M., et al.\ 2007, \pasp, 119, 1403
\bibitem[Jeltema et al.(2007)]{jeltema06} Jeltema, T.~E., Mulchaey, J.~S., Lubin, L.~M., \& Fassnacht, C.~D.\ 2007, \apj, 658, 865 
\bibitem[Jeltema et al.(2008)]{jeltema08} Jeltema, T.~E., Mulchaey, J.~S., \& Lubin, L.~M.\ 2008, \apj, 685, 138 
\bibitem[Jeltema et al.(2009)]{jeltema09} Jeltema, T.~E., Gerke, B.~F., Laird, E.~S., et al.\ 2009, \mnras, 399, 715 
\bibitem[Kaviraj et al.(2008)]{kaviraj08} Kaviraj, S., Khochfar, S., Schawinski, K., et al.\ 2008, \mnras, 388, 67 
\bibitem[Kennicutt et al.(2009)]{kennicutt09} Kennicutt, R.~C., et al.\ 2009, \apj, 703, 1672
\bibitem[Kewley et al.(2004)]{kewley04} Kewley, L.~J., Geller, M.~J., \& Jansen, R.~A.\ 2004, \aj, 127, 2002 
\bibitem[Kocevski et al.(2009a)]{kocevski09a} Kocevski, D.~D., Lubin, L.~M., Gal, R., et al.\ 2009, \apj, 690, 295        
\bibitem[Kocevski et al.(2009b)]{kocevski09b} Kocevski, D.~D., Lubin, L.~M., Lemaux, B.~C., et al.\ 2009, \apj, 700, 901 
\bibitem[Kocevski et al.(2011a)]{kocevski11a} Kocevski, D.~D., Lemaux, B.~C., Lubin, L.~M., et al.\ 2011, \apj, 736, 38 
\bibitem[Kocevski et al.(2011b)]{kocevski11b} Kocevski, D.~D., Lemaux, B.~C., Lubin, L.~M., et al.\ 2011, \apjl, 737, L38 
\bibitem[Krick et al.(2006)]{krick06} Krick, J.~E., Bernstein, R.~A., \& Pimbblet, K.~A.\ 2006, \aj, 131, 168 
\bibitem[Leauthaud et al.(2010)]{leauthaud10} Leauthaud, A., Finoguenov, A., Kneib, J.-P., et al.\ 2010, \apj, 709, 97
\bibitem[Leauthaud et al.(2012)]{leauthaud12} Leauthaud, A., Tinker, J., Bundy, K., et al.\ 2012, \apj, 744, 159 
\bibitem[Lemaux et al.(2009)]{lemaux09} Lemaux, B.~C., et al.\ 2009, \apj, 700, 20 (Lem09)
\bibitem[Lemaux et al.(2010)]{lemaux10} Lemaux, B.~C., Lubin, L.~M., Shapley, A., Kocevski, D., Gal, R.~R., \& Squires, G.~K.\ 2010, \apj, 716, 970 (L10)
\bibitem[Lemaux et al.(2012)]{lemaux12} Lemaux, B.~C., Gal, R.~R., Lubin, L.~M., et al.\ 2012, \apj, 745, 106 
\bibitem[Lemaux et al.(2013)]{lemaux13} Lemaux, B.~C., et al.\ 2013, \emph{in prep}
\bibitem[Lidman et al.(2012)]{lidman12} Lidman, C., Suherli, J., Muzzin, A., et al.\ 2012, \mnras, 427, 550
\bibitem[Lidman et al.(2013)]{lidman13} Lidman, C., Iacobuta, G., Bauer, A.~E., et al.\ 2013, \mnras, 433, 825 
\bibitem[Lin et al.(2008)]{lin08} Lin, L., Patton, D.~R., Koo, D.~C., et al.\ 2008, \apj, 681, 232 
\bibitem[Lin et al.(2013)]{lin13} Lin, Y.-T., Brodwin, M., Gonzalez, A.~H., et al.\ 2013, \apj, 771, 61 
\bibitem[Liu et al.(2009)]{liu09} Liu, F.~S., Mao, S., Deng, Z.~G. et al.\ 2009, \mnras, 396, 2003 
\bibitem[Liu et al.(2012)]{liu12} Liu, F.~S., Mao, S., \& Meng, X.~M.\ 2012, \mnras, 423, 422 
\bibitem[Liu et al.(2013)]{liu13} Liu, F.~S., Guo, Y., Koo, D.~C., et al.\ 2013, \apj, 769, 147
\bibitem[L{\'o}pez-Sanjuan et al.(2012)]{lopez-sanjuan12} L{\'o}pez-Sanjuan, C., Le F{\`e}vre, O., Ilbert, O., et al.\ 2012, \aap, 548, A7
\bibitem[Lubin et al.(2000)]{lubin00} Lubin, L.~M., Brunner, R., Metzger, M.~R., Postman, M., \& Oke, J.~B.\ 2000, \apjl, 531, L5 
\bibitem[Lubin et al.(2009)]{lubin09} Lubin, L.~M., Gal, R.~R., Lemaux, B.~C., Kocevski, D.~D., \& Squires, G.~K.\ 2009, \aj, 137, 4867 
\bibitem[Ludlow et al.(2012)]{ludlow12} Ludlow, A.~D., Navarro, J.~F., Li, M., et al.\ 2012, \mnras, 427, 1322 
\bibitem[Magnelli et al.(2013)]{magnelli13} Magnelli, B., Popesso, P., Berta, S., et al.\ 2013, \aap, 553, A132
\bibitem[Maraston(2005)]{maraston05} Maraston, C.\ 2005, \mnras, 362, 799
\bibitem[Mei et al.(2009)]{mei09} Mei, S., et al.\ 2009, \apj, 690, 42
\bibitem[Merritt(1985)]{merritt85} Merritt, D.\ 1985, \apj, 289, 18 
\bibitem[Milosavljevi{\'c} et al.(2006)]{milosavljevic06} Milosavljevi{\'c}, M., Miller, C.~J., Furlanetto, S.~R., \& Cooray, A.\ 2006, \apjl, 637, L9 
\bibitem[Moran et al.(2007)]{moran07} Moran, S.~M., Ellis, R.~S., Treu, T., et al.\ 2007, \apj, 671, 1503 
\bibitem[Moster et al.(2010)]{moster10} Moster, B.~P., Somerville, R.~S., Maulbetsch, C., et al.\ 2010, \apj, 710, 903 
\bibitem[Mulchaey(2000)]{mulchaey00} Mulchaey, J.~S.\ 2000, \araa, 38, 289 
\bibitem[Murante et al.(2007)]{murante07} Murante, G., Giovalli, M., Gerhard, O., et al.\ 2007, \mnras, 377, 2 
\bibitem[Nelson et al.(2002)]{nelson02} Nelson, A.~E., Simard, L., Zaritsky, D. et al.\ 2002, \apj, 567, 144 
\bibitem[Oemler et al.(2013)]{oemler13} Oemler, A., Jr., Dressler, A., Gladders, M.~G., et al.\ 2013, \apj, 770, 61
\bibitem[Oke \& Gunn(1983)]{oke83} Oke, J.~B., \& Gunn, J.~E.\ 1983, \apj, 266, 713
\bibitem[Oke et al.(1995)]{oke95} Oke, J.~B., et al.\ 1995, \pasp, 107, 375
\bibitem[Oke et al.(1998)]{oke98} Oke, J.~B., Postman, M., \& Lubin, L.~M.\ 1998, \aj, 116, 549 
\bibitem[Osmond \& Ponman(2004)]{osmond04} Osmond, J.~P.~F., \& Ponman, T.~J.\ 2004, \mnras, 350, 1511 
\bibitem[Ostriker \& Tremaine(1975)]{ostriker75} Ostriker, J.~P., \& Tremaine, S.~D.\ 1975, \apjl, 202, L113
\bibitem[Peng et al.(2002)]{peng02} Peng, C.~Y., Ho, L.~C., Impey, C.~D., \& Rix, H.-W.\ 2002, \aj, 124, 266 
\bibitem[Peng et al.(2010)]{peng10} Peng, C.~Y., Ho, L.~C., Impey, C.~D., \& Rix, H.-W.\ 2010, \aj, 139, 2097 
\bibitem[Peterson \& Fabian(2006)]{peterson06} Peterson, J.~R., \& Fabian, A.~C.\ 2006, \physrep, 427, 1
\bibitem[Piffaretti et al.(2011)]{piffaretti11} Piffaretti, R., Arnaud, M., Pratt, G.~W., Pointecouteau, E., \& Melin, J.-B.\ 2011, \aap, 534, A109 
\bibitem[Pipino et al.(2009)]{pipino09} Pipino, A., Kaviraj, S., Bildfell, C., et al.\ 2009, \mnras, 395, 462 
\bibitem[Poggianti et al.(2009)]{poggianti09} Poggianti, B.~M., et al.\ 2009, \apj, 693, 112
\bibitem[Popesso et al.(2007)]{popesso07} Popesso, P., Biviano, A., B{\"o}hringer, H., \& Romaniello, M.\ 2007, \aap, 461, 397 
\bibitem[Postman et al.(2012)]{postman12} Postman, M., Lauer, T.~R., Donahue, M., et al.\ 2012, \apj, 756, 159
\bibitem[Pratt et al.(2009)]{pratt09} Pratt, G.~W., Croston, J.~H., Arnaud, M., B\rm{$\ddot{o}$}umlhringer, H.\ 2009, \aap, 498, 361 
\bibitem[Raymond \& Smith(1977)]{raymond77} Raymond, J.~C., \& Smith, B.~W.\ 1977, \apjs, 35, 419 
\bibitem[Rieke et al.(2004)]{rieke04} Rieke, G.~H., et al.\ 2004, \apjs, 154, 25
\bibitem[Rudick et al.(2011)]{rudick11} Rudick, C.~S., Mihos, J.~C., \& McBride, C.~K.\ 2011, \apj, 732, 48 
\bibitem[Rumbaugh et al.(2012)]{rumbaugh12} Rumbaugh, N., Kocevski, D.~D., Gal, R.~R., et al.\ 2012, \apj, 746, 155 
\bibitem[Rumbaugh et al.(2013)]{rumbaugh13} Rumbaugh, N., Kocevski, D.~D., Gal, R.~R., et al.\ 2013, \apj, 763, 124 
\bibitem[Ruszkowski \& Springel(2009)]{ruszkowski09} Ruszkowski, M., \& Springel, V.\ 2009, \apj, 696, 1094  
\bibitem[Ryan et al.(2012)]{ryan12} Ryan, R.~E., Jr., McCarthy, P.~J., Cohen, S.~H., et al.\ 2012, \apj, 749, 53 
\bibitem[Salim et al.(2007)]{salim07} Salim, S., Rich, R.~M., Charlot, S., et al.\ 2007, \apjs, 173, 267 
\bibitem[Salpeter(1955)]{salpeter55} Salpeter, E.~E.\ 1955, \apj, 121, 161
\bibitem[Seigar et al.(2007)]{seigar07} Seigar, M.~S., Graham, A.~W., \& Jerjen, H.\ 2007, \mnras, 378, 1575 
\bibitem[Sersic(1968)]{sersic68} Sersic, J.~L.\ 1968, Cordoba, Argentina: Observatorio Astronomico, 1968,  
\bibitem[Shankar et al.(2013)]{shankar13} Shankar, F., Marulli, F., Bernardi, M., et al.\ 2013, \mnras, 428, 109 
\bibitem[Smith et al.(2010)]{smith10} Smith, G.~P., Khosroshahi, H.~G., Dariush, A., et al.\ 2010, \mnras, 409, 169 
\bibitem[Springel(2005)]{springel05} Springel, V.\ 2005, \mnras, 364, 1105 
\bibitem[Stott et al.(2011)]{stott11} Stott, J.~P., Collins, C.~A., Burke, C., Hamilton-Morris, V., \& Smith, G.~P.\ 2011, \mnras, 414, 445 
\bibitem[Stoughton et al.(2002)]{stoughton02} Stoughton, C., Lupton, R.~H., Bernardi, M., et al.\ 2002, \aj, 123, 485
\bibitem[Tonini et al.(2012)]{tonini12} Tonini, C., Bernyk, M., Croton, D., Maraston, C., \& Thomas, D.\ 2012, \apj, 759, 43 
\bibitem[Tremonti et al.(2004)]{tremonti04} Tremonti, C.~A., Heckman, T.~M., Kauffmann, G., et al.\ 2004, \apj, 613, 898 
\bibitem[Trujillo et al.(2006)]{trujillo06} Trujillo, I., F{\"o}rster Schreiber, N.~M., Rudnick, G., et al.\ 2006, \apj, 650, 18 
\bibitem[Valentinuzzi et al.(2010)]{valentinuzzi10} Valentinuzzi, T., Poggianti, B.~M., Saglia, R.~P., et al.\ 2010, \apjl, 721, L19 
\bibitem[van Dokkum et al.(2010)]{vandokkum10} van Dokkum, P.G., Whitaker, K.E., Brammer, G., et al.\ 2010, \apj, 709, 1018
\bibitem[Vikram et al.(2009)]{vikram09} Vikram, V., Wadadekar, Y., Kembhavi, A.~K. et al.\ 2009, \mnras, L355
\bibitem[von der Linden et al.(2007)]{vonderlinden07} von der Linden, A., Best, P.~N., Kauffmann, G., \& White, S.~D.~M.\ 2007, \mnras, 379, 867
\bibitem[Wen \& Han(2011)]{wen11} Wen, Z.~L., \& Han, J.~L.\ 2011, \apj, 734, 68
\bibitem[Whiley et al.(2008)]{whiley08} Whiley, I.~M., Arag{\'o}n-Salamanca, A., De Lucia, G., et al.\ 2008, \mnras, 387, 1253
\bibitem[White et al.(2005)]{white05} White, S.~D.~M., Clowe, D.~I., Simard, L., et al.\ 2005, \aap, 444, 365 
\bibitem[Wu et al.(2013)]{pofeng13} Wu, P-F. et al.\ 2013, \emph{in prep}
\bibitem[Xue \& Wu(2000)]{xue00} Xue, Y.-J., \& Wu, X.-P. 2000, \apj, 538, 65
\bibitem[Yan et al.(2006)]{yan06} Yan, R., Newman, J.~A., Faber, S.~M., Konidaris, N., Koo, D., \& Davis, M.\ 2006, \apj, 648, 281 (Y06)
\bibitem[Zabludoff \& Mulchaey(1998)]{zabludoff98} Zabludoff, A.~I., \& Mulchaey, J.~S.\ 1998, \apj, 496, 39 

\end{thebibliography}
\end{document}